\documentclass[11pt,headings=big,numbers=noenddot,DIV=14,a4paper]{article}%

\pdfoutput=1

\usepackage[margin=1 in]{geometry}

\usepackage{graphicx}  
\usepackage{dcolumn}   
\usepackage{bm,relsize}        
\usepackage{amsfonts,amsmath,amssymb,amsthm,mathtools}
\usepackage{slashed}
\usepackage{lipsum, color}
\usepackage[usenames,dvipsnames,svgnames,table]{xcolor}
\usepackage[linktoc=page,bookmarks=false,colorlinks=false,linkbordercolor=RoyalBlue,citebordercolor=ForestGreen,urlbordercolor=CornflowerBlue]{hyperref}
\usepackage[sort&compress,numbers,merge]{natbib}

\def\beq{\begin{equation}}
\def\eeq{\end{equation}}
\def\bea{\begin{eqnarray}}
\def\eea{\end{eqnarray}}

\def\FO			{{\mathsmaller{\rm FO}}}
\def\FD			{F_{\mathsmaller{D}}}
\def\FY			{F_{\mathsmaller{Y}}}
\def\fV			{f_{\mathsmaller{V}}}
\def\GammaV		{\Gamma_{\mathsmaller{V}}}
\def\tauV		{\tau_{\mathsmaller{V}}}
\def\mV			{m_{\mathsmaller{V}}}
\def\MDM		{M_{\mathsmaller{\rm DM}}}
\def\MDMSM		{M_{\mathsmaller{\rm DM}}^{\mathsmaller{\rm SM}}}
\def\Muni		{M_{\rm uni}}
\def\MuniSM		{M_{\rm uni}^{\mathsmaller{\rm SM}}}
\def\MuniSMs	{M_{\rm uni,0}^{\mathsmaller{\rm SM}}}
\def\MPl		{M_{\mathsmaller{\rm Pl}}}
\def\mZ			{m_{\mathsmaller{Z}}}

\def\gD			{g_{\mathsmaller{D}}}
\def\gSM		{g_{\mathsmaller{\rm SM}}}

\def\gSMdec		{g_{\mathsmaller{\rm SM}}^{\rm dec}}

\def\xfo		{x^\FO}
\def\xfoSM		{x^\FO_{\mathsmaller{\rm SM}} }
\def\sSM		{s_{\mathsmaller{\rm SM}}}
\def\SSM		{S_{\mathsmaller{\rm SM}}}
\def\DSM		{D_{\mathsmaller{\rm SM}}}
\def\DSMbar		{\bar{D}_{\mathsmaller{\rm SM}}}
\def\TD			{T_{\mathsmaller{D}}}
\def\TSM		{T_{\mathsmaller{\rm SM}}}
\def\TSMdecay		{T_{\mathsmaller{\rm SM}}^{\mathsmaller{\rm decay}}}
\def\TtildeSM	{\widetilde{T}_{\mathsmaller{\rm SM}}}

\def\rtilde		{\tilde{r}}
\def\gtildeD	{\tilde{g}_{\mathsmaller{D}}}
\def\gtildeSM	{\tilde{g}_{\mathsmaller{\rm SM}}}

\def\aD			{\alpha_{\mathsmaller{D}}}

\def\aSM        {\alpha_{\mathsmaller{\rm SM}}}
\def\ann		{{\rm ann}}
\def\BSF		{\mathsmaller{\rm BSF}}

\def\vrel       {v_{\mathsmaller{\rm rel}}}

\begin{document}

\begin{flushright}
\footnotesize
DESY 18-200\\
Nikhef-2018-055
\end{flushright}
\color{black}

\begin{center}

{\LARGE \bf Homeopathic Dark Matter\\
\vspace{.3 cm}
\large or how diluted heavy substances produce high energy cosmic rays
}

\medskip
\bigskip\color{black}\vspace{0.5cm}

{
{\large Marco Cirelli}$^a$,
{\large Yann Gouttenoire}$^{a,b}$,
{\large Kalliopi Petraki}$^{a,c}$,
{\large Filippo Sala}$^b$
}
\\[7mm]

{\it \small $^a$ LPTHE,  CNRS \& Sorbonne Universit\'e, 4 Place Jussieu, F-75252, Paris, France}\\
{\it \small $^b$ DESY, Notkestra{\ss}e 85, D-22607 Hamburg, Germany}\\
{\it \small $^c$ Nikhef, Science Park 105, 1098 XG Amsterdam, The Netherlands}\\
\end{center}

\bigskip

\centerline{\bf Abstract}
\begin{quote}
\color{black}

We point out that current and planned telescopes have the potential of probing 
annihilating Dark Matter (DM) with a mass of $O(100)$ TeV and beyond.
As a target for such searches, we propose models where DM annihilates into lighter mediators, themselves decaying into Standard Model (SM) particles.
These models allow to reliably compute the energy spectra of the SM final states, and to naturally evade the unitarity bound on the DM mass.
Indeed, long-lived mediators may cause an early matter-dominated phase in the evolution of the Universe and, upon decaying, dilute the density of preexisting relics thus allowing for very large DM masses.
We compute this dilution in detail and provide results in a ready-to-use form.
Considering for concreteness a model of dark $U(1)$ DM, we then study both dilution and the signals at various high energy telescopes observing $\gamma$ rays, neutrinos and charged cosmic rays.
This study enriches the physics case of these experiments,
and opens a new observational window on heavy new physics sectors.
\end{quote}

\clearpage
\noindent\makebox[\linewidth]{\rule{\textwidth}{1pt}} 
\tableofcontents
\noindent\makebox[\linewidth]{\rule{\textwidth}{1pt}}

\section{Introduction}

The much anticipated physics beyond the Standard Model (BSM) has so far not been discovered at experiments, that have then pushed the scale of many related theories beyond the TeV range.
How can we gain an experimental access to such energy scales?
High energy cosmic rays currently constitute a privileged access to that realm, as opposed to the more subtle one offered by various precision measurements.
Many telescopes are indeed now observing such cosmic rays~({\sc Hess~II, Hawc, Veritas, Magic, Taiga, Antares, IceCube, Baikal, Ams, Calet, Dampe}), and many are planned for the near future~({\sc Cta, Lhaaso, Km3net, Herd, Iss-Cream}).
Using this data optimally to advance our knowledge of fundamental physics is therefore of pressing importance.

Dark matter (DM) is arguably the strongest evidence for BSM physics that can be potentially probed by these observations.
New physics (NP) sectors at scales beyond 10-100 TeV, addressing other open questions in our understanding of Nature, have long been known to 
provide DM candidates in that same mass range, see for example~\cite{Dimopoulos:1996gy,Antipin:2014qva}. Moreover, 
the empty-handed searches for weakly interacting massive particles (WIMPs) 
motivate exploring new regimes of WIMP models, including DM masses larger than about 10 TeV, see e.g.~\cite{Antipin:2015xia,DelNobile:2015bqo}.
But masses much larger than this are challenged, in the wide class of DM models where thermal freeze-out sets the relic abundance, by the so-called unitarity bound~\cite{Griest:1989wd}: the DM annihilation cross section is bounded from above by the unitarity of the $S$ matrix, and this translates into an upper bound on the DM mass of $O(100)$~TeV, see e.g.~\cite{vonHarling:2014kha,Baldes:2017gzw}.

However, the injection of entropy in the SM bath after DM freeze-out can dilute its abundance, and thus open the possibility of obtaining thermal relic DM beyond the unitarity bound. Such a dilution of relics takes place in well motivated NP models, see e.g.~\cite{deCarlos:1993wie,Banks:1993en,Moroi:1999zb}. Its consequences for DM have been studied since a long time~\cite{McDonald:1989jd}, also in relation with heavy DM~\cite{Giudice:2000ex} (in the rest of this paper, `heavy DM' will be used as a synonym of  DM with a mass at and beyond 10-100 TeV). Recent years have seen a growing interest in cosmological histories that also yield to the dilution of standard relic DM, e.g. due to late entropy  injection~\cite{Patwardhan:2015kga,Berlin:2016vnh,Berlin:2016gtr,Cirelli:2016rnw,Bramante:2017obj,Hamdan:2017psw}, late DM reannihilations~\cite{Binder:2017lkj,Geller:2018biy}, or both~\cite{Mitridate:2017oky}.

These DM scenarios offer therefore an important physics goal for high energy cosmic ray telescopes, which could test BSM physics at an energy scale where no other experiment has direct access. 
However, in the standard WIMP-inspired DM scenarios, this is complicated by the treatment of electroweak (EW) radiation at large DM masses. The SM spectra from DM annihilations are indeed governed by an expansion in $\sim \alpha_w/\pi \, \log^2(\MDM/m_W)$, that should be resummed when this number is order one (analogoulsy to what is done for QCD radiation), which happens at $\MDM \approx 100$ TeV.
At present this resummation constitutes a technical challenge, and limits the reliability of the interpretation of high energy cosmic ray data in terms of DM.\footnote{
For example, two of the tools often used for this purpose, {\sc Pppc4dmid}~\cite{Cirelli:2010xx} and Pythia~\cite{Sjostrand:2014zea}, respectively include EW radiation only at first order~\cite{Ciafaloni:2010ti} and lack radiation processes among gauge bosons like $W^* \to W \gamma$~\cite{Christiansen:2014kba}, and thus cannot be completely trusted for heavy DM.}

\medskip

As pointed out recently in~\cite{Berlin:2016vnh,Berlin:2016gtr,Cirelli:2016rnw}, a class of DM models that allows to evade the unitarity limit consists of DM annihilating into mediators belonging to a dark sector that themselves decay into SM particles. If their lifetime is long enough and they are sufficiently heavy, the mediators may temporarily dominate the energy density of the Universe, so that, when they eventually decay, they inject significant entropy in the SM bath and thus dilute any relic, including DM. A frozen-out overabundant DM can then be diluted to the observed density. This allows for a smaller cross section at freeze-out time, and therefore a larger DM mass, relaxing the unitarity limit.

We observe that the same class of models circumvents also the challenge of reliably computing the SM spectra from heavy DM annihilations. Indeed, the energy scale relevant for computing the SM spectra is now the mass of the mediator $\mV$, rather than $\MDM$. For mediators lighter enough than $100$~TeV, $\alpha_w/\pi \log^2(\mV/m_W) \ll 1$, so one can use the spectra with EW corrections computed as in~\cite{Cirelli:2010xx,Ciafaloni:2010ti} and then boost them from the mediator rest frame to the DM one. 

From the two points of view we just discussed, these \textit{secluded DM} models appear to constitute an ideal target for telescopes observing high energy cosmic rays, and would allow them to reliably test for the first time annihilating DM beyond 100 TeV. However, to our knowledge,
only two experimental analyses of secluded DM models have been performed, which are specific to neutrinos from DM in the Sun and reach DM masses up to 10~TeV~\cite{Adrian-Martinez:2016ujo,Ardid:2017lry}.
Are current and planned telescopes sensitive to DM models of this kind?
In this paper we show this to be the case, thus enriching the physics case of these experiments, and opening a new window on BSM theories at the high energy frontier.

The discussion is organized as follows. In section~\ref{sec:dilution} we present a quantitative model-independent study of dilution in secluded DM models, and of the limits from Big-Bang Nucleosynthesis (BBN).
For concreteness, we then consider a specific model of DM charged under a dark $U(1)$, which we discuss in section~\ref{sec:U1model}.
We study its signals in section~\ref{sec:pheno}, where we analyse the constraints from observations of CMB, 21cm, neutrinos, gamma rays, antiprotons, electrons and positrons. In section~\ref{sec:outlook} we summarise our results and indicate possible future directions.

\section{Heavy secluded DM and the unitarity limit}
\label{sec:dilution}

In this section we study the dilution of relics in secluded DM models, taking into account both the entropy injection from late decays of the mediator and the effect of different temperatures of the SM and the dark sector -- i.e. the plasma of DM and the particles into which it annihilates.

In secluded DM models, the DM particles $\left( X,{\bar X} \right)$ annihilate dominantly into mediators $V$. For simplicity we will work in the case where only one mediator species exists.
Like in the usual narrative, when $\TD \simeq \MDM/25$, where $\TD$ is the temperature of the dark sector, the DM annihilation rate becomes smaller than the expansion rate of the Universe, and the comoving DM number density freezes-out. At that same time, the comoving number density of the mediators also freezes out, since the annihilation and pair creation processes  $X{\bar X} \leftrightarrow VV$ are the only significant $V$ number changing interactions.

The dark mediators are assumed to interact with the SM such that they can decay into SM particles. 
However, if they are sufficiently heavy and long-lived, they can become non-relativistic and dominate the energy density of the Universe as a cold relic {\it before} they decay, because the ratio of the energy density of a non-relativistic species over the energy density of the SM radiation increases linearly with the scale factor. In this case, when the mediators eventually decay into relativistic SM particles, the entropy generation leads to a large dilution of the comoving DM number density, as we describe
in subsection~\ref{sec:dilution_DarkU1}.\footnote{The large dilution that is inherent in these DM scenarios leads us to dub them `Homeopathic Dark Matter', hence the title of this work.
We are relieved that our findings (see subsections~\ref{sec:max_homeopathy} and \ref{sec:UniBound}) show that, unlike in non-allopathic medicine~\cite{homeopathy}, there is a built-in limit to homeopathy in these scenarios.
}

The entropy injected by the mediator decay depends of course on the temperature of the dark sector. It is unknown whether the SM and the dark plasma were in thermal equilibrium at high temperatures or not.
Our discussion here will encompass both possibilities, and only assume that the dark sector and the SM were {\it not} in kinetic equilibrium when the SM plasma had a temperature $\TSM < \TtildeSM$.\footnote{
$\TtildeSM \gtrsim$~few~MeV could be justified by the requirement to not ruin BBN, depending on the specific model. It is not our purpose here to elaborate on this observation.}
The subsequent evolution of $\TD$ and $\TSM$ is controlled by the conservation of the entropy density of the two sectors separately; their ratio reads
\begin{equation}
\label{eq:r_ratio_decay}
r \equiv \frac{\TD}{\TSM}
=\left( \frac{\gSM}{\gD}
\frac{\gtildeD}{\gtildeSM}\right)^{\!\!\frac{1}{3}}\,
\rtilde\,,
\end{equation}
where $\gSM$ and $\gD$ are the numbers of relativistic degrees of freedom in the SM and in the dark sector, and the $\sim$ on top of a symbol denotes that the corresponding quantity is evaluated at $\TSM = \TtildeSM$.
$\rtilde = 1$ corresponds to the case where the SM and the dark plasma were in thermal equilibrium for $\TSM > \TtildeSM$.

\subsection{Dilution by entropy injection}
\label{sec:dilution_DarkU1}

We first review the dilution via entropy injection. When a cold relic (in our case, the mediator) decays into SM particles, its energy density is transferred to the SM bath. The non-relativistic degrees of freedom of the mediator are converted into relativistic degrees of freedom of the SM. This heat production increases the total entropy of the Universe and dilutes the frozen-out comoving DM number density $n_{X}/s_{\rm tot}$.
We shall use the superscripts `before' and `after' to denote the times before $V$ decays but after DM freezes-out ($t \ll \tauV$),  and after $V$ decays ($t \gtrsim \tauV$), respectively. 
Then the comoving DM density today $n^{0}_{X}/s_{\rm tot}^{0}$ is related to the comoving DM density at freeze-out $n^\FO_{X}/s^\FO_{\rm tot}$ through\footnote{
As is conventional, the uppercase letters $S$ and $N$ refer to the comoving entropy and number densities whereas the lowercase $s$ and $n$ refers to the entropy and number densities, such that $S=s a^{3}$ and $N=n a^{3}$ where $a$ is the scale factor.
}
\beq
\frac{n^{0}_{X}}{s_{\rm tot}^{0}}
= \frac{n^{0}_{X}}{\sSM^{0}}
= \frac{N^{\rm after}_{X}}{\SSM^{\rm after}} 
= \frac{1}{\DSM}\frac{N^{\rm before}_{X}}{\SSM^{\rm before}} 
=   \frac{1}{\DSM}  \frac{n^\FO_{X}}{\sSM^\FO} 
=   \frac{1}{D}  \frac{n^\FO_{X}}{s_{\rm tot}^\FO} \,,
\label{eq:dilution_factor}
\eeq
where we used $N_X^{\rm after} = N_X^{\rm before}$ and defined the dilution factors
\begin{equation}
\label{eq:dilution_factor_definition}
\DSM \equiv \SSM^{\rm after}/\SSM^{\rm before}
\qquad \text{and} \qquad
D \equiv \SSM^{\rm after}/S_{\rm tot}^{\rm before} \,,
\end{equation}
which are of course related via
\begin{equation}
D = \frac{\DSM}{1+ (\gtildeD/\gtildeSM) \rtilde^3} \,. 
\label{eq:D_and_DSM}
\end{equation}
$D = 1$ corresponds to no dilution of the DM density. We report below expressions for $\DSM$ because they are simpler, and refer to appendix~\ref{app:dilution_DP_relativistic} for more details on their relation.

\paragraph{The dilution factor.}
The dilution factor has first been computed in \cite{Scherrer:1984fd}. Under the assumption that the dark mediators have become non-relativistic before they decay, it reads
\begin{equation}
\label{eq:dilution_factor_expression}
\DSM = \left(1 + 
\frac{4}{3}
\frac{\fV \mV}{(\gSM^{\rm before})^{1/3} \, \TSM^{\rm before}}
\int_{x^{\rm before}}^{x^{\rm after}} dx 
\, \gSM^{1/3}(x) \, z(x) \, e^{-x}  
\right)^{3/4},
\end{equation}
where $x=t/\tauV$ is the time in units of the mediator lifetime and $z(x)=a(x)/a^{\rm before}$ is the normalized scale factor.
$\fV$ is the ratio of the mediator number density over the SM entropy density before decay
\begin{equation}
\fV \equiv \frac{n_V^{\rm  before}}{\sSM^{\rm before}}
= \frac{45 \, \zeta(3)}{2\pi^{4}}\frac{\gD^{\rm before}}{\gSM^{\rm before}} r_{\rm before}^{3}
= \frac{45 \, \zeta(3)}{2\pi^{4}} \frac{\gtildeD}{\gtildeSM} \, \rtilde^3 \simeq 0.0169 \left(\frac{\gtildeD}{6.5}\right)  
\rtilde^3  \,,
\label{eq:fV}
\end{equation}
where in the second equality we assumed that the mediators were still relativistic at DM freeze-out\footnote{
When the mediators are instead non-relativistic at the time of freeze-out, their number density is exponentially suppressed, and in turn also the dilution. We include this effect in all our results, see Appendix~\ref{app:dilution_DP_non_relativistic_FO} for more details. \label{footnote:DPnrFO}
},
and in the third equality we have used eq.~\eqref{eq:r_ratio_decay}.
In the fourth and last equality of eq.~(\ref{eq:fV}),  for definiteness we have assumed $\TtildeSM$ large enough that $\gtildeSM = 106.75$, and normalised to the value of $\gtildeD$ of the concrete model we will study in the next sections.\footnote{Here and in the following we work with the number of relativistic degrees of freedom in energy equal to that in entropy, which is an excellent approximation for the temperatures we are interested in.}

To compute the dilution factor from  eq.~(\ref{eq:dilution_factor_expression}), we first solve the Friedmann equation for $z(x)$, in presence both of the SM radiation and of the mediator decaying into SM radiation. Since $z(x) \, e^{-x}$ is maximal at $x \approx 1$, and $\gSM^{1/3}$ is a slow-varying function, we define 
$\gSMdec =\gSM (x=1)$ 
and use the excellent approximation 
$\int dx\,\gSM^{1/3}(x) \, z(x) \, e^{-x} \simeq (\gSMdec)^{1/3} \int  dx\,z(x) \, e^{-x}$.
We refer the reader to appendix~\ref{app:dilution} for more details of our computation, and in particular for a quantitative appraisal of the 
goodness of this and other approximations that are found in the literature.
We report here an expression for the dilution factor 
($\MPl \simeq 2.4\times10^{18}$ GeV, $\GammaV=1/\tauV$)
\begin{equation}
\DSM = 
\left[ 1 + 0.77 \; (\gSMdec)^{1/3} \; \fV^{4/3} 
\left(\frac{ \mV^2 }{\GammaV\,\MPl}\right)^{\!\! \frac{2}{3}} \right]^{\!\! \frac{3}{4}},
\label{eq:dilution_factor_expression_VD_dom_approx}
\end{equation}
which assumes that the mediator dominates the energy density of the Universe when it starts to decay, thus setting $z \propto x^{2/3}$ for a matter-dominated Universe.
This is the case in all the parameter space where dilution is important, and yields the expression~(\ref{eq:dilution_factor_expression_VD_dom_approx}) that is both very precise (see appendix~\ref{app:dilution}) and ready-to-use in 
any situation of dilution from decay of relics.

\paragraph{The effect of the dark sector temperature on the dilution.}
From eq.~(\ref{eq:dilution_factor_expression_VD_dom_approx}) we see that for a given mass and lifetime of the mediator, the dependence of the dilution factor on the thermodynamics of the dark sector -- its degrees of freedom and its temperature -- is dominantly encoded in $\fV$. We neglect the dependence on $\rtilde$ introduced by $\gSMdec$, because it is much milder than that arising from $\fV$, and because here we do not aim at an extremely precise study but rather at a simple analytical understanding.
We may then rewrite eq.~(\ref{eq:dilution_factor_expression_VD_dom_approx}) as
\beq
\DSM \simeq \left[1 + \left(\DSMbar^{4/3}-1 \right) \, \rtilde^4 
\right]^{3/4},
\label{eq:DSM_rtilde}
\eeq
where $\DSMbar$ encapsulates all the dependence on the model parameters $\mV$, $\GammaV$, $\gtildeD$ and $\gtildeSM$, %
\begin{equation}
\DSMbar \equiv 
\left[1+ 
\left(0.23 
\:(\gSMdec)^{1/4} 
\: \frac{\gtildeD}{\gtildeSM} 
\frac{\mV}{\sqrt{\GammaV \MPl}}
\right)^{4/3}
\right]^{3/4}
\,,
\label{eq:DSM_rtilde1}
\end{equation}
where $\DSMbar = \DSM^{\rtilde=1}$. For large values of $\DSMbar$ and/or the temperature ratio $\tilde{r}$, eq.~\eqref{eq:DSM_rtilde} implies the simple scaling 
$\DSM \propto \rtilde^3$,  with $D$ consequently saturating to a constant value, $D \simeq (\gtildeSM/\gtildeD) \DSMbar \simeq 0.23 (\gSMdec)^{1/4}  \mV  /  \sqrt{\GammaV \MPl}$ [cf.~eq.~\eqref{eq:D_and_DSM}].

\paragraph{Cross-section.} Ultimately, we are interested in the couplings needed to  reproduce the correct relic abundance, since they determine the DM signals. Let $\langle \sigma \vrel \rangle^\FO$ be the thermally averaged effective annihilation cross-section times relative velocity around the time of freeze-out, which is of course determined by these couplings.\footnote{
We use the specification ``$\FO$" because if $\sigma \vrel$ depends on $\vrel$, as is the case with Sommerfeld enhanced processes that are relevant for heavy DM, then its value during freeze-out is of course different than at later times, when the DM indirect detection signals are generated. Moreover, we use the term ``effective annihilation cross-section" because the DM density may be depleted via processes other than direct annihilation -- in particular, the formation and decay of unstable bound states -- as we shall see in section~\ref{sec:U1model}.}  
Large dilution implies that a smaller $\langle \sigma \vrel \rangle^\FO$ is needed. 
From eq.~(\ref{eq:DSM_rtilde}), it is evident that the dilution is larger the warmer the dark sector is, since more energy is stored into the mediators and transferred to the SM once they decay. However, $r_\FO \neq 1$ also affects $\langle \sigma \vrel \rangle^\FO$ directly, i.e.~even in the absence of any significant dilution.  We shall now incorporate this effect.

In the presence of a dark plasma, the Hubble parameter is
$ H/H_\mathsmaller{\rm SM} = \sqrt{1+ (\gD / \gSM) \,r^4}$, 
where $H_\mathsmaller{\rm SM}$ is the Hubble parameter in the absence of the dark sector. The comoving DM energy density today is
\beq
\frac{\MDM n_X^0}{\sSM^0} 
= \frac{\MDM}{\DSM} \frac{n_X^\FO}{\sSM^\FO }
\simeq  
\frac{\MDM }{\DSM \, \sSM^\FO} \ 
\frac{H^\FO}{\langle \sigma \vrel \rangle^\FO} 
=
\frac{\MDM  H_{\mathsmaller{\rm SM}}^\FO}{\sSM^\FO}
\frac{\sqrt{1+ (\gD^\FO / \gSM^\FO) \,r_\FO^4}}{\DSM} 
\frac{1}{\langle \sigma \vrel \rangle^\FO}
\,,
\eeq
where in the first step we used eq.~\eqref{eq:dilution_factor}, and in the second step we used the approximate freeze-out condition 
$n_X^\FO \langle \sigma \vrel \rangle^\FO \simeq H^\FO$. The combination 
$\MDM H_{\mathsmaller{\rm SM}}^\FO / \sSM^\FO$ is proportional to 
$\MDM /\TSM = r_\FO x^\FO$, where $x^\FO \equiv \MDM/\TD^\FO$ (not to be confused with $x=t/\tauV$ defined earlier) itself depends on $r_\FO$. From the above, we deduce that 
\beq
\frac
{\langle \sigma \vrel \rangle^\FO} 
{\langle \sigma \vrel \rangle^\FO_{\mathsmaller{\rm SM}}}
\simeq 
\frac{\xfo}{\xfoSM}\,
\frac{r_\FO \sqrt{1+ (\gD^\FO/\gSM^\FO)r_\FO^4}}{\DSM} \,,
\label{eq:sigmav_fo}
\eeq
where $\xfoSM$ and $\langle \sigma \vrel \rangle^\FO_{\mathsmaller{\rm SM}}$ denote respectively the time parameter $\xfo$ and the cross-section needed to establish the observed DM density if DM were annihilating directly into SM particles (in which case we identify $\TD$ with $\TSM$ in the definition of $\xfo$). $\DSM$ can be computed from  eqs.~\eqref{eq:DSM_rtilde} and \eqref{eq:DSM_rtilde1}, and the dark-to-SM temperature ratio at freeze-out $r_\FO$ is related to $\rtilde$ via eq.~\eqref{eq:r_ratio_decay}. 
Note that eq.~\eqref{eq:sigmav_fo} holds even when comparing the required cross-sections for DM with different masses, $\MDM$ and $\MDMSM$, where $\MDMSM$ is the mass of the DM particle in the standard (no dilution) case. 
Following the standard treatment\footnote{\label{foot:xfo}
For $\langle \sigma \vrel \rangle	= \sigma_0 x^{1/2}$, 
the freeze-out temperature can be estimated by 
(see e.g.~\cite{Kolb:1990vq}, and ref.~\cite{Baldes:2017gzw} for generalization to dark sector freeze-out)
\beq 
\xfo \simeq \ln \left[
0.095
\frac{g_{\mathsmaller{\rm DM}}}{\sqrt{\gSM^\FO}} 
\frac{ r_\FO^2 }{ \sqrt{1+ (\gD^\FO/\gSM^\FO) r_\FO^4}}
M_{\rm Pl} \MDM \sigma_0 
\right]  \,,
\nonumber
\eeq
where $g_{\mathsmaller{\rm DM}}$ are the DM degrees of freedom and $M_{\rm Pl}$ is the reduced Planck mass.
\label{footnote:xfoKolbTurner}
}
and using eq.~\eqref{eq:sigmav_fo}, we find 
\beq
\xfo - \xfoSM  \simeq
\frac{1}{2}\ln \left( \xfo / \xfoSM \right)
+ \ln (\MDM/\MDMSM)
- \ln \DSM 
+ 3 \ln r_\FO\,,
\label{eq:xfo}
\eeq
where we assumed that the effective annihilation cross-section scales as $\sigma \vrel \propto 1/\vrel$ around the time of freeze-out, as is the case for Sommerfeld-enhanced processes and the unitarity bound (cf.~sec.~\ref{sec:UniBound}). Any departure from this scaling results only in logarithmic corrections to $\xfo$.
Note finally that eq.~(\ref{eq:xfo}) assumes the DM was non-relativistic during freeze-out, therefore it becomes inconsistent for extremely large dilution, $\ln(\DSM/r_\FO^3) =O(\xfoSM)$.

\begin{figure}
\centering
\includegraphics[height=0.45\textwidth]{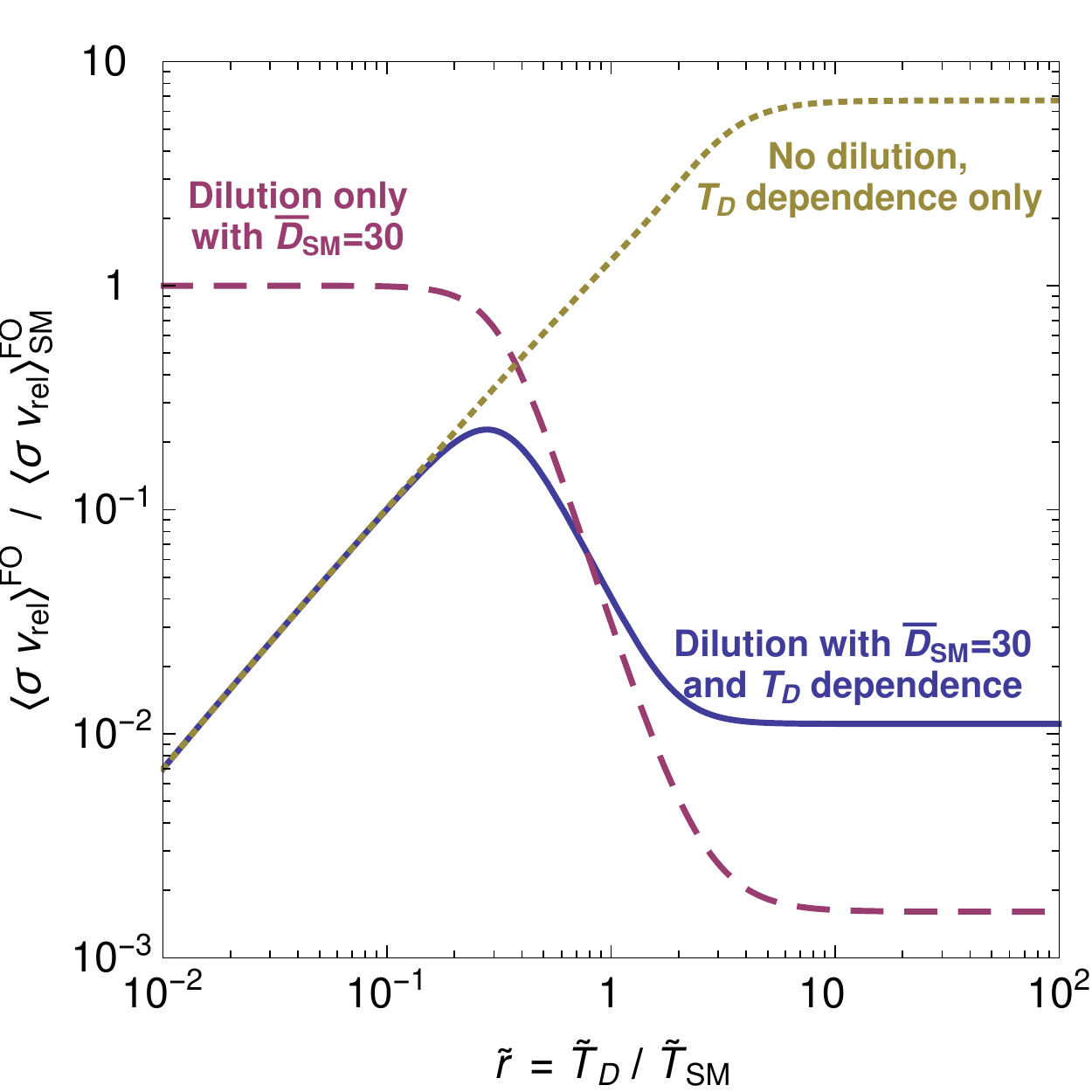}
~~~
\includegraphics[height=0.45\textwidth]{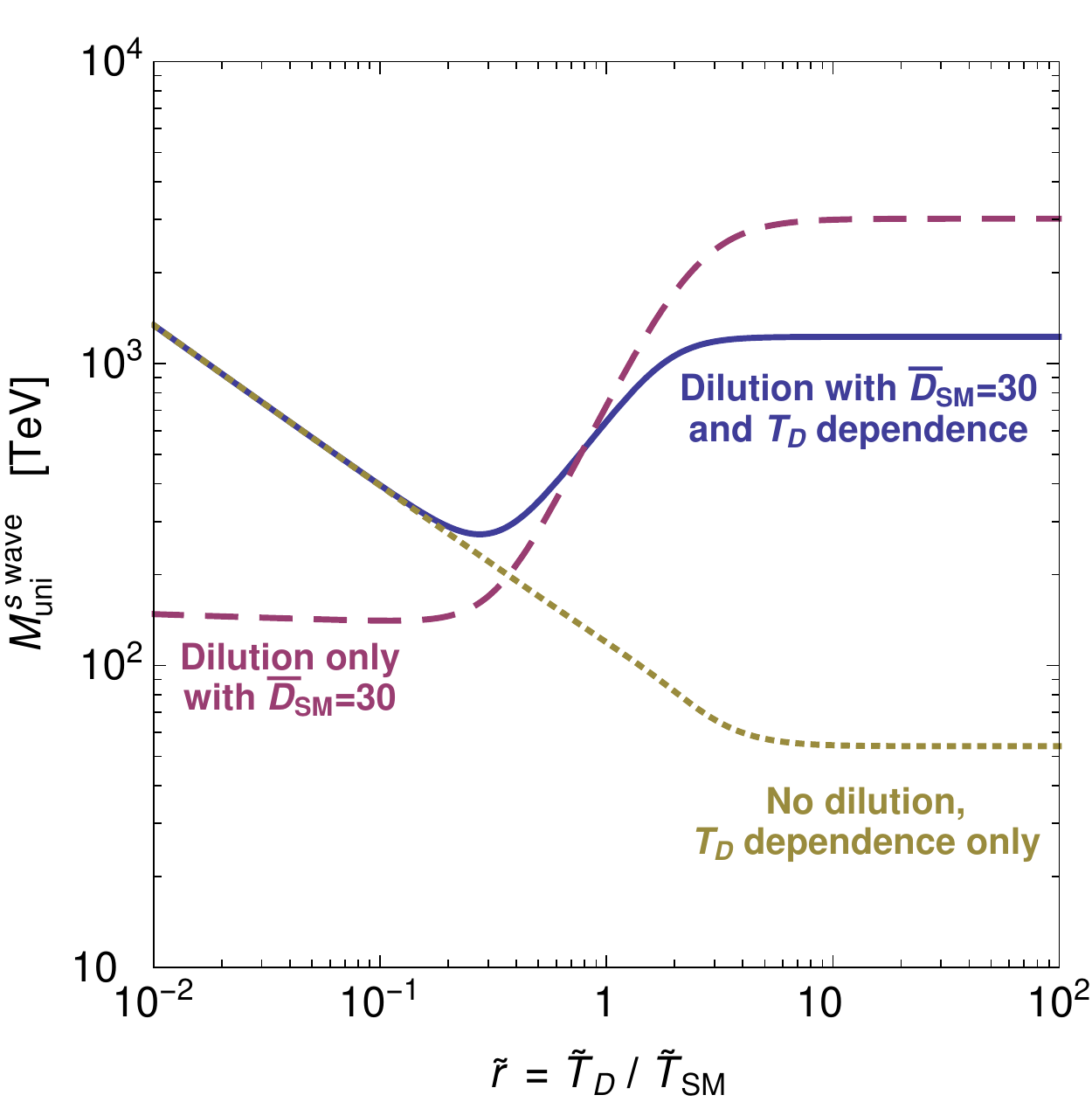}
\caption{\it \small \label{fig:sigmaFOandMuni} \it \small
Dependence of the effective annihilation cross-section at freeze-out required to attain the observed DM density (left) and of the $s$-wave unitarity limit on the mass of thermal relic DM (right), on the dark-to-SM temperature ratio and on the dilution due to the decay of the dark mediators. 
$\langle \sigma \vrel\rangle^\FO$ is normalized to the corresponding value for DM of the same mass annihilating directly into SM particles with no dilution occurring, $\langle \sigma \vrel\rangle_{\mathsmaller{\rm SM}}^\FO$.  The continuous blue lines display the combined effect of the dilution and the temperature ratio.
To ease its understanding we also show the effect without dilution, namely $D=1$ (dotted yellow), and of the dilution only, namely we retain the dependence on $\rtilde$ of $\langle \sigma \vrel\rangle_{\mathsmaller{\rm SM}}^\FO$ (dashed purple).
Note that higher partial waves may contribute significantly to the depletion of DM in the early Universe, thereby raising the unitarity bound on the DM mass. 
For definiteness, we have used $\gtildeSM = 106.75$, $\gtildeD = 6.5$, $\gD^\FO =3$ (cf.~sec.~\ref{sec:U1model}) and for the left panel we assumed $\xfoSM =29$.
}
\end{figure}

Let us now discuss the features of eq.~\eqref{eq:sigmav_fo}. For small $\rtilde$, the dilution factor is close to unity and eq.~\eqref{eq:sigmav_fo} implies $\langle \sigma \vrel \rangle_\FO \propto r_\FO$. 
This is expected since for smaller $\tilde{r}$ the DM freeze-out is happening earlier, thus $\TSM^\FO$ is larger. This implies that
\begin{itemize}
\item[$\circ$] the dilution due to the subsequent Hubble expansion is larger, so that the cross-section must be reduced by $r_\FO^3$ in order to compensate for the deficit in the DM abundance,
\item[$\circ$] the Hubble expansion rate during freeze-out is faster, so that the cross-section must be enhanced by $r_\FO^{-2}$ in order to keep $\TD^\FO$ and then the DM abundance fixed.
\end{itemize} 
As a result $\langle \sigma \vrel \rangle_\FO \propto r_\FO$.
For larger $\rtilde$, the dilution factor becomes more significant, thus suppressing $\langle \sigma \vrel \rangle_\FO$. However, for sufficiently large $\rtilde$, the dark sector energy density dominates the Universe and the Hubble expansion rate becomes independent of $r_\FO$, resulting in the suppression of $\langle \sigma \vrel \rangle_\FO$ to saturate to the value
\beq
\left[\frac
{\langle \sigma \vrel \rangle^\FO} 
{\langle \sigma \vrel \rangle^\FO_{\mathsmaller{\rm SM}}}
\right]_{\min} 
\simeq
\frac{\xfo}{\xfoSM}
\frac{1}{\DSMbar} \, \frac{\gtildeD/\gtildeSM}{\sqrt{\gD^\FO/\gSM^\FO}}
\simeq
4.35~\frac{\xfo}{\xfoSM}
\frac{\sqrt{\GammaV \MPl}}{\mV}
\frac{(\gSM^\FO/\gD^\FO)^{1/2}}{(\gSMdec)^{1/4}} \,.
\label{eq:sigmaFO_Min}
\eeq
The interplay between $\rtilde$ and $\DSMbar$ in determining $\langle \sigma \vrel \rangle^\FO$ is depicted in fig.~\ref{fig:sigmaFOandMuni} (left panel).

\subsection{A limit to `homeopathy': on the maximal dilution in secluded models}
\label{sec:max_homeopathy}

\begin{figure}[!t]
\begin{center}
\includegraphics[width=0.48\textwidth]{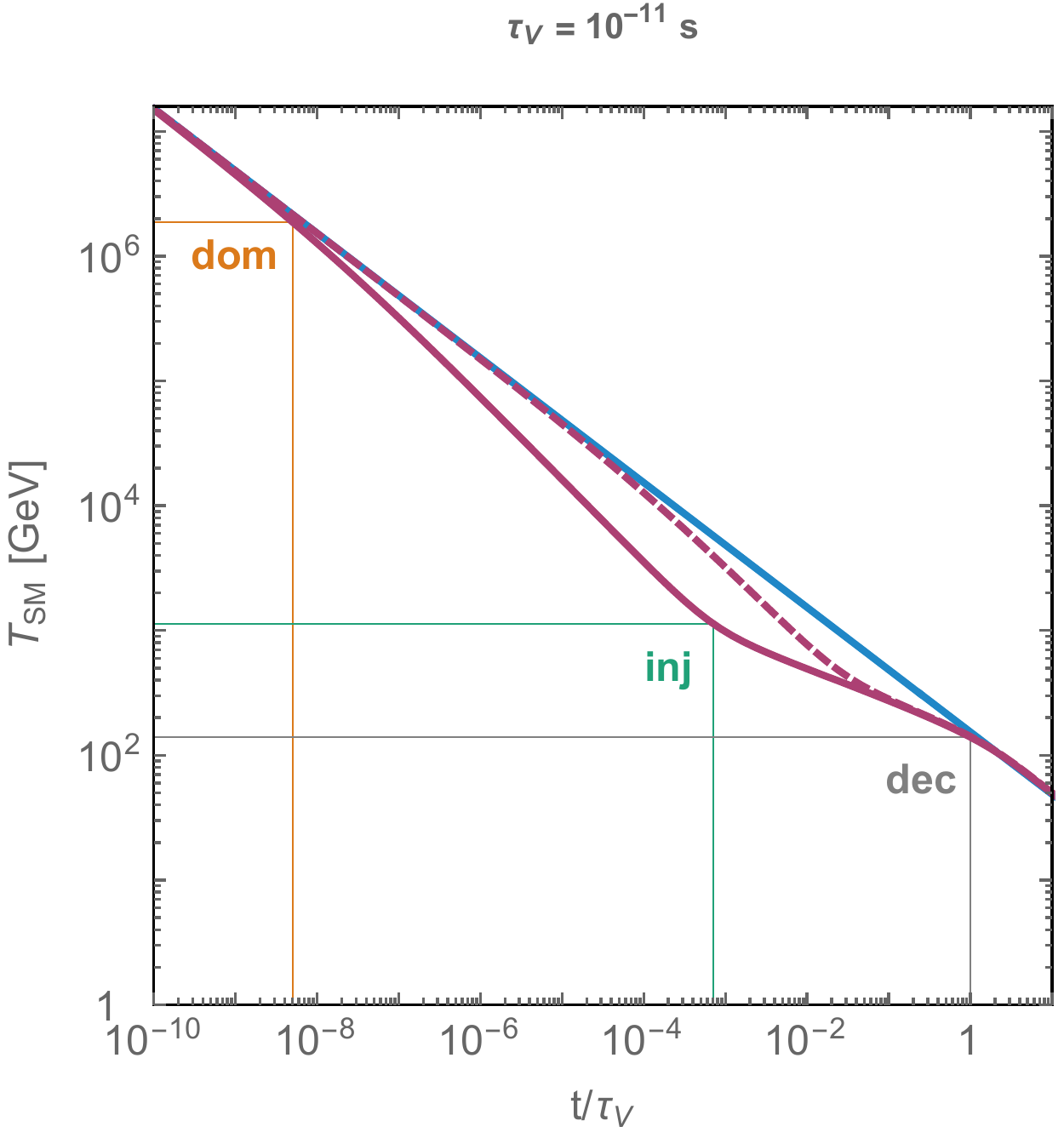}
\includegraphics[width=0.48 \textwidth]{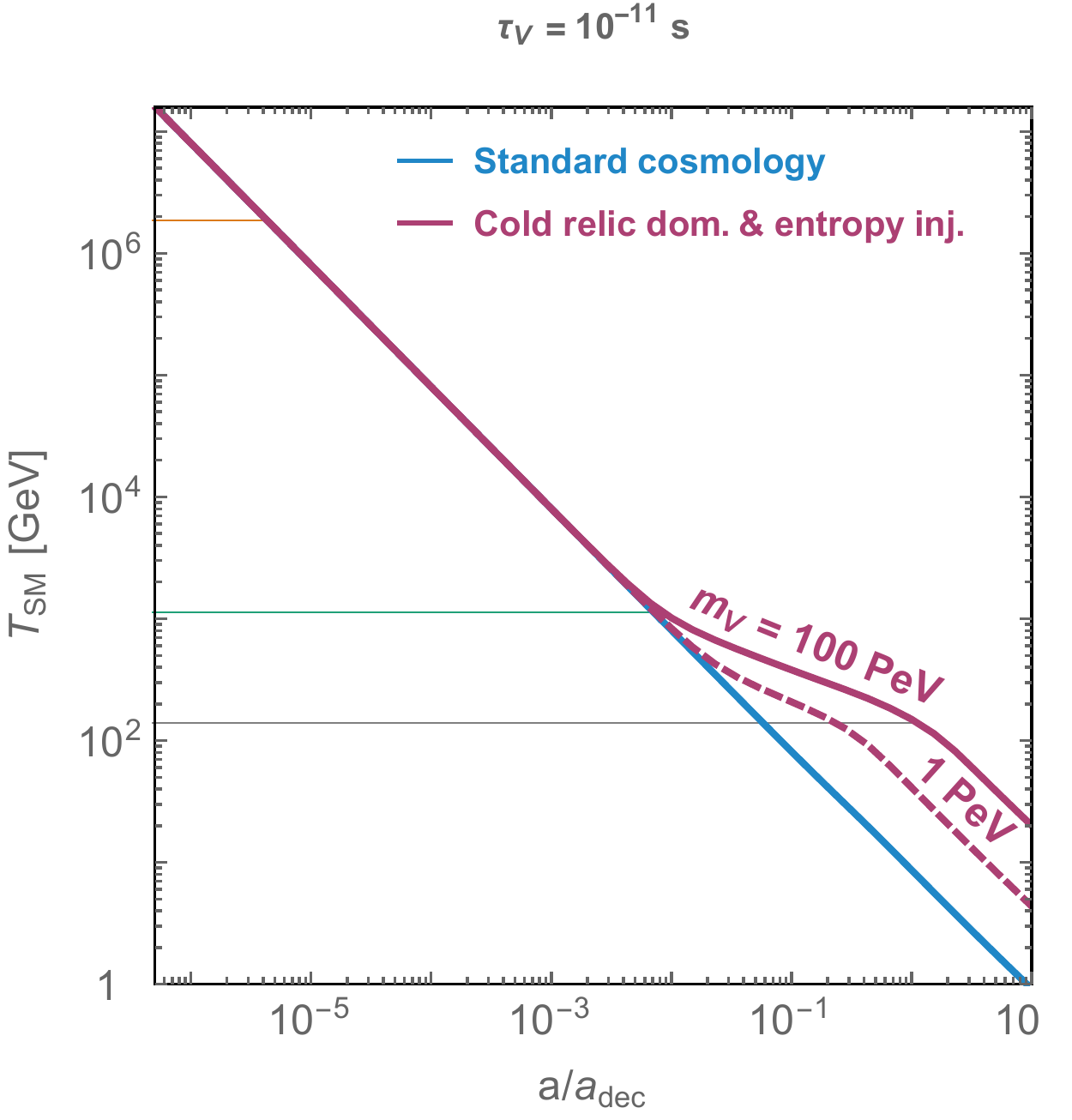}
\end{center}
\caption{\it \small
Evolution of the SM temperature with time (left) and with scale factor (right), for $\mV = 100~\text{PeV}$ (continuous purple) and 1~PeV (dashed purple), computed after integrating the full Friedmann equation~\eqref{eq:Friedmann_eq}. The evolution of the SM temperature in the absence of dilution is shown as a continuous blue line for comparison. 
Just after the cold relic dominates the energy density of the universe (`dom'), the temperature decreases with time as $T \propto t^{-2/3}$, i.e. faster than in a radiation-dominated Universe for which $T \propto t^{-1/2}$.
However, when entropy injection starts to become significant (`inj'),
the temperature decreases more slowly $T \propto t^{-1/4}$.
The faster redshift during the period of matter domination is counterbalanced by the slower redshift during the period of entropy injection such that the temperature of the SM at the time of decay (`dec') is independent of the mass of the relic and is the same as in standard cosmology.
Concerning the scale factor, the temperature starts to depart from the standard evolution,  $T \propto a^{-3/8}$ instead of $T \propto a^{-1}$, only after the entropy injection has started.
The three moments described above, `dom' `dec' and `inj', are indicated for the case $\mV=100$~PeV with thin lines, obtained analytically as described in the text.
}
\label{fig:temperature_evolution}
\end{figure}

In order to preserve the light element abundance, the mediator must decay before BBN starts. This places an upper limit on the possible dilution. The study in \cite{Jedamzik:2006xz} shows that the lifetime of an abundant particle, decaying $30\%$ or more hadronically, must be smaller than $0.03$~s.
In order to incorporate the modification of the cosmology due to the mediator dominating the energy density of the universe and injecting entropy, we rewrite the upper bound on the lifetime, $\tauV < 0.03$ s, as a lower bound on the SM temperature at the time of decay, $\TSMdecay > 5$ MeV. Therefore, we shall now describe the evolution of the SM temperature in our non-standard cosmology.

The evolution of the SM temperature $\TSM(z)$ in presence of the cold relic is driven by a combination of redshift and entropy injection
\begin{equation}
\label{eq:T_vs_z}
\TSM(z) = \frac{\TSM^{\rm before}}{z}  \left( \frac{\gSM(\TSM^{\rm before})}{\gSM\left(\TSM(z)\right)} \right)^{\!1/3}  \left(  \frac{\SSM(x)}{\SSM^{\rm before}}  \right)^{\!1/3},
\end{equation}
where $z(x) \equiv a(x)/a^{\rm before}$, $x=t/\tauV$ and~\cite{Scherrer:1984fd}
\begin{equation}
\label{eq:entropy_ratio}
\frac{\SSM(x)}{\SSM^{\rm before}} = \left(1 + 
\frac{4}{3}
\frac{\fV \mV}{(\gSM^{\rm before})^{1/3} \, \TSM^{\rm before}}
\int_{0}^{x} d\tilde{x} 
\, \gSM^{1/3}(\tilde{x} ) \, z(\tilde{x} ) \, e^{-\tilde{x} }  
\right)^{3/4}.
\end{equation}
Integrating 
the Friedmann equation (see eq.~\eqref{eq:Friedmann_eq} in Appendix~\ref{app:dilution}) assuming that the cold relic dominates the universe energy density, gives
\begin{equation}
\label{eq:z_vs_x_matter_domination}
z(x)\simeq\left( \frac{3}{2}\frac{x}{x_\text{before}}\right)^{\!2/3},
\end{equation}
where $x_\text{before} = \GammaV \sqrt{3\MPl^{2}/\rho_V^{\rm before}}$ and $\rho_V^{\rm before} = \mV \fV (2\pi^{2}/45) \gSM^{\rm before}T_{\rm SM, before}^{3}$ is the energy density in the mediators before decay.
For $x \leq 1$, for simplicity we use $e^{-\tilde{x}} \simeq 1$ in eq.~\eqref{eq:entropy_ratio}.
Using eq.~\eqref{eq:z_vs_x_matter_domination}, eq.~\eqref{eq:T_vs_z} then becomes
\begin{equation}
\label{eq:T_vs_x}
\TSM(x) \simeq 1.25\,\sqrt{\MPl\GammaV}\;\frac{(\gSMdec)^{\!1/12}}{\gSM^{\!1/3}}
\frac{x_\text{inj}^{5/12}}{x^{2/3}}\left( 1 + \Big( \frac{x}{x_{\rm inj}} \Big)^{\! 5/3}  \right)^{\!1/4} ,
\end{equation}
where 
\begin{equation}
x_\text{inj} \simeq 1.43 \left(\frac{\MPl^2}{\gSMdec \, \fV^{4} \, \mV^4 \, \tauV^2} \right)^{1/5}.
\end{equation}
$x_\text{inj}$ corresponds to the time where entropy injection becomes significant, which we define as $\SSM(x_\text{inj})/\SSM^{\rm before} = 2^{3/4}$, see eq.~\eqref{eq:entropy_ratio}. Note that eq.~\eqref{eq:T_vs_x} implies $\TSM \propto x^{-2/3}$ for $x \ll x_\text{inj}$, as expected for a matter-dominated Universe, and $\TSM \propto x^{-1/4}$ for $x \gg x_\text{inj}$, as expected for a matter-dominated Universe during a period of entropy dilution (in which $T \propto a^{-3/8}$).
We emphasise that, in deriving eq.~\eqref{eq:T_vs_x}, we have assumed a matter dominated universe. That is an excellent approximation for $x \ll 1$. However, around $x\simeq 1$, the radiation energy content of the universe is non-negligible, and therefore our result for the temperature at time of decay, $\TSMdecay$, is a priori less accurate.
We also determine the temperature evolution by solving the Friedmann equations (with decay) numerically, and show our results in fig.~\ref{fig:temperature_evolution}.
Our analytical determination of $x_{\rm inj}$ and $\TSM^{\rm inj}$ reproduces very well the point where the numerical solution changes slope, as also shown in fig.~\ref{fig:temperature_evolution}.
Setting $x=1$ in eq.~\eqref{eq:T_vs_x} and assuming a significant entropy injection $x_{\rm inj}\ll1$, we get the well known temperature below which standard cosmology is recovered~\cite{Scherrer:1984fd}
\begin{equation}
\label{eq:TSMdecay}
\TSMdecay \approx 1.25 \frac{(\gSMdec)^{\!1/12}}{\gSM^{\!1/3}}\sqrt{\MPl\GammaV}.
\end{equation}
We first note that $\TSMdecay$ reproduces fairly well its numerical determination, see fig.~\ref{fig:temperature_evolution}.
$\TSMdecay$ is independent of the mass of the relic (except via the dependence of $\GammaV$ on $\mV$), its number of degrees of freedom or its initial temperature. %
Furthermore, under the very good approximation where $\gSMdec=\gSM$ (see Appendix~\ref{app:dilution}), $\TSMdecay$  is almost equal to the temperature of a radiation-dominated universe with age $\tauV$ (set by $H(\TSM)=1/2\tauV$),
$
T_{\mathsmaller{\rm rad}}(\tauV) \simeq  1.23 \sqrt{\MPl\GammaV}/\gSM^{1/4}
$.\footnote{This corroborates the validity of the approximation, common in the literature, where it is assumed that the cold relic decays instantaneously when $H = \Gamma_{V}$.}
The net result is that if a cold relic has dominated the energy density of the universe and has subsequently injected entropy during its decay, then the temperature of the universe at decay time, $\TSMdecay$, corresponds roughly to that in the standard cosmology (namely without early matter domination nor entropy injection) at time $t=\tauV$. The only difference with respect to the standard case is that the universe has expanded more: this is the dilution effect.

Therefore, for the purpose of setting BBN constraints, we impose $\tauV <0.03$ s for all values of the mass $\mV$, i.e. even for those leading to matter domination and entropy injection.
By putting an upper limit on the lifetime of the mediator, BBN then constrains the maximal dilution to be (see eqs~\eqref{eq:D_and_DSM},\eqref{eq:DSM_rtilde},\eqref{eq:DSM_rtilde1})
\begin{equation}
D^\text{max} \simeq 5.6 \times 10^{9}~\frac{\mV}{100~\text{PeV}} ~ \left( \frac{\tauV}{0.03~s  } \right)^{1/2},
\label{eq:DSM_Max}
\end{equation}
independent of $\rtilde$ and $\gtildeD$.
As we will show in the next subsection, the unitarity of DM annihilations effectively imposes an upper limit on $\mV$, and in turn on~$D$.

\subsection{Impact on unitary bound  \label{sec:UniBound}}

\begin{figure}[t!]
\begin{center}
\includegraphics[width=0.48\textwidth]{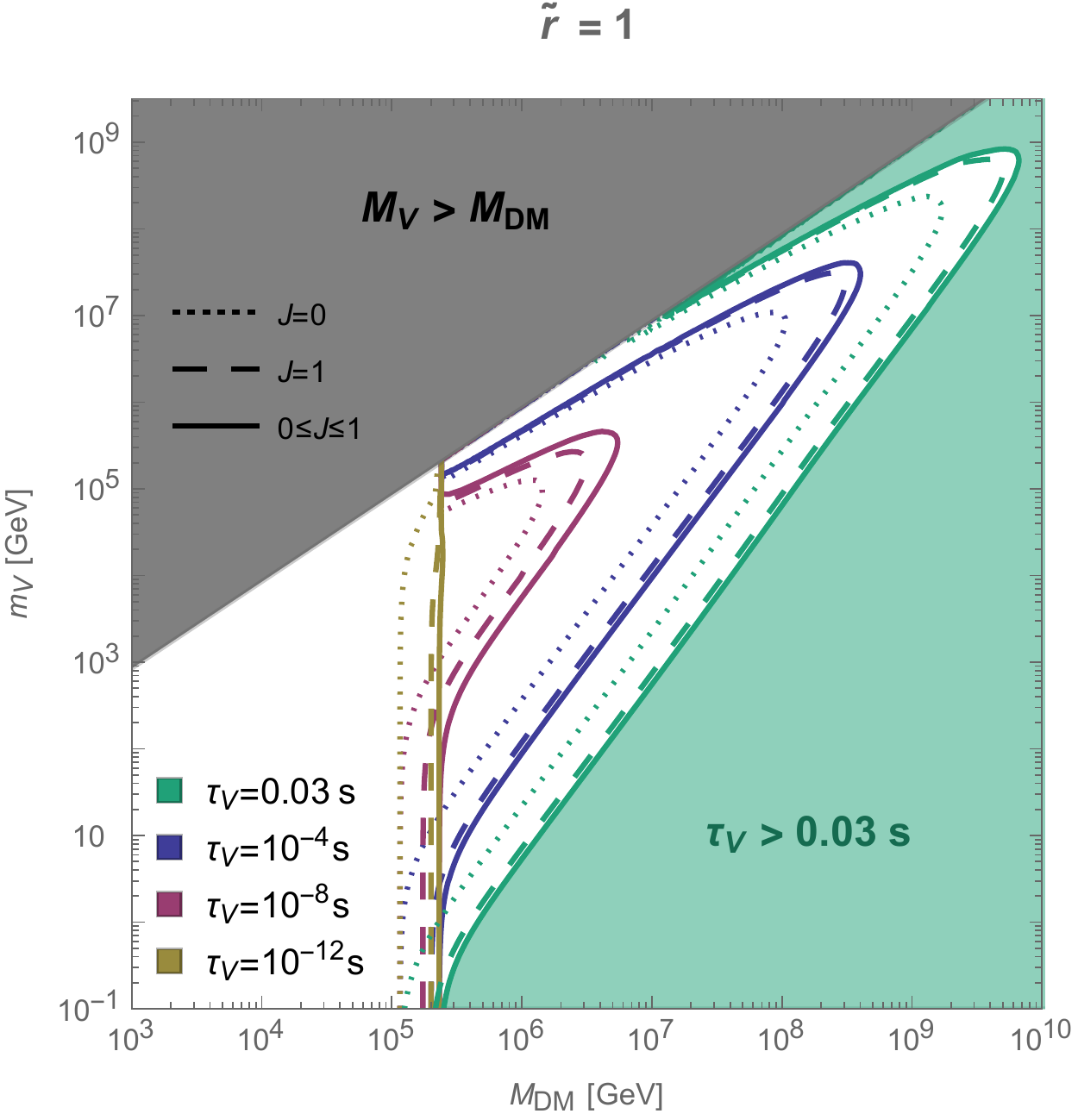}
~~~
\includegraphics[width=0.48\textwidth]{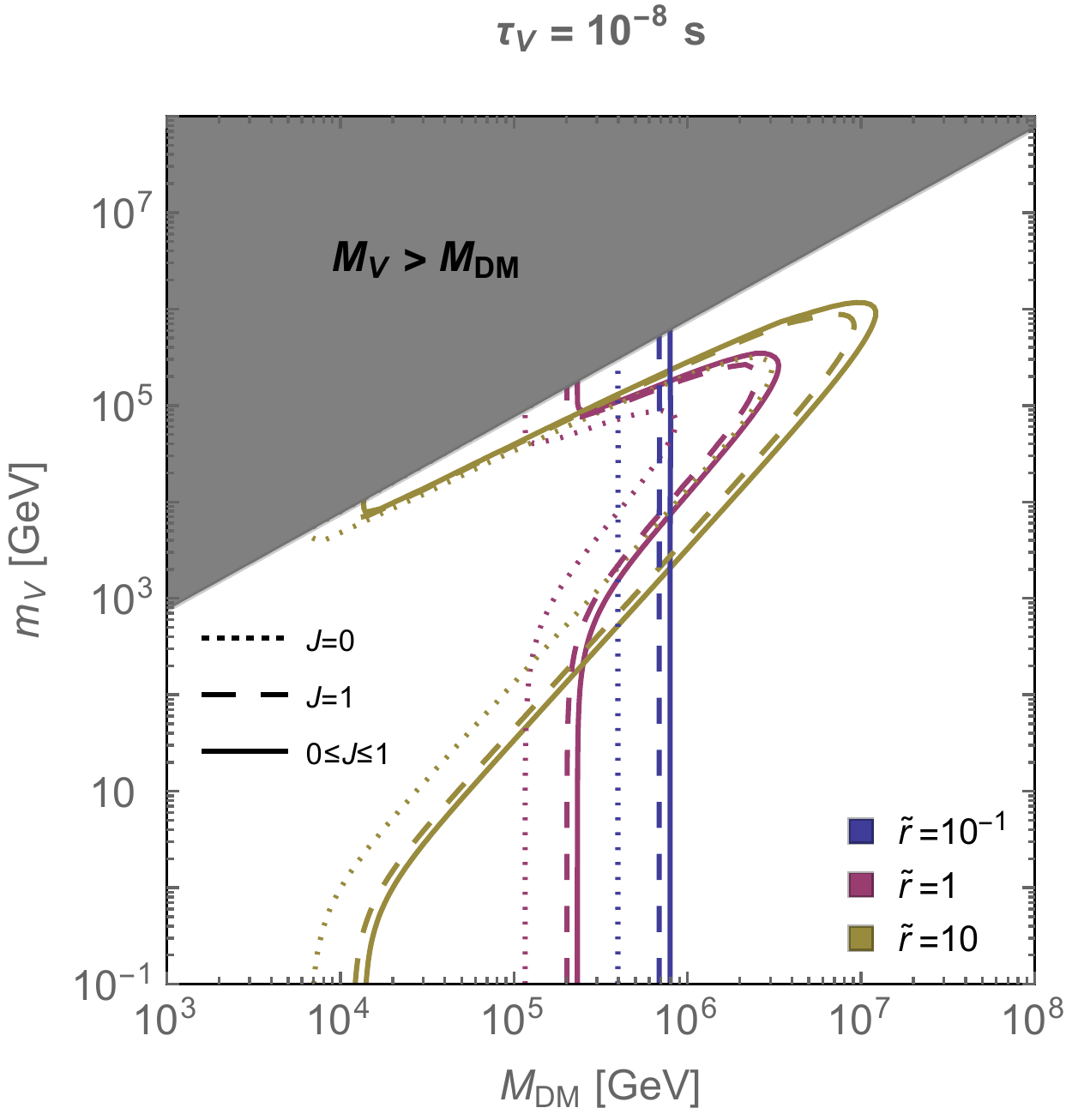}
\end{center}
\caption[]{\it \small
The dependence of the upper bound on the mass of thermal relic DM implied by unitarity, $\MDM \leqslant \Muni$ (coloured lines), on the mediator mass $\mV$ and lifetime $\tauV$, and the dark-to-SM temperature ratio $\tilde{r}$. For definiteness, we have used $\gtildeSM = 106.75$, $\gtildeD = 6.5$ and $\gD^\FO =3$ (cf.~sec.~\ref{sec:U1model}).
\emph{Left:} The dark plasma and the SM are assumed to have the same temperature at early times, $\rtilde =1$. Large $\mV$ and/or $\tauV$ imply that the cosmological energy density carried by the mediators before their decay is large, leading to significant dilution of the frozen-out DM abundance upon their decay and raising $\Muni$. However, a lifetime longer than $0.03$~s
is in conflict with BBN (green-shaded area). Also, the dilution factor gets exponentially suppressed with $\mV$ when $\mV$ is close to $\MDM$ (footnote 4), thus bringing $\Muni$ back to its value in the standard scenario (for the appropriate $\rtilde$).
\emph{Right:} For fixed $\tauV$ and small $\mV$ the dilution is negligible; then, larger $\rtilde$ implies less efficient depletion of DM (due to less Universe expansion from DM freeze-out, which happens later, till today) and more stringent $\Muni$. 
However, for larger $\mV$, the dilution due to the mediator decay is significant and increases with $\tilde{r}$, thereby raising $\Muni$. Note that for sufficiently large $\rtilde$, $\Muni$ becomes independent o{\tiny }f $\rtilde$ (see text for discussion).
}
\label{fig:UnitaryLimit}
\end{figure}

The unitarity of the $S$ matrix implies an upper bound on the partial-wave inelastic cross-section, which physically corresponds to maximal probability for inelastic scattering. In the non-relativistic regime, this is~\cite{Griest:1989wd}
\begin{equation}
\label{eq:Unitarity_sigmav}
\sigma_{\rm inel}^{(J)} v_{\rm rel} \leqslant
\sigma_{\rm uni}^{(J)} v_{\rm rel} =
\frac{4\pi (2J+1)}{\MDM^{2}v_{\rm rel}} \,,
\end{equation}
where $J$ denotes the partial wave. 
As is well known, DM that thermalised in the early Universe and decoupled while non-relativistic must annihilate efficiently enough in order not to overclose the Universe. Since the upper limit \eqref{eq:Unitarity_sigmav} on the inelastic cross-section  decreases with $\MDM$, this implies an upper bound on the mass of thermal-relic DM, $\MDM \leqslant \Muni$~\cite{Griest:1989wd}.

While this conclusion is model-independent, it is important to identify in what theories or parametric regimes the unitarity limit~\eqref{eq:Unitarity_sigmav} --- and consequently the upper bound on the mass of thermal-relic DM --- may be approached or realised. The scaling of $\sigma_{\rm uni} \vrel$ with $\vrel$ is characteristic of attractive long-range interactions, i.e.~interactions mediated by force carriers that are much lighter than the interacting particles. In this case, the long-range force distorts the wavepacket of the incoming particles and enhances their interaction probability at low velocities; this is well-known as the Sommerfeld effect~\cite{Sommerfeld:1931,Sakharov:1948yq}. (We shall return to this in more detail in sec.~\ref{sec:U1model}.)
It has been argued that the parametric dependence of $\sigma_{\rm uni} \vrel$ on $\MDM$ and $\vrel$ indicates that $\sigma_{\rm uni} \vrel$ may be approached or realised in theories with long-range interactions; conversely, the underlying interactions manifest as long-range in the parametric regime where $\sigma_{\rm inel}$ approaches  $\sigma_{\rm uni}$~\cite{Hisano:2002fk,vonHarling:2014kha,Baldes:2017gzw}.

This has several important implications for determining $\Muni$. 
\begin{itemize}

\item[$\circ$]
The velocity dependence of eq.~\eqref{eq:Unitarity_sigmav} ought to be taken into account in computing the thermally averaged DM annihilation rate in the early Universe. Consequently, the latter is enhanced at low temperatures, resulting in more efficient depletion of DM~\cite{vonHarling:2014kha,Baldes:2017gzw}. 

\item[$\circ$]
Long-range interactions imply that higher partial waves may contribute significantly in the DM depletion rate, either via direct annihilation or via the radiative capture into and the subsequent decay of unstable bound states~\cite{Baldes:2017gzw}.\footnote{
In the original computation of $\Muni$ in ref.~\cite{Griest:1989wd}, the physical significance of the velocity dependence of eq.~\eqref{eq:Unitarity_sigmav} was not recognised and $\sigma_{\rm uni} \vrel$ was computed at a fixed velocity typical of the DM velocity during freeze-out. Moreover, it was argued that the lowest partial wave $J=0$ dominates and suffices to determine $\Muni$. The proper thermal average of eq.~\eqref{eq:Unitarity_sigmav} and the participation of higher partial waves result in larger $\Muni$.}

\item[$\circ$] 
In theories with long-range interactions, DM annihilates dominantly into the light force mediators to which it couples. The mediators may then decay into SM particles. As discussed above, if they are sufficiently massive and long-lived, their decay dilutes the DM density, thereby decreasing the effective annihilation cross-section needed to obtain the observed abundance and raising $\Muni$.  Moreover, the dark plasma may be at a different temperature than the SM during DM freeze-out, which also alters $(\sigma \vrel)_\FO$ [cf.~eq.~\eqref{eq:sigmav_fo}].

\end{itemize}

Collecting all the above considerations, 
we substitute in eq.~\eqref{eq:sigmav_fo} the thermally averaged unitarity cross-section, 
$
\langle\sigma \vrel \rangle^\FO =
\langle\sigma_{\rm uni} \vrel \rangle
=4 (2J+1) \sqrt{\pi x^\FO} / \Muni^2
$, and obtain
\begin{equation}
\Muni \simeq \MuniSMs 
\left(\frac{\xfoSM}{\xfo} \right)_{\rm \! \! uni}^{1/4} 
\left[
\frac{\DSM}
{r_\FO \sqrt{1+ (\gD^\FO/\gSM^\FO)r_\FO^4}}
\right]^{1/2}  
\times \left\{
\begin{alignedat}{10}
\sqrt{2J+1},	& \quad \text{solely}~J
\\
J_{\max}+1,		& \quad 0\leqslant J \leqslant J_{\max}
\end{alignedat}
\right.
\,,
\label{eq:Muni}
\end{equation}
where $\MuniSMs \simeq 135$~TeV is the $s$-wave unitarity limit in the absence of any dilution and if DM is annihilating directly into the SM plasma.
The parameters $\xfo$ and $\xfoSM$ at the unitarity limit, appearing in  eq.~\eqref{eq:Muni}, can be determined by the standard formula (cf.~footnote~\ref{foot:xfo}), using the unitarity cross-section; we find 
$\xfoSM \simeq 31 + (1/2) \ln [(2J+1) / \gSM^\FO]$ and 
$\xfo - \xfoSM =-\ln(\Muni/\MuniSM) 
+ \ln (r_\FO^2 / \sqrt{1+ (\gD^\FO/\gSM^\FO)r_\FO^4})$.

We recall that $r_\FO$ and $\DSM$ can be computed using eqs.~\eqref{eq:r_ratio_decay}, \eqref{eq:DSM_rtilde} and \eqref{eq:DSM_rtilde1}. This result is consistent with previous results in the limit of no dilution and $\rtilde = 1$~\cite{vonHarling:2014kha,Baldes:2017gzw}.
Evidently, $\Muni$ depends on the thermodynamics of the dark sector, its temperature and its degrees of freedom; in the presence of dilution, it also depends on the properties of the mediator, its lifetime and mass. In figs.~\ref{fig:sigmaFOandMuni} (right panel) and \ref{fig:UnitaryLimit}, we illustrate these relations.

To conclude this section, we find it interesting to report the maximal mass that DM could in principle reach in such models, $\Muni^\text{max}$, being agnostic on the way this limit would be realised. We obtain $\Muni^\text{max}$ from eq.~\eqref{eq:Muni} upon maximising $\DSM$ as in eq.~\eqref{eq:DSM_Max} (we recall that EeV $= 10^3$~PeV)
\beq
\Muni^\text{max} \simeq 1.1~\text{EeV} \, \Big(\gD^\FO \frac{\mV}{100~\text{PeV}} \left( \frac{\tauV}{0.03~\text{s}}\right)^{\!\frac{1}{2}}\Big)^{\!\frac{1}{2}}
\frac{r_\FO}{\left( 1+0.009\,\gD^\FO\,r_\FO^4\right)^{\!\frac{1}{4}}}
\times \left\{
\begin{alignedat}{10}
\sqrt{2J+1},	& \quad \text{solely}~J
\\
J_{\max}+1,		& \quad 0\leqslant J \leqslant J_{\max}
\end{alignedat}
\right.\,,
\label{eq:Muni_Max}
\eeq
where for simplicity of the exposition we have omitted the weak log dependence.\footnote{For completeness, it amounts
to multiplying eq.~(\ref{eq:Muni_Max}) by 
\beq
\left[1 - 0.025 \log \left( r_\FO^{-2} \; \gD^\FO \, \frac{\mV}{\text{100~\text{PeV}}} \, \left(\frac{\tauV}{0.03~\text{s}}\right)^{\!1/2} \left( 1+0.009\,\gD^\FO\,r_\FO^4  \right)^{1/2} \right)\right]^{-1/4}
\eeq
}
In the dark $U(1)$ model, with $\gtildeD = 6.5$, $\rtilde=1$, $\gD^\FO=3$, $s+p$ wave, this translates to
\beq
\Muni^\text{max} \simeq 2.4~\text{EeV}\,\left(\frac{\mV}{100~\text{PeV}}\right)^{\!1/2}\left( \frac{\tauV}{0.03~\text{s}}\right)^{\!1/4},
\label{eq:mDMmax_darkU1}
\eeq
where we have normalized $\mV$ to a value which guarantees that freeze-out happens when $V$ is relativistic (otherwise eq.~(\ref{eq:mDMmax_darkU1}) does not hold).

\section{The dark $\boldsymbol{U(1)}$ model as a case of study  \label{sec:U1model}}

\subsection{Setup}
We consider Dirac fermionic DM $X$, charged under a dark gauge group $U(1)_D$, under which all the SM particles are neutral. We assume that the dark sector communicates with the SM via kinetic mixing between $U(1)_D$ and the hypercharge group $U(1)_Y$. The Lagrangian then reads
\beq
\mathcal{L}= - \frac{1}{4} {\FD}_{\mu\nu} \FD^{\mu\nu} - \frac{\epsilon}{2\,c_{w}} {\FD}_{\mu\nu} \FY^{\mu\nu}
 + \bar{X}(i \slashed{D} - \MDM)X,
 \label{eq:L}
 \eeq
where $D_\mu = \partial_\mu + i \gD V_\mu'$ is the covariant derivative, with $V_\mu'$ being the dark gauge field. (We reserve the symbol $V^\mu$ for the state arising after the diagonalisation of the kinetic terms, as described below.)  $F_{\mathsmaller{D},\mathsmaller{Y}}$ are the dark and hypercharge field strength tensors, and $c_w$ is the cosine of the weak angle. We define the dark fine structure constant $\aD~\equiv~\gD^2/(4 \pi)$.

We further assume that the dark photon $V^\mu$ obtains a mass $\mV$, either via the St\"{u}ckelberg 
or the Higgs mechanisms. We shall remain agnostic about which of the two is realized, and only note that in case of the Higgs mechanism, the extra scalar can be decoupled (and thus made irrelevant for the phenomenology discussed in this paper) by choosing its charge to be much smaller than one.
Dark photons described by eq.~(\ref{eq:L}) and with a St\"{u}ckelberg mass arise, for example, in string theory constructions, where the predicted ranges for $\epsilon$ and $\mV$ include those of interest in our study, see e.g.~\cite{Jaeckel:2010ni,Essig:2013lka}.

\subsection{Dark photon}
\label{sec:darkphoton}

The dark photon interacts with SM particles via its kinetic mixing $\epsilon$ with the hypercharge gauge boson. Upon diagonalisation of the kinetic and mass terms\footnote{
This has an effect also on the mass and couplings of the $Z$ boson, that is phenomenologically irrelevant for the small values of $\epsilon$ that we will be interested in, see e.g.~\cite{Curtin:2014cca} for the related constraints.}, 
its couplings may be written as
\beq
\mathcal{L} \supset
\epsilon\,e \left( \frac{1}{1-\big(\frac{\mV}{\mZ}\big)^{\!2}}\,J^\mu_{\rm em}
-\frac{1}{c_w^2} \frac{\big(\frac{\mV}{\mZ}\big)^{\!2}}{1-\big(\frac{\mV}{\mZ}\big)^{\!2}}\,J^\mu_{Y}
+O(\epsilon^2) \!\!\right) \! V_\mu
\label{eq:L_DP}
\eeq
where $J^\mu_{\rm em}$ and $J^\mu_Y$ are the usual electromagnetic and hypercharge currents, and where we have denoted by $V$ the mass eigenstate dominated by the gauge eigenstate $V'$.
Equation~\eqref{eq:L_DP} makes transparent both the $\epsilon$-suppression of the $V$ couplings to SM particles, and the physical limits where $V$ couples as the photon ($\mV \ll \mZ$) or as the hypercharge gauge boson ($\mV \gg \mZ$). The expansion of eq.~\eqref{eq:L_DP} in $\epsilon$ is not valid for $\mV \simeq \mZ$, where $V$ couples as the $Z$ boson. In our study we use the full expressions valid also in that limit.

\begin{figure}[t]
\begin{center}
\includegraphics[width=0.6 \textwidth]{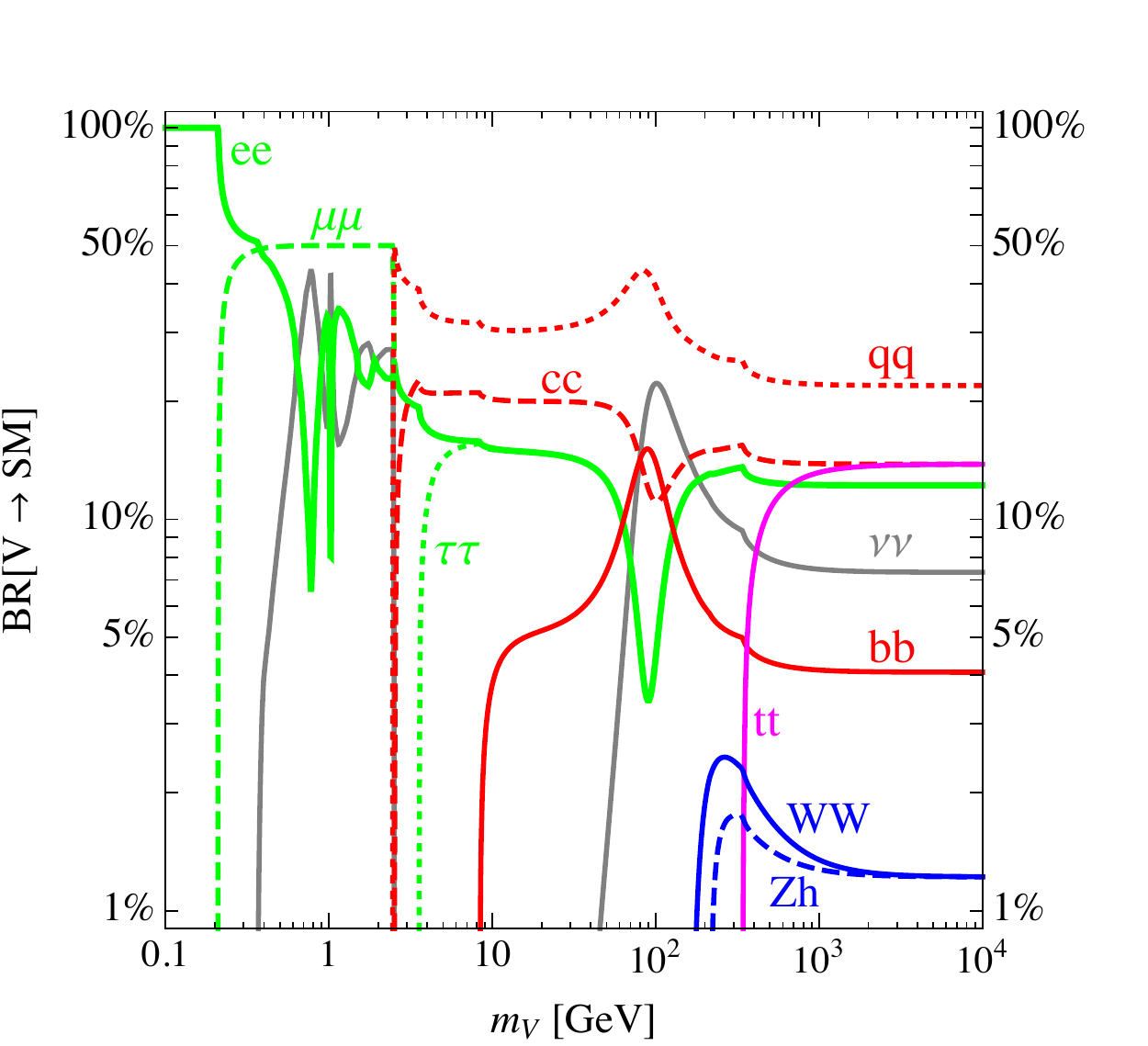}
\end{center}
\caption{\it \small Branching ratios of the dark photon.  This figure updates figure 4 of~\cite{Cirelli:2016rnw}.}
\label{fig:BRs}
\end{figure}

In this work we will be interested in dark photon masses larger than 100~MeV, so that tree-level decays to at least a pair of SM particles are always open. We compute the decay widths of the dark photon to all possible two-body final states, and present them in appendix~\ref{app:DP}. For $\mV > \Lambda_{\rm QCD}$, the hadronic decay channels are open, but for $\mV < {\rm few~GeV}$ they cannot be described perturbatively.
In that interval, which for definiteness we take to be $350~{\rm MeV} < \mV < 2.5~{\rm GeV}$, we determine the total decay width of $V$ from measurements of $e^+ e^- \to$~hadrons at colliders (see e.g.~\cite{Curtin:2014cca,Buschmann:2015awa}). Following~\cite{Cirelli:2016rnw}, we assume this width to consist half of $\mu$ and half of $\nu$ pairs, since these constitute the dominant final states from hadronic decays. 
We show the resulting branching ratios in fig.~\ref{fig:BRs} and refer to appendix~\ref{app:DP} for more details. 

Throughout this paper, we will be interested in the region $\MDM > \mV$. The phenomenology of DM is then driven by its annihilations into dark photons, whose tree-level cross section scales as $\aD^2$. The tree-level cross section for DM annihilation into SM pairs scales as $\epsilon^2\,\aD \aSM$, where $\aSM$ generically stands for an SM EW coupling, and is therefore negligible for the values of $\epsilon$ we will be interested in.

\subsection{Sommerfeld effect and DM bound states}

If the dark photon is sufficiently light, then it mediates a long-range interaction between DM particles. In the non-relativistic regime, and to leading order in $\aD$, the interaction between a dark particle-antiparticle pair is described by the attractive Yukawa potential
\beq 
V(r) = - \aD \exp(-\mV\,r)/r \,,  
\label{eq:YukawaPotential} 
\eeq 
which amounts to the resummation of the one-dark-photon exchange diagram. The long-range nature of the interaction gives rise to the Sommerfeld effect and to bound states, as we summarize in the following.

\paragraph*{\bf Sommerfeld effect.}
The long-range interaction distorts the wavefunctions of the unbound pairs of particles, which are consequently not well described by plane waves. This is the well-known `Sommerfeld effect'~\cite{Sommerfeld:1931,Sakharov:1948yq}, whose significance for the phenomenology of heavy DM has been known since~\cite{Hisano:2002fk,Hisano:2003ec}. For an attractive interaction, it can lead to a dramatic enhancement of the DM annihilation cross section with respect to the tree-level computation.

For $s$-wave annihilation, the cross sections times relative velocity factorises (at leading order) into the hard scattering contribution, $\sigma_0 \vrel = \pi \aD^2 / \MDM^2$, and the Sommerfeld factor $S_{\ann} = |\psi_{\bf k}(r=0)|^2$.  Here, $\psi_{\bf k}({\bf r})$ is the wavefunction of a particle-antiparticle pair, with ${\bf k} = \MDM {\bf v}_{\rm rel}/2$ being the expectation value of the momentum of each particle in the center-of-momentum (CM) frame. $\psi_{\bf k}({\bf r})$ is determined by solving the Schr\"odinger equation with the static potential~\eqref{eq:YukawaPotential} and energy eigenvalue $E = \MDM \, \vrel^2/4$~\cite{ArkaniHamed:2008qn,Pospelov:2008jd,Cassel:2009wt}. Note that only the $\ell=0$ mode of $\psi_{\bf k}$ contributes to $S_\ann$. (Indeed, $\psi_{{\bf k}, \, \ell\ne 0} (0)=0$.) The annihilation cross-section is given by
\beq
\sigma_{\rm ann} \vrel = 
\frac{\pi\aD^2}{\MDM^2} S_{\ann} \Pi_{\ann} \,,
\label{eq:sigmaAnn}
\eeq
where
$\Pi_{\ann} =  
\left( 1-\mV^2/\MDM^2 \right)^{3/2}/ 
\left[ 1-\mV^2/(2\MDM^2) \right]^2$.

For massless mediators, the Sommerfeld effect becomes significant at low velocities, when the momentum of the particles in the CM frame, $\MDM \vrel/2$, is comparable to or smaller than the Bohr momentum, $\MDM\aD/2$, i.e. for
\beq
\vrel \lesssim \aD  \,.
\label{eq:SommerfeldRegime}
\eeq
For a massive mediator, in addition to the above, the range of the interaction must be comparable to or larger than the Bohr radius, 
\beq
\mV^{-1} \gtrsim 2/(\aD \MDM) \,.
\label{eq:mVlessthanBohrMom}
\eeq 
In fact, the scattering state wavefunction $\psi_{\bf k}$ exhibits parametric resonances, which depend on the mode $\ell$ and occur at the thresholds for the existence of bound-state levels with the same angular momentum $\ell$. They appear at discrete values of the Bohr-momentum-to-mediator-mass ratio, with the first resonance ($\ell=0$) at $\aD \MDM / \mV \simeq 1.68$ (see e.g.~\cite{Petraki:2016cnz}).
Note that only the $\ell=0$ resonances affect the $s$-wave annihilation cross-section of eq.~\eqref{eq:sigmaAnn}.

When the average momentum transfer between the $X-\bar{X}$ pair, $(\MDM/2) \vrel$, is larger than the mediator mass, i.e.~at
\beq
\vrel \gtrsim 2 \mV / \MDM \,,  
\label{eq:CoulombLimit}
\eeq
then the system is insensitive to the finite range of the potential. This realizes the so-called ``Coulomb limit'', where $S_{\rm ann}$ can be computed analytically,
\beq
S_{\rm ann}^{\rm C} = 
\frac{2\pi/\epsilon_v}{1-e^{-2\pi/ \epsilon_v}} \,,
\label{eq:Sann_Coulomb}
\eeq
with $\epsilon_v \equiv \vrel/\aD$. The resonances grow at lower velocities, outside the regime \eqref{eq:CoulombLimit}.

\paragraph*{\bf Bound state formation and decay.}
Above the thresholds for their existence, dark particle-antiparticle bound states may form. These are weakly-coupled unstable bound states, analogous to positronium~\cite{Deutsch:1951zza} and true-muonium~\cite{Brodsky:2009gx}. They decay via the same channels that contribute to the DM annihilation, i.e.~primarily into dark photons. This two-step process effectively opens a new DM annihilation channel,\footnote{
Note that on threshold, where the parametric resonances occur, the corresponding bound states have zero binding energy. The formation of bound states discussed here, refers to the capture into levels of finite binding energy. This is a distinct process from annihilation, with different final state particles and cross-section~\cite{vonHarling:2014kha,Petraki:2015hla}.} 
which depletes the DM density in the early Universe~\cite{vonHarling:2014kha} and contributes to the DM indirect signals~\cite{Pospelov:2008jd,An:2016gad,Cirelli:2016rnw,Baldes:2017gzu}.

The bound-state formation (BSF) cross section depends on the overlap of the scattering and bound state wavefunctions and the radiative vertex~\cite{Petraki:2015hla}. The bound state wavefunctions $\psi_{n\ell m} ({\bf r})$ obey the Schr\"odinger equation with the non-relativistic potential \eqref{eq:YukawaPotential} and a discrete set of energy eigenvalues determined by the requirement that $\lim_{r\to \infty} \psi_{n\ell m} ({\bf r}) =0$. Here, $\{n \ell m\}$ stand for the standard principal and angular momentum quantum numbers. The radiative part of the process is computed perturbatively, at leading order in $\aD$. We refer the reader to~\cite{Petraki:2016cnz} for the full computation of the BSF cross-section in this model (see also~\cite{An:2016gad}), and summarise below a few important features.

Similarly to the annihilation, BSF is also enhanced by the Sommerfeld effect, which becomes efficient in the regime specified by eq.~\eqref{eq:SommerfeldRegime}. 
Outside this regime, BSF is rather suppressed (in contrast to $s$-wave annihilation, which tends to its perturbative value). 
The capture from a scattering state into a bound-state level necessitates the dissipation of energy, which occur typically radiatively, via the emission of a force mediator, provided that this is kinematically possible. The energy available to be dissipated in the capture process is the relative kinetic energy of the initial-state particles plus (the absolute value of) the binding energy of the bound state. Thus, for capture into the ground state ($n=1$), BSF via emission of a dark photon is possible if
\beq
\mV \leqslant \MDM \, (\vrel^2 + \aD^2) / 4 \,,
\label{eq:BSF_condition}
\eeq
which simplifies to $\mV  \lesssim \MDM \aD^2/4$ in the regime where BSF is important ($\vrel < \aD$). Here, we have set the binding energy equal to that of a bound state with $\mV = 0$. For $\mV>0$, the binding energy is lower and the upper limit on $\mV$ strengthens; however, the above kinematic threshold already ensures that we are well within the regime of~\eqref{eq:mVlessthanBohrMom}, which in turn implies that the bound state wavefunction and the corresponding binding energy are very close to their Coulomb limit. (Note however that this condition does not ensure that BSF can be approximated by its Coulomb value.)

We will consider only capture into the ground state, with quantum numbers $\{n \ell m\} = \{100\}$. Capture to other bound states is possible for a smaller range of $\mV$, but is always subdominant either with respect to capture to the ground state or to the DM annihilation~\cite{Petraki:2016cnz}. For BSF via emission of a dark photon, the bound-state angular momentum $\ell$ differs by one unit with respect to that of the scattering state. Therefore, the capture into the $\{100\}$ state is a $p$-wave process. This is important for the velocity dependence of $\sigma_\BSF$ at small DM velocities, as we discuss below. The BSF cross-sections can be expressed as
\beq
\sigma_{\BSF} \vrel = \frac{\pi\aD^2}{\MDM^2}  S_{\BSF} \Pi_{\BSF},
\label{eq:sigmaBSF}
\eeq
where  $\Pi_{\BSF}$ includes both the phase-space suppression due to $\mV >0$, and the enhancement due to the third polarization of the dark photon, $\Pi_{\BSF} = s_{\rm ps}^{1/2} (3-s_{\rm ps})/2$, with $s_{\rm ps} = 1-16 \mV^2/[\MDM (\vrel^2 + \aD^2)]^2$. 
$S_\BSF$ can be computed analytically only in the Coulomb regime, which is specified again by the condition \eqref{eq:CoulombLimit} for the scattering state wavefunction; as noted above, the condition \eqref{eq:BSF_condition} ensures that the bound-state wavefunction is well approximated by its Coulomb limit. In the Coulomb approximation~\cite{Petraki:2016cnz},
\beq
S_{\BSF}^{\rm C} =  
\frac{2\pi / \epsilon_v}{1-e^{-2\pi/ \epsilon_v}} \,
\frac{2^9}{3}\,\frac{e^{-4 {\rm arctan}(\epsilon_v)/\epsilon_v}}{(1+\epsilon_v^2)^2} \,.
\label{eq:SBSF_Coulomb}
\eeq
Because it is a $p$-wave process, $\sigma_{\BSF}$ exhibits resonances at the thresholds for the existence of $\ell = 1$ bound states, which grow outside the Coulomb regime. 

\medskip

Working at lowest order in $\aD$, and in the non-relativistic regime, the spin configuration of the scattering state is the same as that of the bound state. The multiplicity of the initial spin states then implies that spin-1 (ortho-) bound states $B_{\uparrow\uparrow}$ form 3/4 of the time, and that spin-0 (para-) ones $B_{\uparrow\downarrow}$ form 1/4 of the time.
Angular momentum conservation implies that $B_{\uparrow\uparrow}$ decays dominantly to three $V$ and $B_{\uparrow\downarrow}$ to two $V$. The related widths read
\begin{align}
\Gamma(B_{\uparrow\uparrow} \to VVV) 
&= \dfrac{4(\pi^2 - 9)}{9 \pi}\dfrac{\aD^6}{2} \MDM \,,
\\
\Gamma(B_{\uparrow\downarrow} \to VV) 
&= \dfrac{\aD^5}{2} \MDM  \,,
\end{align}
where we used again the Coulomb approximation for the bound state wavefunction. For the values of $\aD$ and $\MDM$ that we will work with, the associated lifetimes are at most of $O(10^{-2})$ seconds.
As we will see in section~\ref{sec:DMabundance}, these finite lifetimes play a role in the computation of the DM relic abundance~\cite{vonHarling:2014kha}.
On the contrary, the associated decay lengths are orders of magnitude smaller than any astrophysical scale of interest for the signals we will study in section~\ref{sec:pheno}.

\paragraph*{\bf The role of the DM velocity.} 

For $2 \mV/\MDM \lesssim \vrel \lesssim \aD$, both $S_{\ann}^{\rm C}$ and $S_{\BSF}^{\rm C}$ scale as $\sim 1/\vrel$, and their ratio is
$S_{\BSF}^{\rm C}/S_{\rm ann}^{\rm C} \simeq 3.13$, so that BSF dominates over the direct DM annihilation. However, for $\vrel \lesssim 2 \mV/\MDM$, $S_{\rm ann}$ and $S_{\BSF}$ depend also on $\aD \MDM/\mV$, and display a resonant structure along this parameter, as mentioned above~\cite{Petraki:2016cnz}. In this regime, and away from resonances, the $1/\vrel$ behaviour in the Coulomb regime switches to $\vrel^{2\ell}$, so that $S_{\rm ann}$ saturates to a constant value while $S_{\BSF}$ becomes insignificant. 

Close to resonances, the cross sections instead grow with decreasing velocity as $1/\vrel^2$, i.e.~faster than in the Coulomb limit. 
This scaling can eventually cause an unphysical violation of the partial-wave unitarity bound on the total inelastic cross section, $\sigma_{\rm inel}^{(\ell)} \vrel < 4 \pi (2\ell+1)/(\vrel \MDM^2)$~\cite{Griest:1989wd}. The unitarity limit is also seemingly violated for very large $\aD$. 
To ensure that our computations do not violate unitarity, we impose the unitarity bound for $\ell=0$ and $\ell=1$, as a hard cutoff on the annihilation and BSF cross-sections respectively.

\subsection{DM relic abundance and dilution}
\label{sec:DMabundance}

\begin{figure}[t]
\centering
\includegraphics[width=0.49\textwidth]{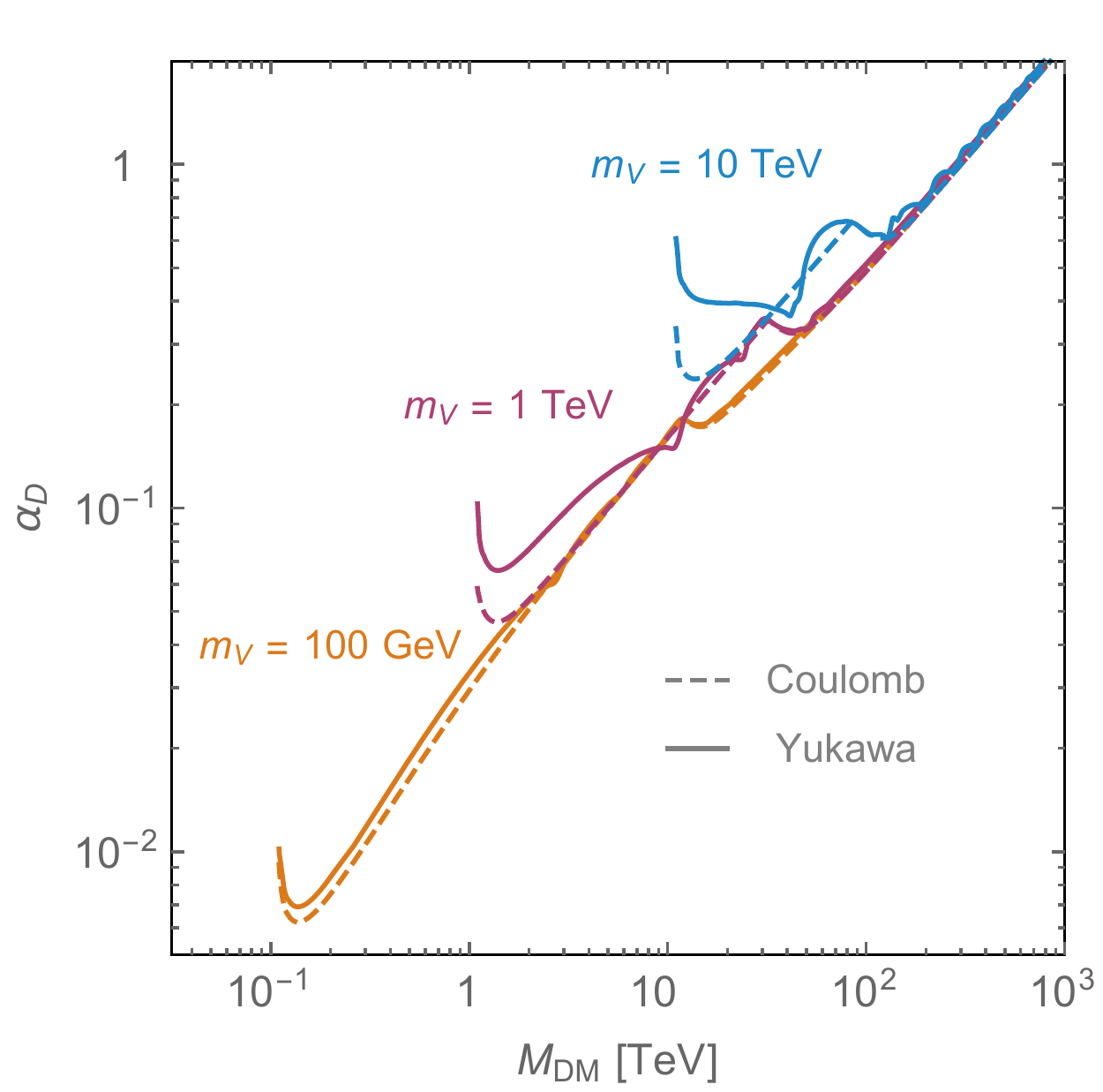}\;
\includegraphics[width=0.49\textwidth]{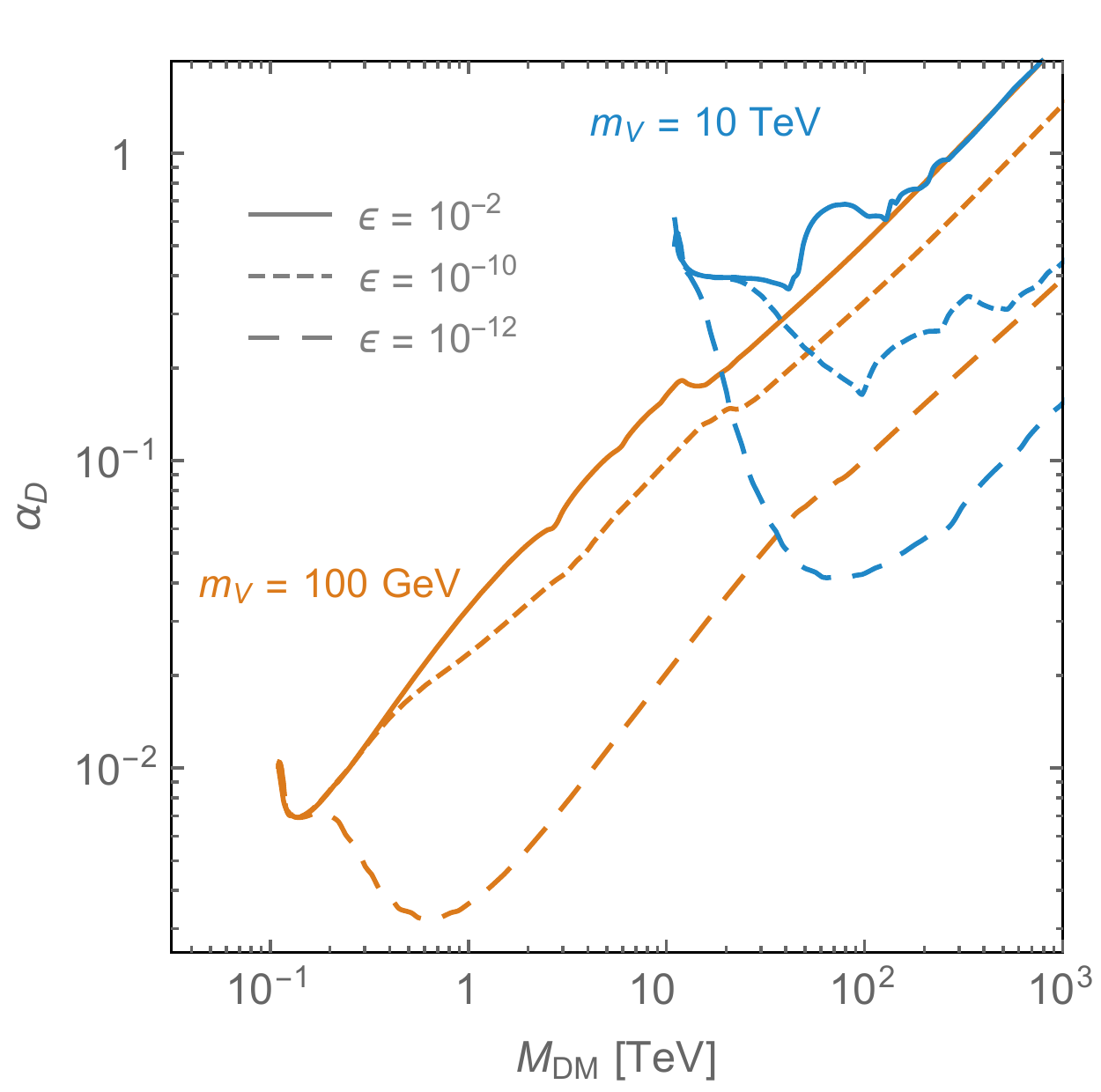}
\caption{\it \small \label{fig:alphaD}
Values of $\aD$ that yield the observed DM abundance as a function of the DM mass, for some reference values of the dark photon mass. Left: our results including the full Yukawa potential (continuous) are compared with the ones in the Coulomb approximation (dashed).
Right: comparison of cases with different kinetic mixing~$\epsilon$ and therefore with different dilution.
The dilution is exponentially suppressed when $\mV$ is close to $\MDM$ (footnote \ref{footnote:DPnrFO}).}
\end{figure}

For each values of the masses $\MDM$ and $\mV$, we compute\footnote{
We use the latest value of the DM relic abundance today measured by Planck~\cite{Patrignani:2016xqp},  $\Omega_{\rm DM}h^{2}=0.1186 \, \pm \, 0.0020$ and the value of the effective number of neutrinos computed in~\cite{Mangano:2005cc}, $N_{\rm eff}=3.046$.}
the value of the dark fine structure constant $\aD$ which leads to the correct value of the DM relic abundance today. 
We use the procedure in~\cite{vonHarling:2014kha} which takes into account both the Sommerfeld enhancement of the direct annihilation and BSF. The novelties of the present work with respect to the most recent analogous computations~\cite{vonHarling:2014kha,Cirelli:2016rnw} are the following:
\begin{itemize}
\item[$\circ$] 
We take into account the full Yukawa potential for computing the Sommerfeld enhancement and BSF, as opposed to the Coulomb limit only. 
This is important at large values of the dark photon mass, as seen in the left panel of fig.~\ref{fig:alphaD} (we remind that the cross section relevant for the signals scales as $\aD^3$ away from resonances).
\item[$\circ$] We study the impact of the entropy dilution due to the dark photon decay, whose effect is shown in the right panel of fig.~\ref{fig:alphaD} for some reference values of the kinetic mixing~$\epsilon$ (note $\GammaV \propto \epsilon^2$ and see eq.~(\ref{eq:dilution_factor_expression_VD_dom_approx}) for the dilution).
\end{itemize}
Equipped with the dark coupling constant that gives the observed DM relic density, we next compute DM annihilation and BSF rates relevant for the cosmic ray signals.

\subsection{DM signals}
\label{sec:DMsignals}

\begin{figure}[!t]
\begin{center}
\includegraphics[width=0.48 \textwidth]{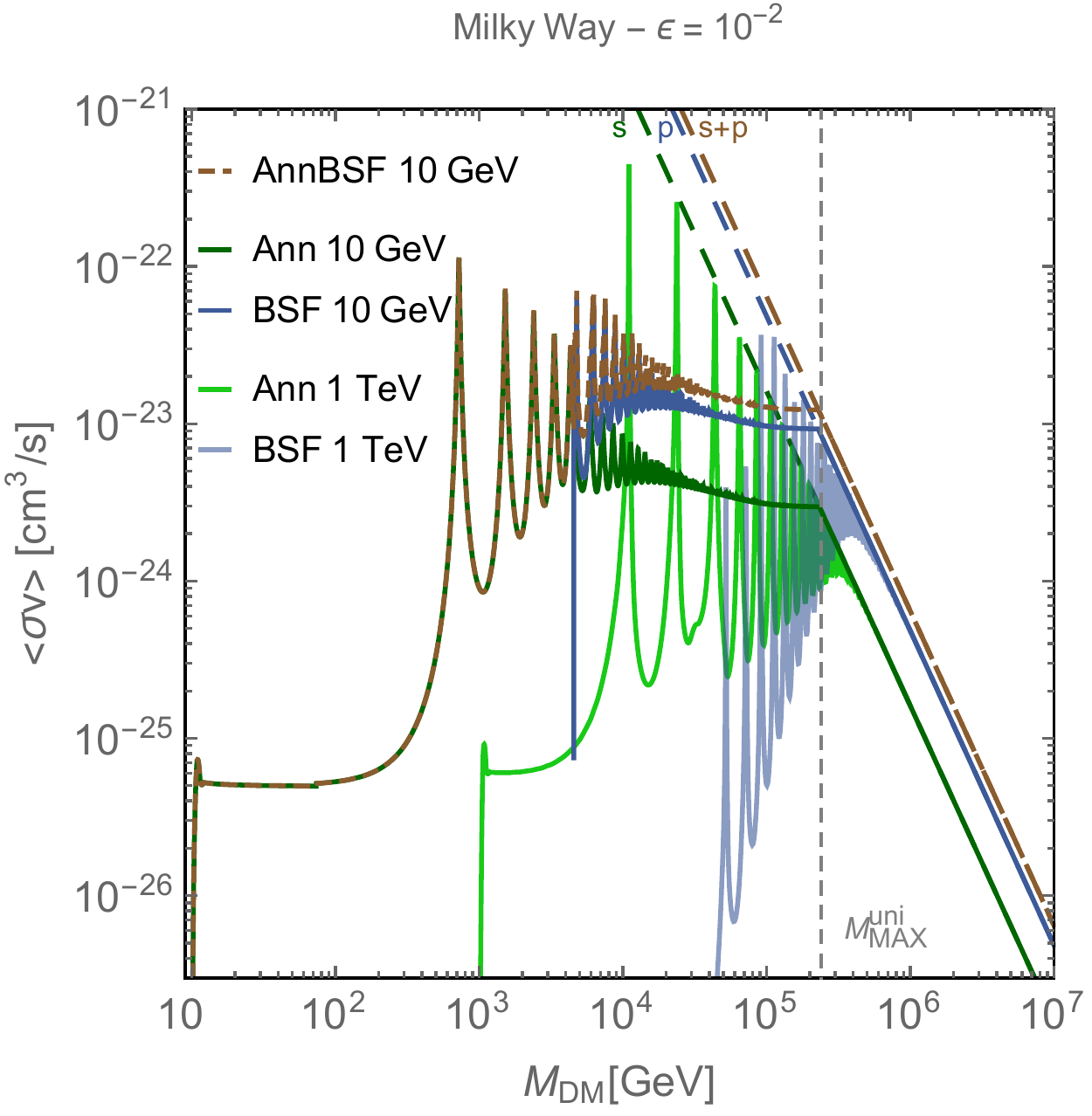} \;
\includegraphics[width=0.48 \textwidth]{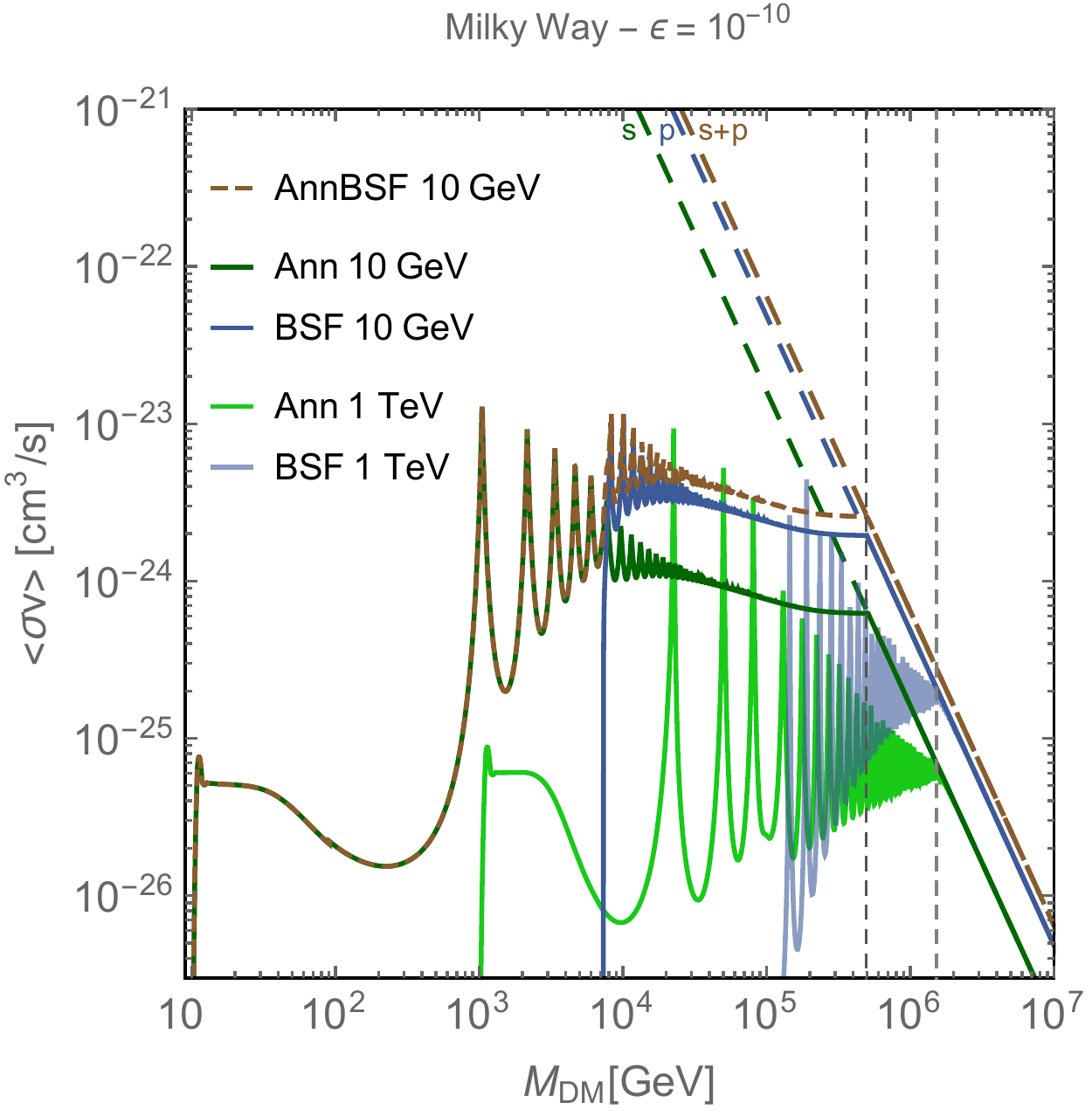} 
\end{center}
\begin{center}
\includegraphics[width=0.48 \textwidth]{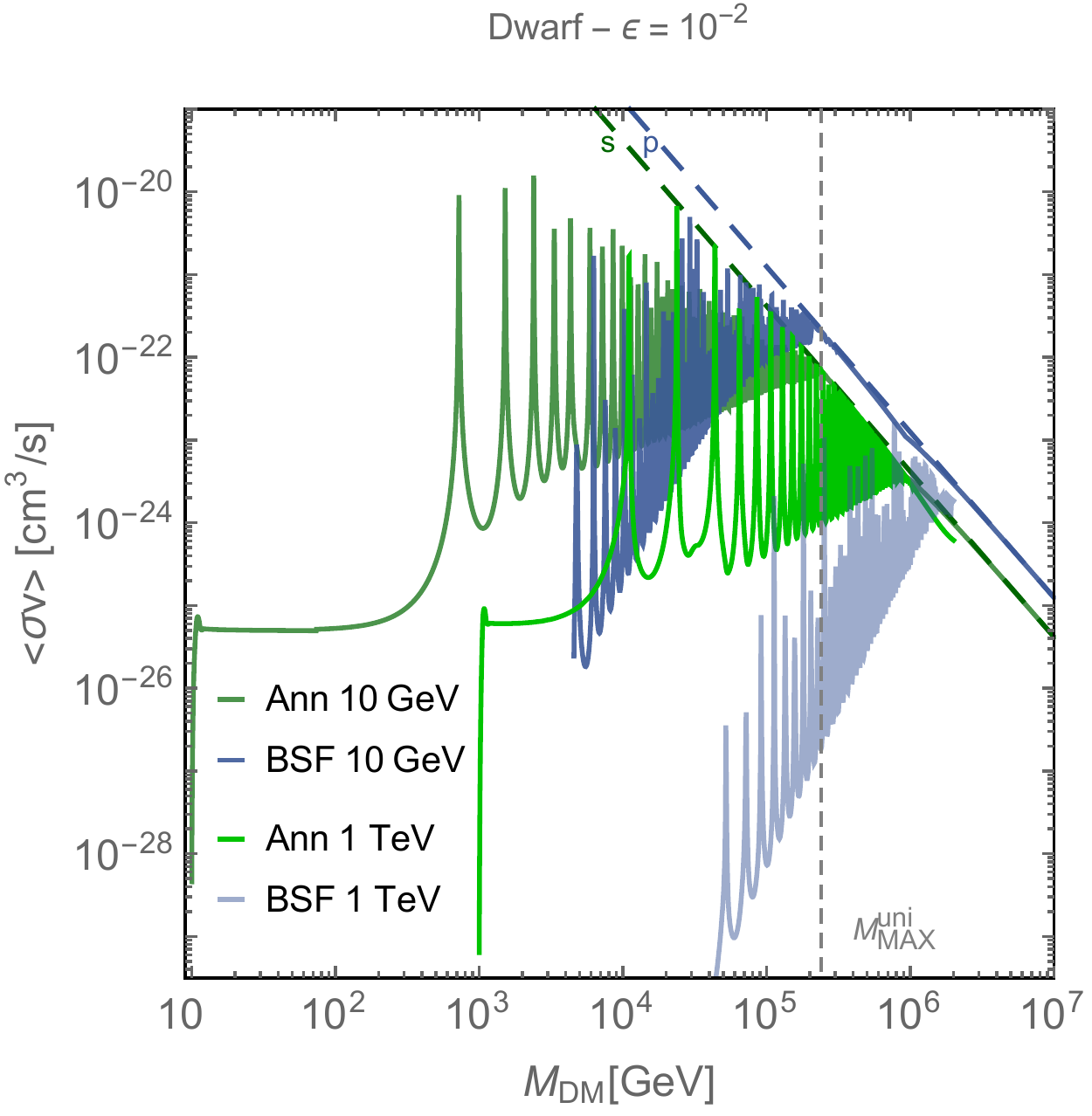} \;
\includegraphics[width=0.48 \textwidth]{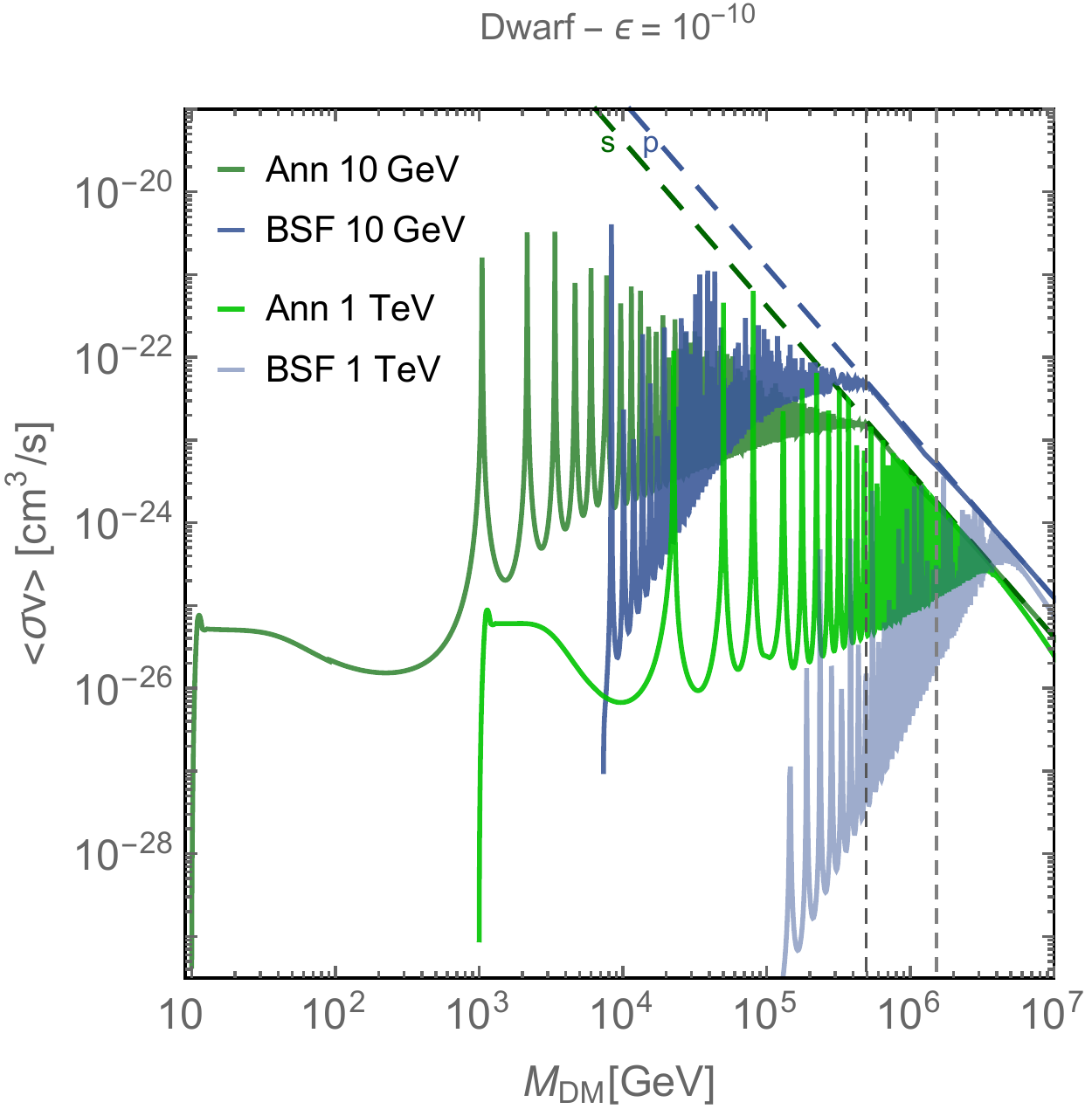} 
\end{center}
\caption{\it \small \label{fig:sigmav_MW_unitarity_mVD_10GeV}
Cross-section averaged over the velocity distribution of the DM in the Milky Way and in Dwarfs versus the dark photon mass. The dashed straight lines are the maximum cross-section values allowed by unitarity for $s$, $p$ and $s+p$ partial waves. The first column is without entropy dilution whereas in the second column, entropy dilution is present with kinetic mixing $\epsilon=10^{-10}$. For $\mV$ close to $\MDM$, the dilution is exponentially suppressed (footnote \ref{footnote:DPnrFO}).}
\end{figure}

We take both  the direct DM annihilation and BSF into account in the computation of the DM relic abundance (and of the unitarity bound), and in the estimation of the DM signals during the cosmic microwave background (CMB), 21 cm, at reionization, in the Milky Way (MW) and in dwarf spheroidal galaxies (dSph).

\paragraph*{\bf MW and dSph.}
For the MW and dSph we compute $S_{\rm ann}$ and $S_{\BSF}$ numerically as in~\cite{Petraki:2016cnz}. We then assume a Maxwellian DM velocity distribution $f(\mathbf{v}) = N \, \theta (v - v_{\rm esc}) \, e^{-v^2/v_0^2}$ for each of the interacting particles, where $N$ is fixed by imposing $\int d^3 v \,f(\mathbf{v}) = 1$.  The velocity $\mathbf{v}$ and the distribution $f(\mathbf{v})$ should not be confused with $\vrel$ and the distribution over $\vrel$; the latter is derived from
$\int f(v_1) f(v_2) \, d^3v_1 d^3v_2$ after carrying out the integration over the mean velocity $(\mathbf{v_1}+\mathbf{v_2})/2$. We choose $v_{\rm esc} = 533$~km/s and $v_0 = 220$~km/s for the MW~\cite{Piffl:2013mla}, and $v_{\rm esc} = 15$~km/s and $v_0 = 10$~km/s for dSph's~\cite{McConnachie:2012vd,Burkert:2015vla}.

The resulting cross sections $\langle \sigma_{\ann} \vrel \rangle$ and $\langle \sigma_{\BSF} \vrel \rangle$ are shown in fig.~\ref{fig:sigmav_MW_unitarity_mVD_10GeV} as a function of $\MDM$, for some reference values of $\mV$. We have fixed $\aD$ to the value that reproduces the correct DM relic abundance, and which depends on $\MDM$ and very mildly on $\mV$ (see section~\ref{sec:DMabundance}).
Figure~\ref{fig:sigmav_MW_unitarity_mVD_10GeV} displays the features discussed in the previous subsection.
Going from smaller to larger $\MDM$, the total cross section first coincides with the tree-level one, then the Sommerfeld enhancement of $\sigma_{\ann}\vrel$ becomes relevant, and for even larger $\MDM$ also BSF becomes important. Both $\sigma_{\ann}\vrel$ and $\sigma_{\BSF}\vrel$ display resonances, and reach the Coulomb limit at a larger value of $\MDM$, which is different between the MW and dSph because the two systems are characterised by different DM velocities. The change of slope around $\MDM \simeq 200$~TeV reflects our prescription for not violating the unitarity limit.

\paragraph*{\bf CMB and 21 cm.}
The evolution of the DM temperature implies that, for the values of the parameters of interest for this paper, at redshifts relevant for CMB constraints ($z \approx 600$, see e.g.~\cite{Finkbeiner:2011dx}) one has $\vrel \ll 10^{-10}$~\cite{Cirelli:2016rnw}. 
Therefore
\begin{itemize}
\item[i)] 
the capture into the ground state is negligible because of the $\vrel^2$ suppression, 
and
\item[ii)] 
the Sommerfeld enhancement of the annihilation processes is well within a saturated regime, so we do not need to compute the velocity average.
\end{itemize}
At even smaller redshifts, inhomogeneities grow and DM structures start to form, so that the DM velocity acquires a component that depends on the gravitational potential of such structures, that could be larger than the value of $\vrel$ relevant for CMB. However, at redshifts relevant for the 21~cm signal ($z \approx 17$, see e.g.~\cite{Furlanetto:2015apc}), the largest DM halos formed are expected to have masses several orders of magnitude smaller than MW-size galaxies, thus inducing low DM velocities. In addition, the dominant part of the signal from DM annihilations comes from the smallest halos, see e.g.~\cite{Lopez-Honorez:2016sur,Liu:2016cnk}.
Therefore, the DM velocities relevant for the 21~cm limits are not expected to be so large that they invalidate i) and ii).

The small values of the velocity also imply that the numerical calculation of $S_{\ann}$ is impractical.
Therefore, to derive the CMB and 21cm constraints in section~\ref{sec:ID_early}, we work with the Hulth\'en potential $V_H = - \aD m_* e^{-m_* r}/(1-e^{-m_* r})$, that allows for an analytic solution $S_{\ann}^H$  (see e.g.~\cite{Cassel:2009wt,Slatyer:2009vg}).
$S_{\ann}^H$ approximates well $S_{\ann}$ off-resonance, it correctly reproduces the fact that the resonances become denser at larger values of $\aD\MDM/\mV$, and we choose $m_* = 1.68 \,\mV$ such that the position of the first resonance coincides with the position of the first resonance in the standard Yukawa case.
For both the CMB and 21cm constraints we work with $v = 10^{-11}$.\footnote{We have explicitly verified that the regions excluded by CMB do not change for smaller $v$ values, and that those excluded by 21cm observations and allowed by unitarity do not change up to $v \simeq 10^{-8}$~(see section~\ref{sec:ID_early}).}

\section{Phenomenology \label{sec:pheno}}
\begin{figure}[p]
\begin{center}
\includegraphics[width=0.43\textwidth]{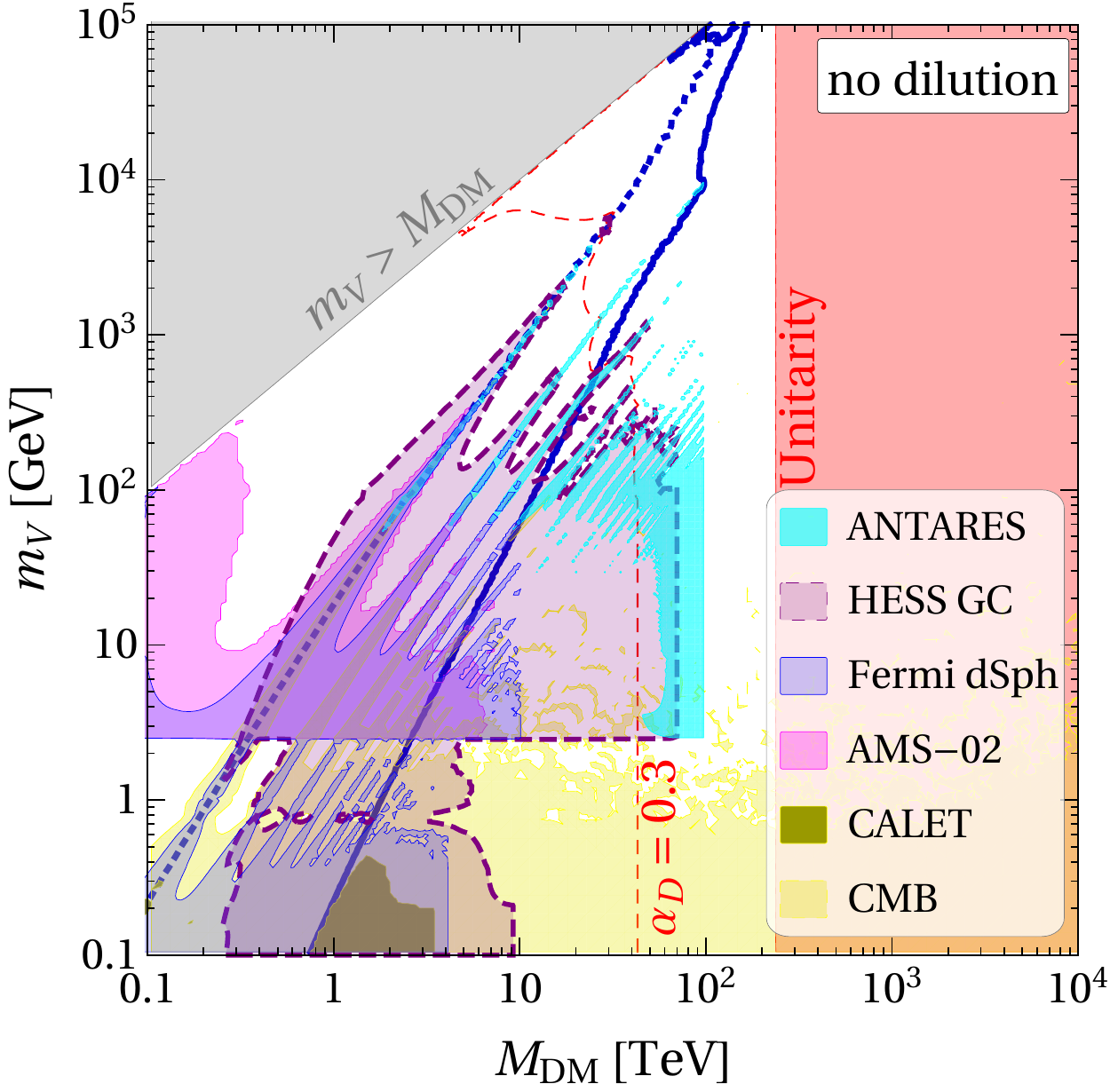} \quad
\includegraphics[width=0.43\textwidth]{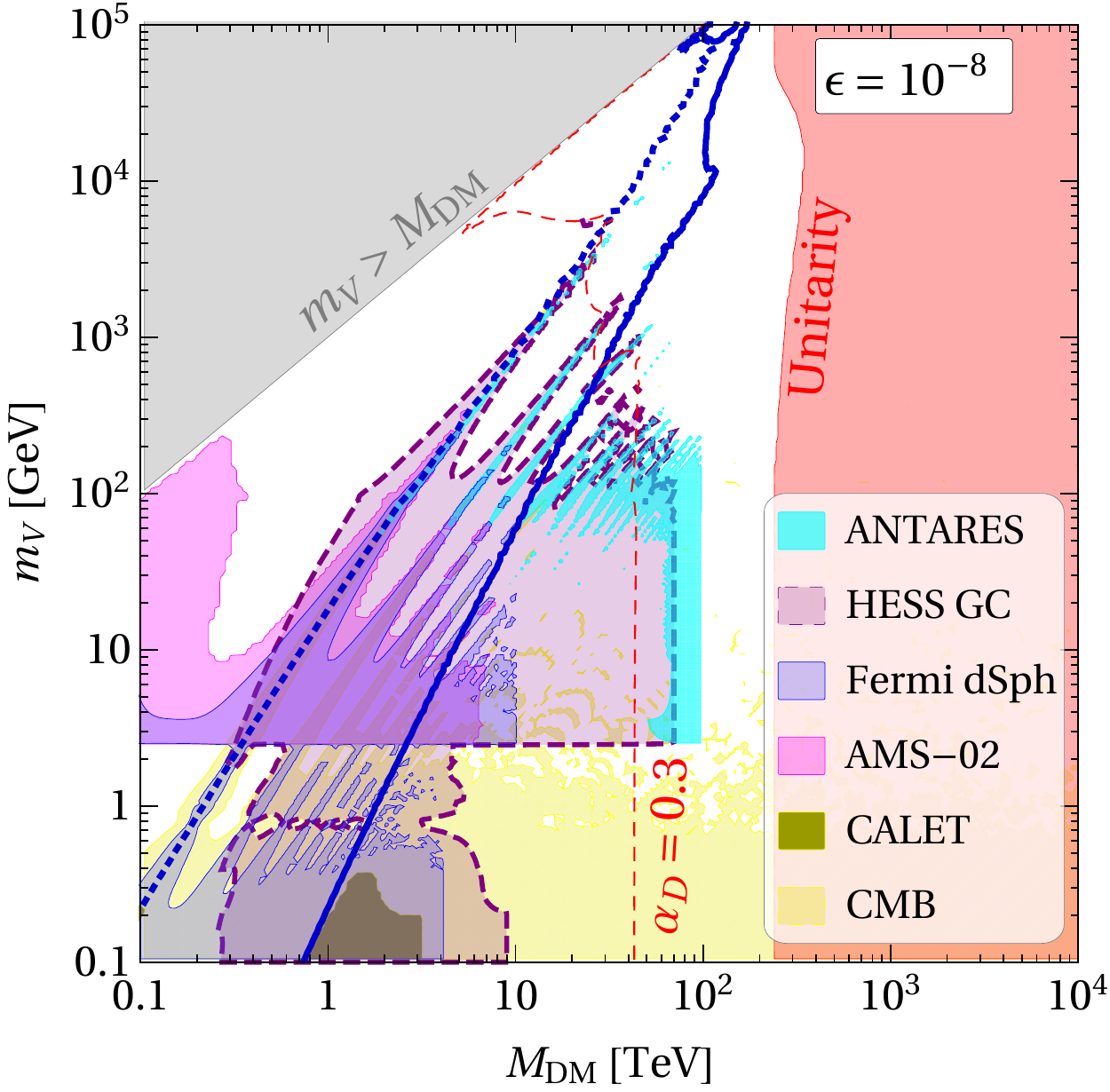}\\
\includegraphics[width=0.43\textwidth]{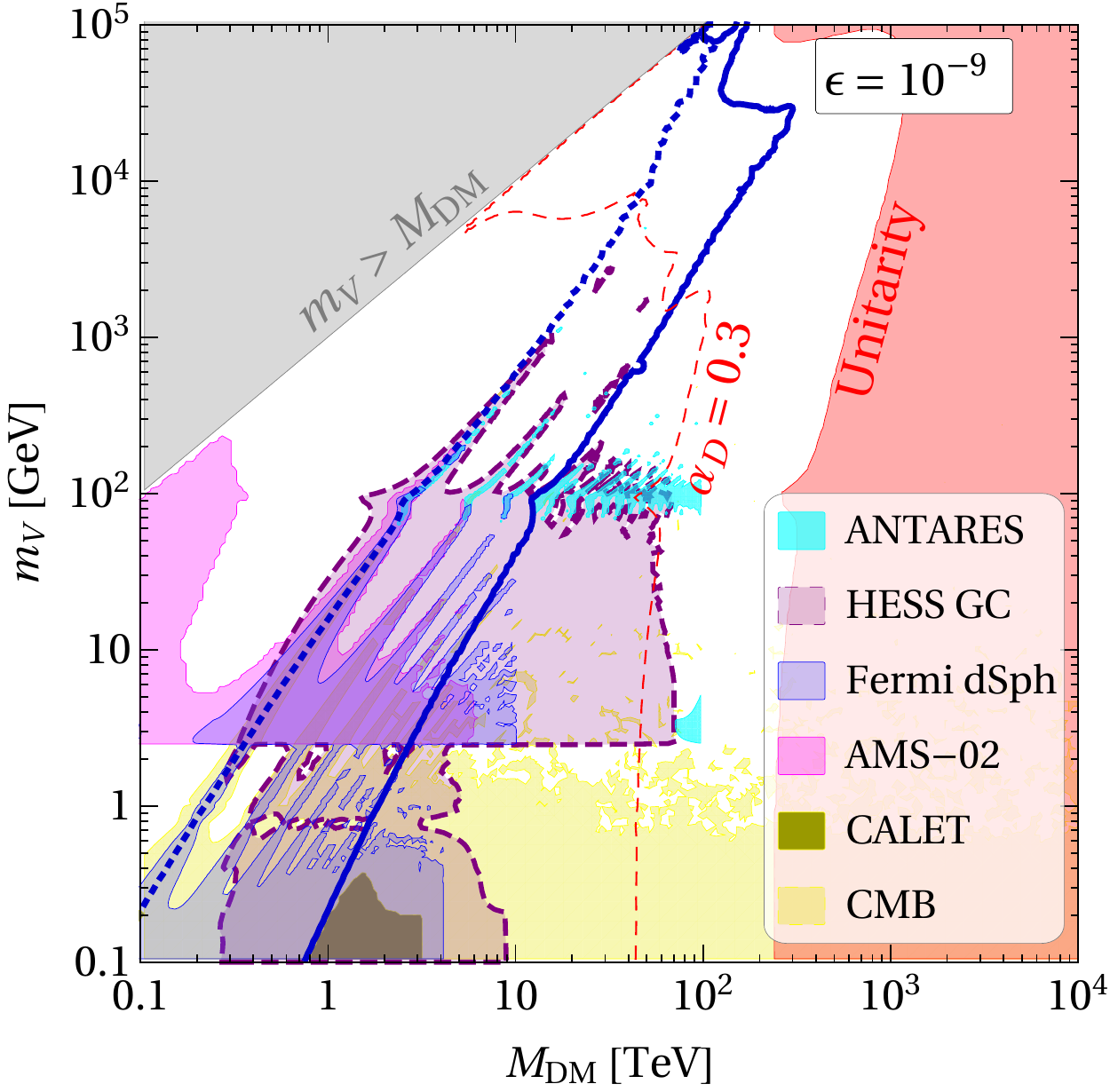}\quad
\includegraphics[width=0.43\textwidth]{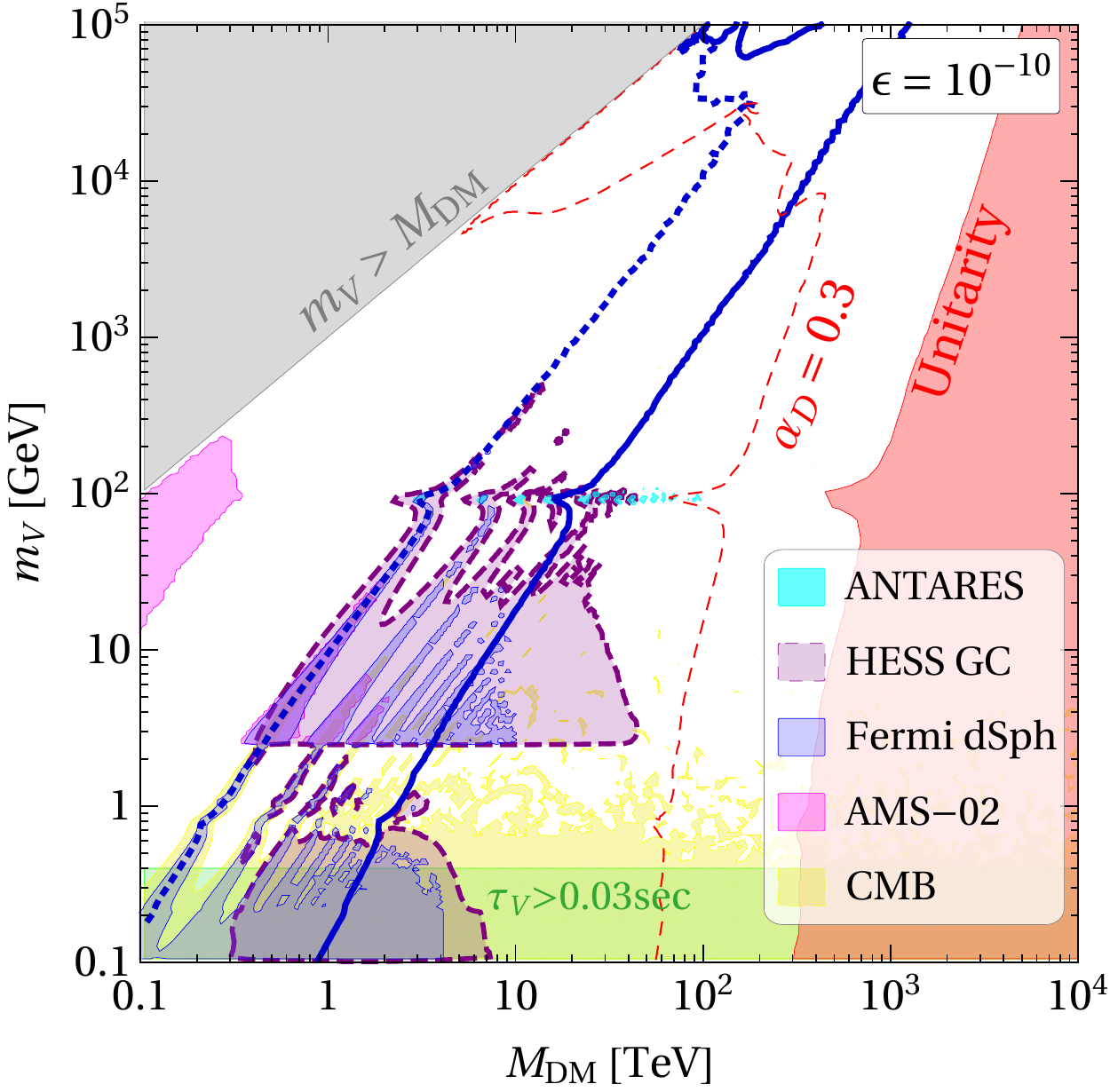}\\
\includegraphics[width=0.43\textwidth]{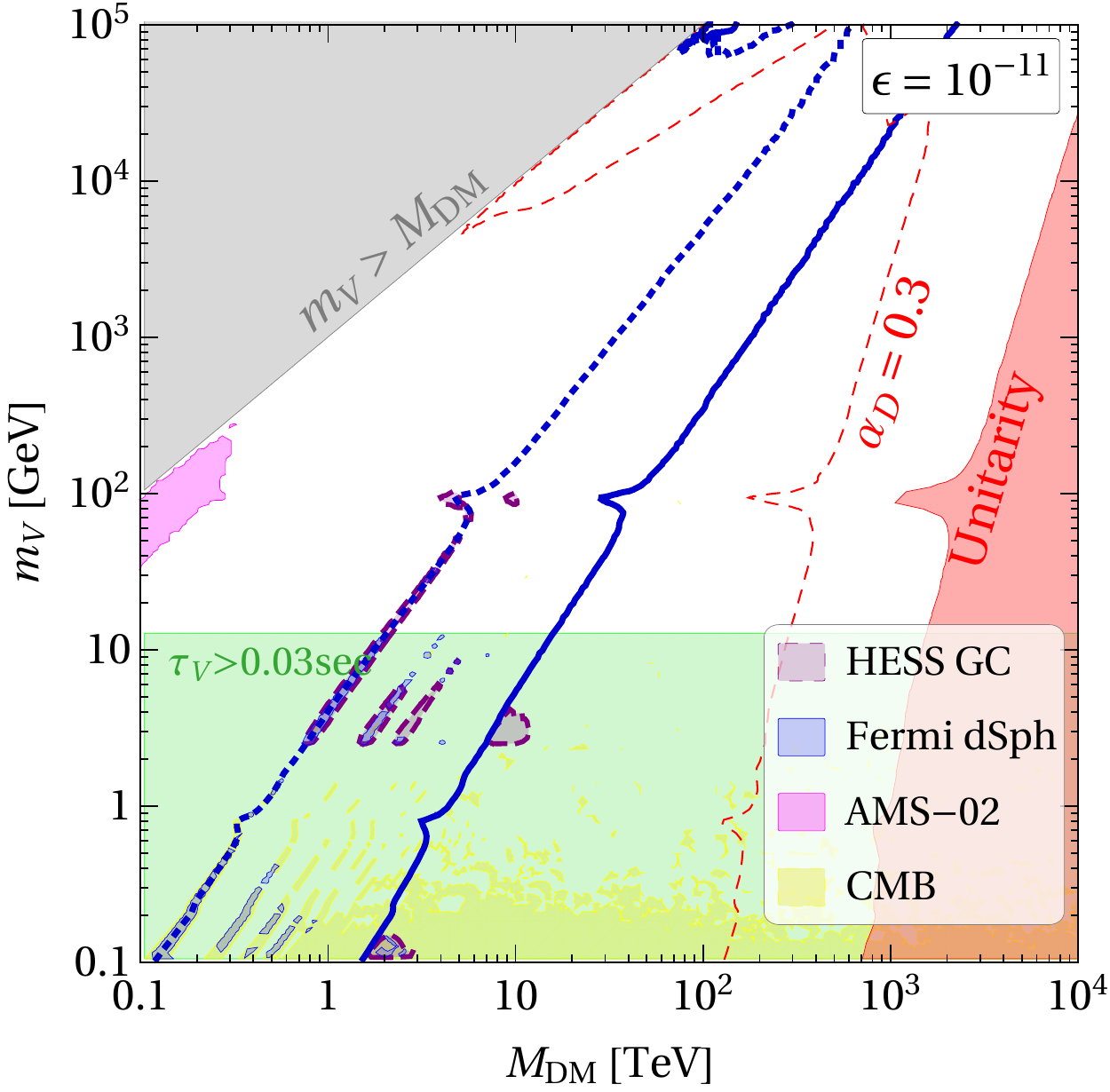}\quad
\includegraphics[width=0.43\textwidth]{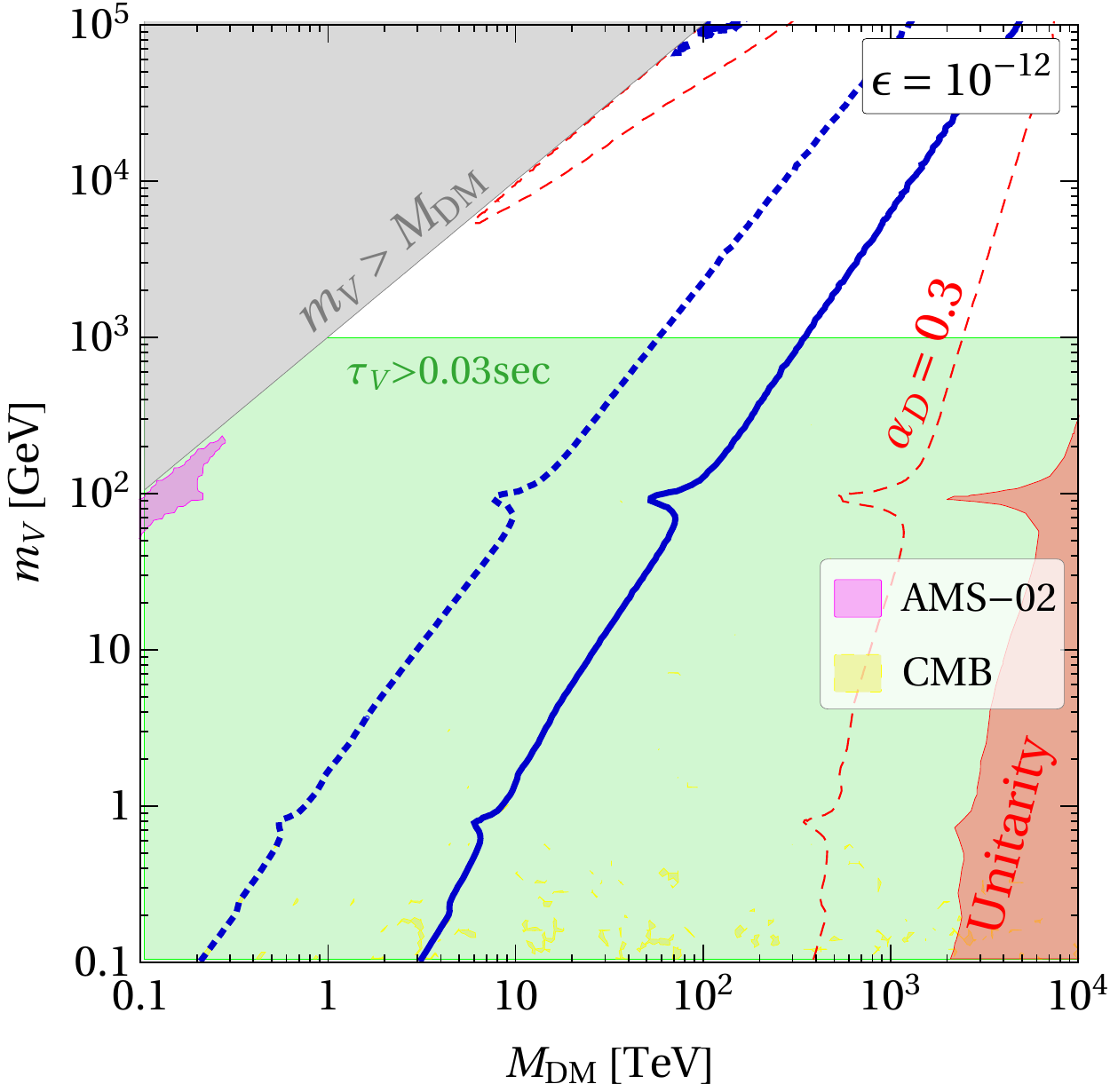}
\end{center}
\caption{\it \small  \label{fig:ID_summary} Indirect detection limits.
Shaded: excluded by {\sc Antares} (cyan), {\sc Ams} (magenta), {\sc Calet} (ocra), {\sc Planck} (light yellow), {\sc Fermi} (blue), within {\sc Hess} reach (magenta), violates s+p wave unitarity bound (red), disfavoured by BBN (green), DM annihilation to dark photons kinematically forbidden (gray). On the right of the lines: bound states exist (blue dotted), bound states can form (blue), $\aD > 0.3$ (red dashed, as a rough indication of where higher order in $\aD$ are expected to become important). From top-left to bottom-right: no DM dilution, and then decreasing values of $\epsilon$ corresponding to increasing DM dilution. 
Note the wider extension in $\MDM$ and $\mV$ in the bottom right panel.
}
\end{figure}

In this Section we present the constraints on the parameter space of the model imposed by the different signals that we consider.
For the ease of the reader, we first present a summary of the results in fig.~\ref{fig:ID_summary}, where each panel corresponds to decreasing values of $\epsilon$, i.e.~increasing dilution ($\GammaV \propto \epsilon^2$ and see eq.~(\ref{eq:dilution_factor_expression_VD_dom_approx}) for the dilution). A few comments are in order.
\begin{itemize}
\item[$\diamond$]
The various indirect detection probes are complementary and collectively cover very large portions of the parameter space. It is remarkable that some of them (neutrinos, $\gamma$-ray and CMB searches) are sensitive to DM with a mass beyond $O(10)$~TeV.
Multimessenger astronomy is therefore a concrete possibility also for the quest of heavy DM, adding motivation to indirect detection in this mass range.
\item[$\diamond$] The fact that the {\sc Antares} and {\sc Hess} regions end abruptly at $\MDM \simeq 100$~TeV is only a consequence of the largest masses considered in the papers of the experimental collaborations. Our study provides a strong motivation for these experiments to extend their current searches to heavier DM masses, as this would likely test unexplored regions of parameter space.
In addition, we point out that  our analysis is partly a reinterpretation of the studies performed by the collaborations themselves and as such suffers from the fact that the SM spectra from cascade decays predicted in secluded models differ from the ones used by these collaborations. The above difficulties would be circumvented if public access to the data was possible.

\item[$\diamond$] As expected, the dilution weakens the sensitivities of the various indirect detection probes, since the cross sections needed to obtain the correct DM abundance are smaller than in the standard case.
It is interesting to note, however, that for increasing dilution the parameter space accessible to indirect detection shrinks because of BBN constraints. This gives a well-defined window of parameter space to aim at, with current and future telescopes.\footnote{Another challenge to test large $\MDM$ with gamma ray telescopes is posed by the non-transparency of the Milky Way to photons energies of hundreds of TeV, see e.g.~\cite{Blanco:2017sbc} for a recent related study.}
From the model-building point of view, this shows explicitly that the upper limit on dilution sets a lower limit on the communication with the SM of secluded DM models.
\end{itemize}
Collider and direct detection searches have instead no power in testing these models, because of the large DM mass and of the small communication with the SM, parameterized by $\epsilon$. We now move to  illustrate in more details the phenomenology of this model.

\subsection{Constraints on the kinetic mixing}
\label{sec:kinetic_mixing}

For $\epsilon \lesssim 10^{-9}$ and $\mV \gtrsim$ GeV, the $\epsilon - \mV$ parameter space is constrained by the requirement that decays of the dark photon do not spoil BBN.
We use here the results of the study~\cite{Jedamzik:2006xz}, that derived BBN constraints on neutral particles decaying at early times. 
It found that, for particles decaying more than 30\% hadronically (like the dark photon in the mass range of our interest), all constraints evaporate for lifetimes of the particle shorter than 0.03 seconds.
Lifetimes as large as $10^2$ seconds become allowed as soon as the abundance of the dark photon, at the time of its decay, drops below $O(0.1)$ times the critical density.
As presented in section~\ref{sec:dilution_DarkU1}, the region we are interested in is that where dark photons dominate the energy density of the Universe just before decaying, to realise sufficient entropy injection. Therefore, we conservatively label as `disfavoured by BBN'  the regions where
$\tauV > 0.03$~sec (see section~\ref{sec:max_homeopathy} for more details).
Note that for mediator masses below the GeV range, where hadronic decay modes close, the BBN constraints weaken significantly~\cite{Hufnagel:2018bjp}.
A more detailed analysis, e.g. with a proper mass dependence of the constraints, goes beyond the purpose of this paper.

For smaller $\mV$ and/or larger $\epsilon$, the existence of dark photons is severely constrained also by DM Direct Detection, observations of the supernova SN1987A, beam dump experiments, and neutrino limits from the Sun. As that region is not the focus of interest of this paper, we do not discuss these constraints here, and refer the interested reader to the discussion in~\cite{Cirelli:2016rnw} for the first three, and to~\cite{Adrian-Martinez:2016ujo,Ardid:2017lry} for constraints from DM annihilation in the Sun.

\subsection{DM constraints from the Early Universe}
\label{sec:ID_early}

\begin{figure}[!t]
\begin{center}
\includegraphics[width=0.44\textwidth]{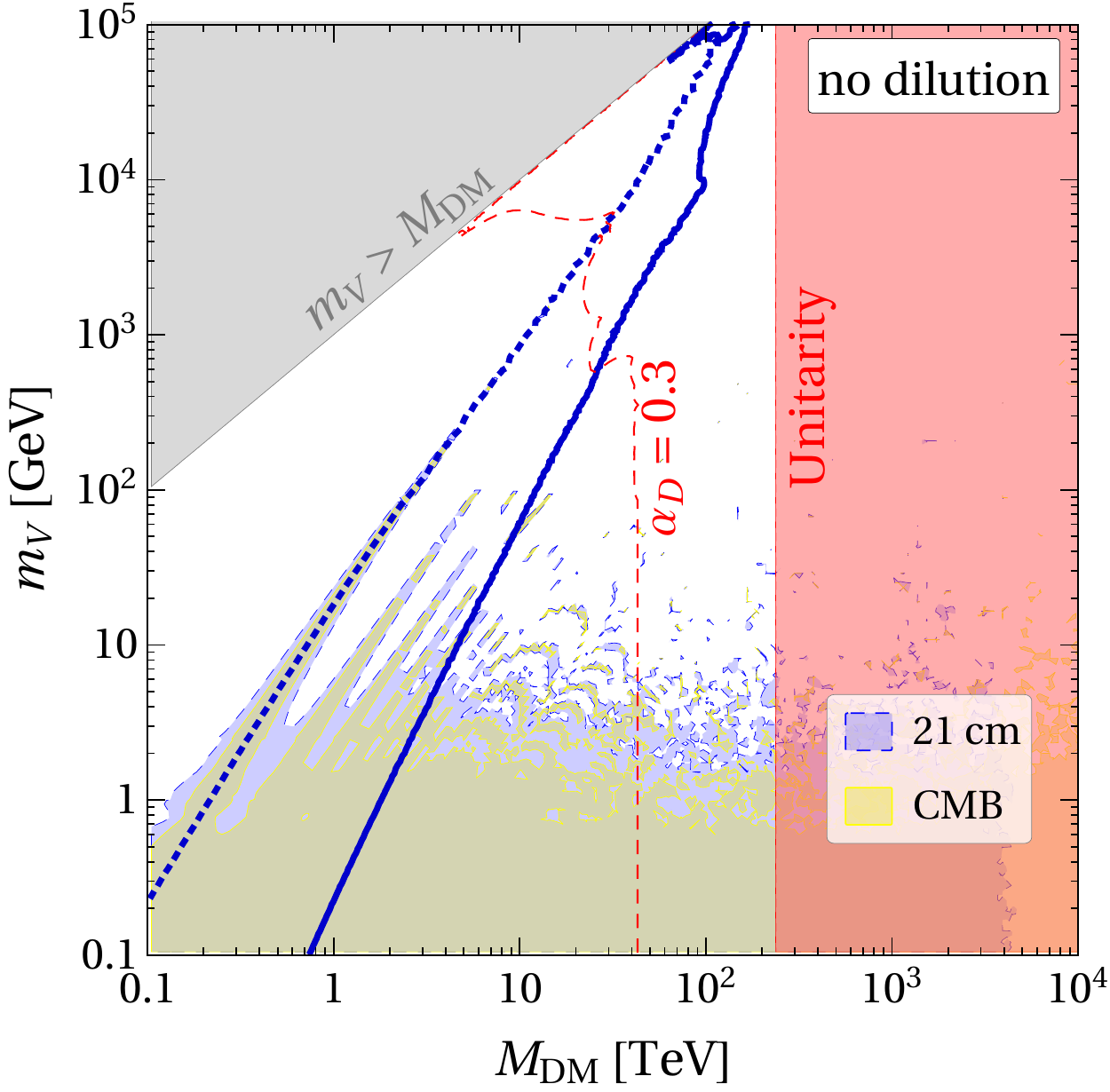}\quad
\includegraphics[width=0.44\textwidth]{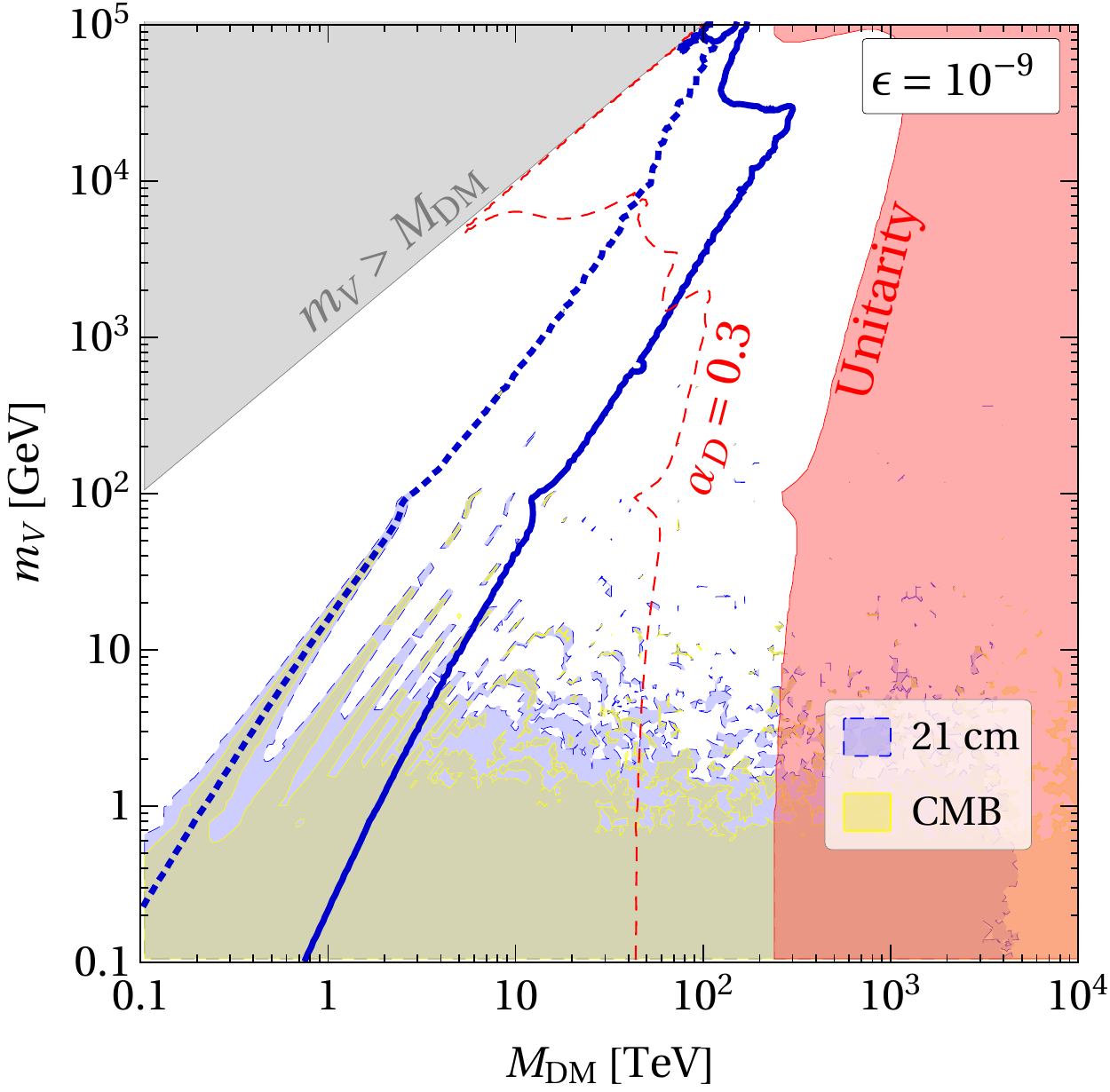}\\
\vspace{.3 cm}
\includegraphics[width=0.44\textwidth]{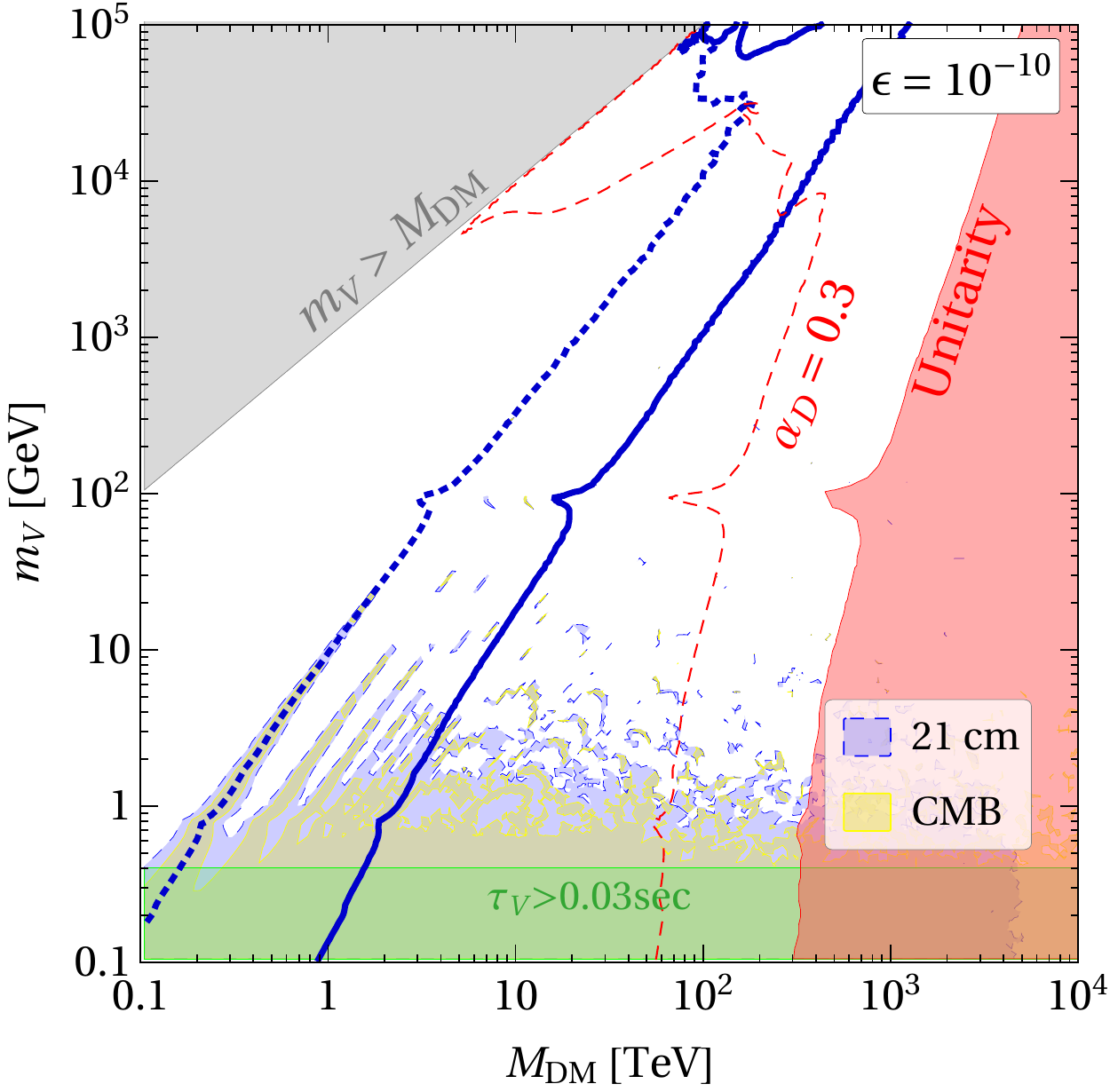}\quad
\includegraphics[width=0.44\textwidth]{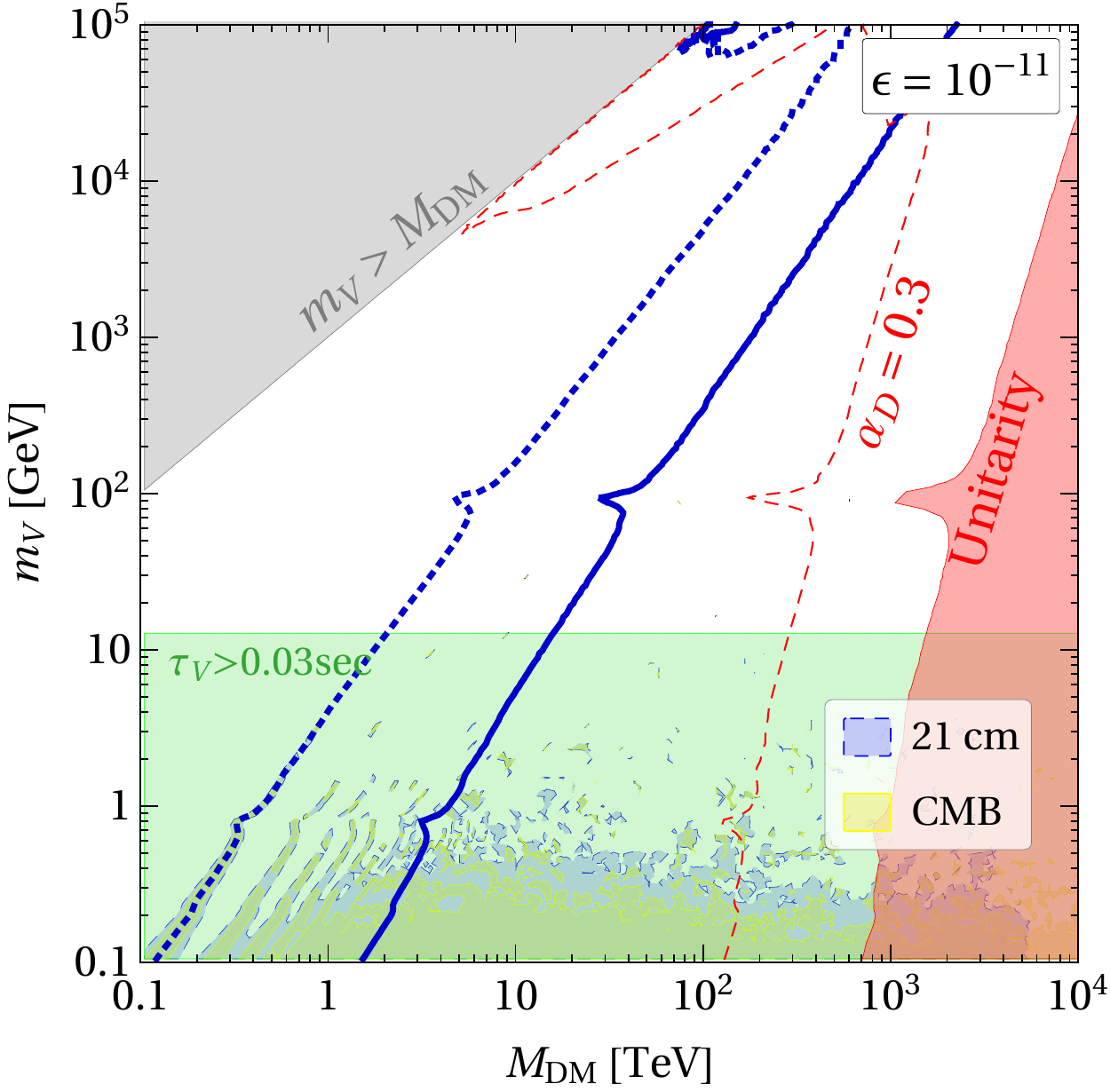}\\
\end{center}
\caption{\it \small \label{fig:EU} Early Universe limits. From top-left to bottom-right: no DM dilution, and then decreasing values of $\epsilon$ corresponding to increasing DM dilution.
Shaded: excluded by CMB (green) and by 21 cm (yellow). The other shaded areas and lines are as in fig.~\ref{fig:ID_summary}. }
\end{figure}

\subsubsection{CMB}
DM annihilations at the time of CMB inject energy in the SM bath and could therefore alter the observed CMB spectrum, resulting in stringent limits~\cite{Ade:2015xua}.
Such limits are driven by the ionizing power of the SM final state $i$, which can be encoded in efficiency factors $f^i_{\rm eff}(\MDM)$ that we take from~\cite{Slatyer:2015jla,Slatyer:2015kla}. Since these depend mostly on the total amount of energy injected, such limits do not depend on the number of steps between the DM annihilation and the final SM products~\cite{Elor:2015bho}.
Following also our discussion in section~\ref{sec:DMsignals}, we then place limits as follows
\beq
\sum_{i=e\bar{e}, u\bar{u},\dots} \!\! \langle \sigma_{\rm ann}^H v\rangle \, {\rm BR}(V\to i) \, f^i_{\rm eff} < 8.2 \times 10^{-26} \,\frac{{\rm cm}^3}{\rm sec}\,\frac{\MDM}{100~{\rm GeV}}\,,
\label{eq:CMB}
\eeq
where on the left-hand side the dependence on $\MDM$ and $\mV$ is implicit.
The resulting exclusion is displayed in figure~\ref{fig:EU} for various possibilities of DM dilution.

\subsubsection{21 cm}
$\Lambda$CDM predicts an absorption signal in the radio band at 21 cm, associated with the relative occupation number of the singlet and triplet hyperfine states of the hydrogen atom at $z\simeq 17$, see e.g.~\cite{Furlanetto:2015apc}.
The {\sc Edges} collaboration has recently reported~\cite{Bowman:2018yin} the observation of such a signal, although with an amplitude larger than expected in $\Lambda$CDM.
Because extra injection of energy in the SM bath would make that amplitude smaller than expected, this observation allows to place limits on DM annihilations and/or decays, that turn out to be competitive with CMB ones~\cite{DAmico:2018sxd,Liu:2018uzy,Cheung:2018vww,Clark:2018ghm,Mitridate:2018iag}.
Here we do not aim at explaining why the amplitude is observed to be larger than predicted, but rather we simply impose a limit on the DM annihilation following ref.~\cite{DAmico:2018sxd}, as
\beq
\sum_{i=e\bar{e}, u\bar{u},\dots} \!\! \langle \sigma_{\rm ann}^H v\rangle \, {\rm BR}(V\to i) \, f^i_{\rm eff} < 6.3 \times 10^{-26} \,\frac{{\rm cm}^3}{\rm sec}\,\frac{\MDM}{100~{\rm GeV}}\,,
\label{eq:21cm}
\eeq
where we have taken into account that DM in our model is not self-conjugate.
Among the limits presented in~\cite{DAmico:2018sxd}, we have chosen the more conservative one, derived by imposing that DM does not reduce the standard amplitude by more than a factor of 4 (`$T_{21} \gtrsim 50$~mK'), and using the more conservative boost due to substructures (`Boost~1').

It is important to stress that the 21~cm limit in eq.~(\ref{eq:21cm}) is derived assuming an immediate absorption of the energy injected from the SM products of DM annihilations (so that, again, the number of steps in the cascade does not matter). This is however not the case especially when they are very energetic, thus for large DM masses.
The effects of delayed energy deposition have been taken into account e.g. in~\cite{DAmico:2018sxd,Cheung:2018vww}, however the resulting limits have not been reported for DM masses beyond~10 TeV.
Up to those masses and choosing as an example the $b\bar{b}$ final state, the limits derived as in eq.~(\ref{eq:21cm}) are stronger than those that take into account delayed energy deposition by a factor of a few ($\MDM \simeq 10$~GeV) to a few tens ($\MDM \simeq 10$~TeV)~\cite{DAmico:2018sxd}.
We expect a difference in the same ballpark also in our model, because of the extra softening of the spectra from the extra-steps in the cascade, and because the dominant SM final states (see figure~\ref{fig:BRs}) result in an electron and photon spectra of energies much lower than $\MDM$, in qualitative agreement with $b\bar{b}$.
Such differences in the limits (between eq.~(\ref{eq:21cm}) and those that include delayed energy depositions) are of the same order of the uncertainties induced by assumptions like the boost model and the precise value of the bound to apply, see e.g.~\cite{DAmico:2018sxd,Cheung:2018vww}.
Therefore, despite the caveats discussed, we find it interesting to compare CMB constraints on the dark $U(1)$ DM model with the 21~cm ones from eq.~(\ref{eq:21cm}). They are displayed in figure~\ref{fig:EU} for various possibilities of DM dilution.
Waiting for further experimental confirmations of the signal, and for a precise assessment of the 21~cm limit to be settled, we refrain from putting the 21~cm limits on the same footing of the CMB ones, and we show only the latter ones in the summary in Fig.~\ref{fig:ID_summary}.

\subsection{DM constraints from the Local Universe}
\label{sec:ID_local}

Dark Matter interactions (annihilation, formation and decay of bound states) in the local Universe may induce signals at space- and ground-based telescopes, whose observations therefore put limits on the parameter space of our model.
Here we study constraints from telescopes observing neutrinos (section~\ref{sec:ID_local_nu}), gamma rays (\ref{sec:ID_local_gamma}), antiprotons (\ref{sec:ID_local_pbar}) and electrons and positrons (section~\ref{sec:ID_local_epem}).

In our model, DM interactions result into 2 (from annihilations and from decays of $B_{\uparrow\downarrow}$) or 3 (from decays of $B_{\uparrow\uparrow}$) energetic dark photons, themselves decaying to SM pairs according to the branching ratios in Fig.~\ref{fig:BRs}.
The resulting `one-step' SM energy spectra are therefore moved to lower energies compared to the standard case of direct DM annihilation into SM pairs.
This constitutes an important input in deriving the constraints discussed in this section, and will be discussed case by case.
The additional and much less energetic $V$, that is emitted to form the bound state, will be discussed separately in section~\ref{sec:ID_DPemittedBS}.

\subsubsection{Neutrinos}
\label{sec:ID_local_nu}

\begin{figure}[!t]
\begin{center}
\includegraphics[width=0.44\textwidth]{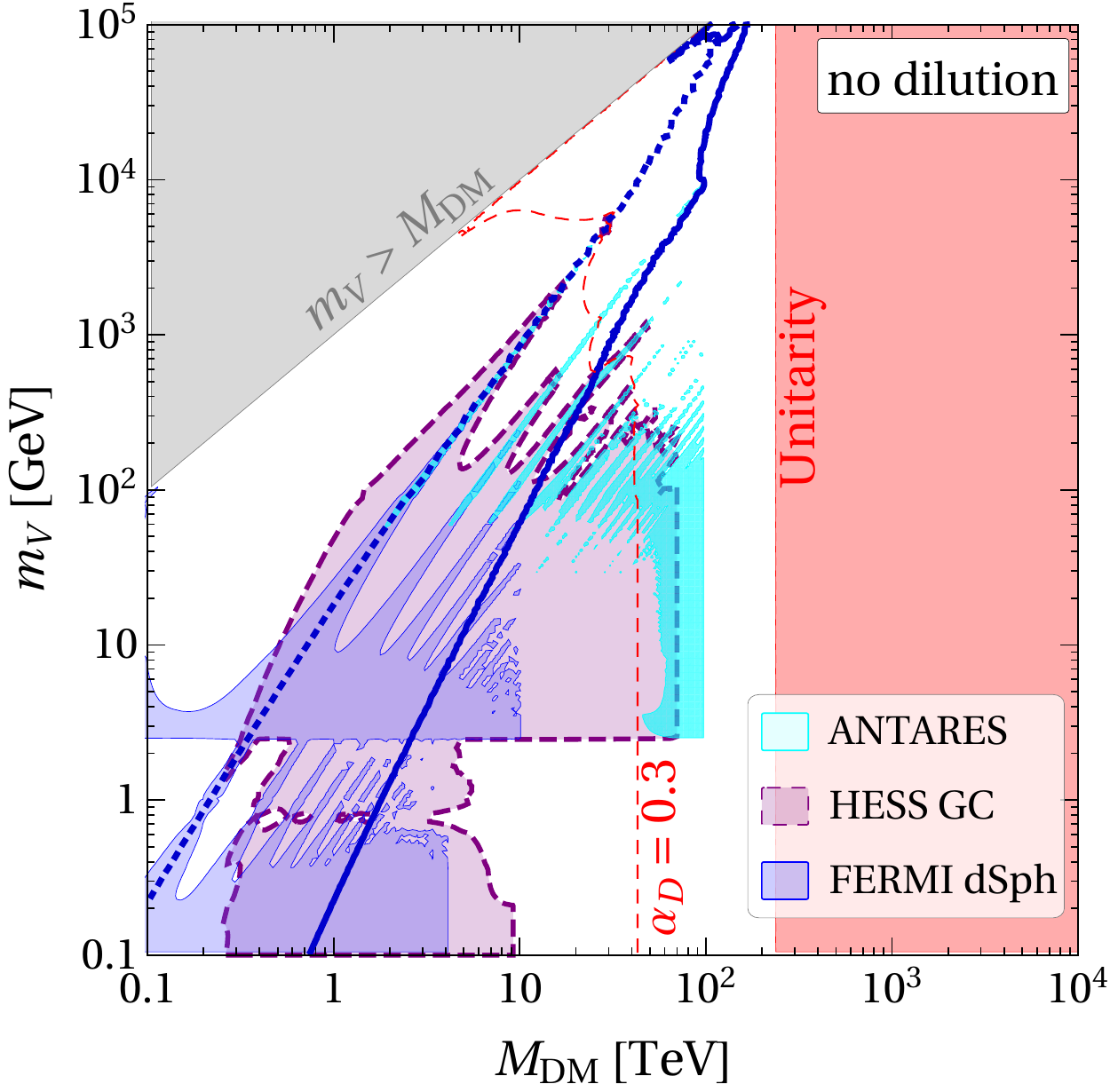}\quad
\includegraphics[width=0.44\textwidth]{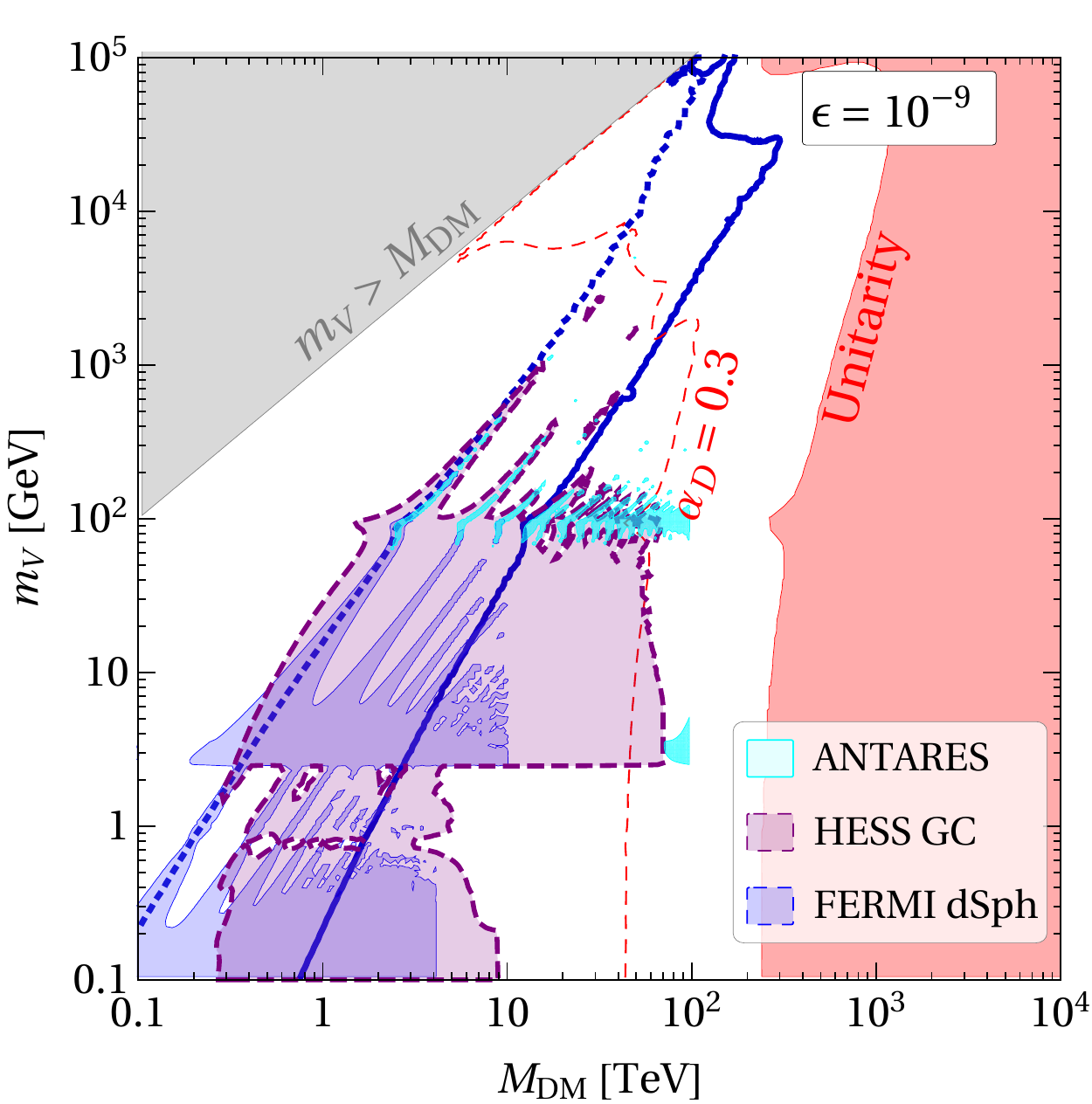}\\
\vspace{.3 cm}
\includegraphics[width=0.44\textwidth]{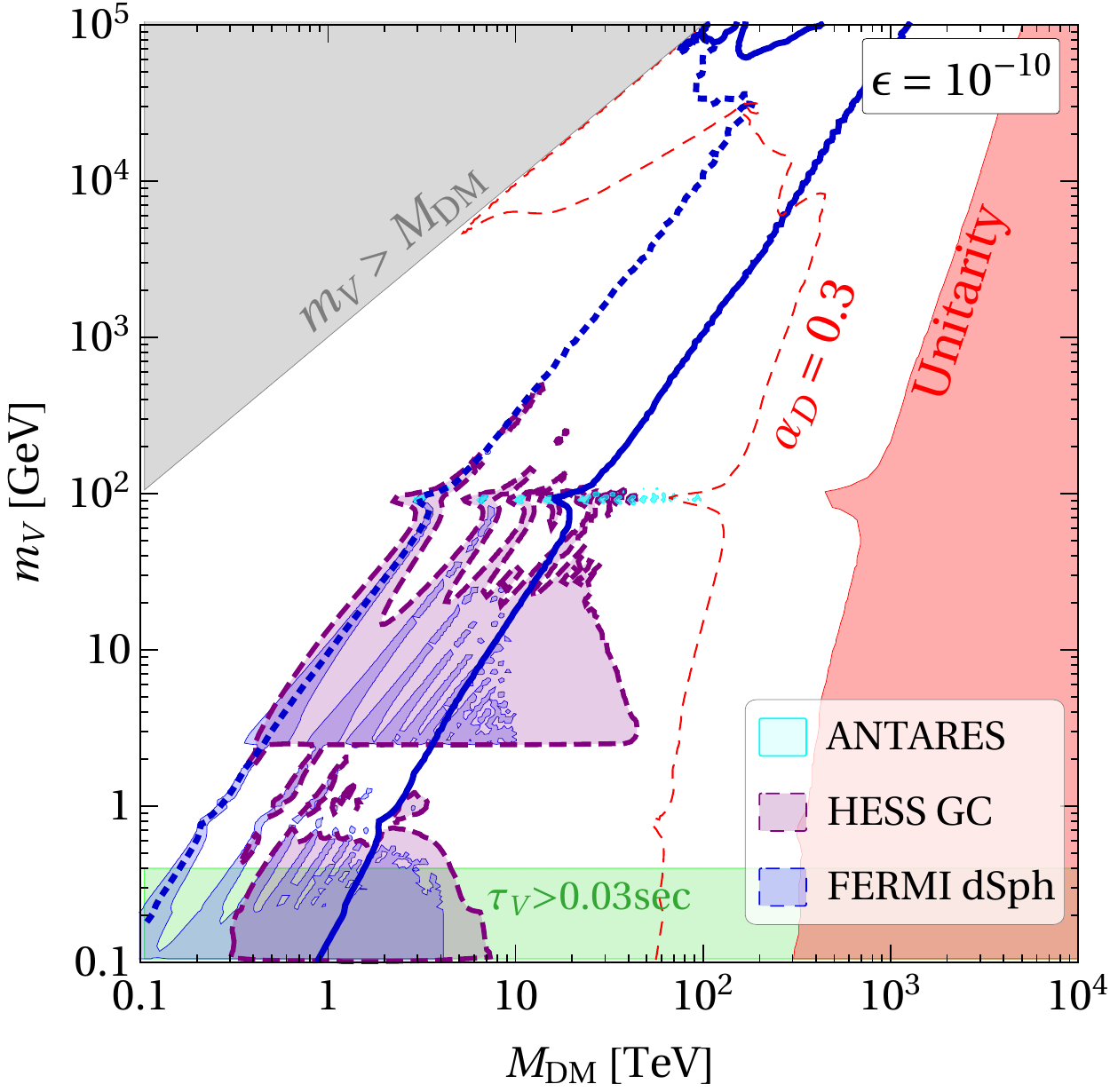}\quad
\includegraphics[width=0.44\textwidth]{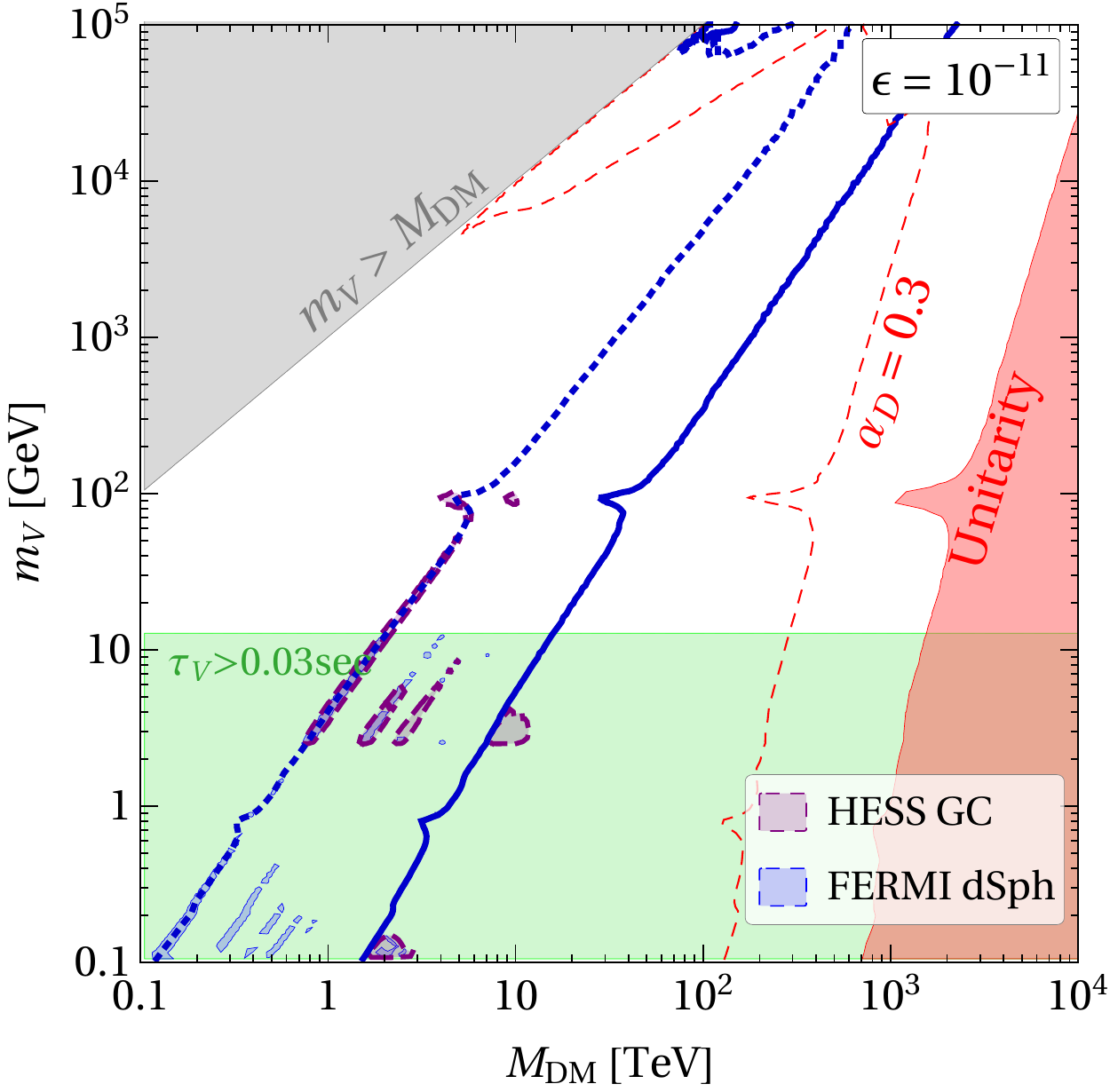}
\end{center}
\caption{\it \small \label{fig:LU_neutral} Local Universe limits from neutral messangers (neutrinos and photons). 
From top-left to bottom-right: no DM dilution, and then decreasing values of $\epsilon$ corresponding to increasing DM dilution. 
Shaded: excluded by {\sc Antares} (cyan), by {\sc Fermi} (blue) and within {\sc Hess} reach (purple). The other shaded areas and lines are as in fig.~\ref{fig:ID_summary}.
}
\end{figure}

The {\sc Antares} collaboration derived constraints from the non-observation of excesses in muon neutrino fluxes coming from the Milky Way halo~\cite{Albert:2016emp}.
Those limits were cast assuming direct DM annihilations into various SM pairs, and therefore cannot be directly applied to our case.
We circumvent this limitation following~\cite{Baldes:2017gzu}, that observed that the {\sc Antares} limits are driven by the most energetic part of the neutrino spectra from DM. Therefore they can roughly be applied to one-step signals that result in quarks (up to bottom), because their zero- and one-step neutrino spectra are very similar for neutrino energies close to $\MDM$.
They can also be applied to one-step signals that result in neutrinos, as is the case in our model for $\mV$ larger than a few tens of GeV, with a further caveat that we now discuss.
The resulting `one-step' neutrino spectrum is spread to energies lower than $\MDM$, while direct DM annihilation into neutrino pairs would induce a spike at $E_\nu \simeq \MDM$. However, the {\sc Antares} finite energy resolution of O(50\%) causes a sizeable fraction of `one-step' neutrinos to fall in the same energy range of those originating from a spike.
To summarize, we exclude a point in our parameter space if it does not respect the limit
\beq
\langle \sigma_{\rm tot} \vrel \rangle {\rm BR}(V \to \bar{f} f) < 2\,C\,\langle \sigma_{{\rm DM\,DM} \to f \bar{f}}\, \vrel \rangle_{\rm limit}^{\textsc{Antares}}
\label{eq:ID_ANTARES}
\eeq
for at least one of the two final states i) $\bar{f} f = \bar{b} b +\bar{c} c +\dots +\bar{u} u$, that we compare with the $\bar{b} b$ limit from {\sc Antares} because of the very similar spectrum, and ii) $\bar{f} f = \bar{\nu}_\mu \nu_\mu$, where we include the effect of oscillations ({\sc Antares} presents limits for DM annihilation into $\bar{\nu}_\mu \nu_\mu$, so that the effective flux of muon neutrinos at the telescope is reduced to roughly 40\% of the initial one).
The factor of 2 in eq.~(\ref{eq:ID_ANTARES}) accounts for the fact that, in our model, DM is not self-conjugate.
We also choose $C=1$ for quark final states and $C=2$ for neutrino ones, to very roughly account for the neutrinos that will be properly recognized to have $E_\nu < \MDM$.\footnote{This last prescription for neutrinos, together with the precise values of the dark photon branching ratios, are the only aspects in which this analysis departs from the one in~\cite{Baldes:2017gzu}.}

The resulting limits are shown in figure~\ref{fig:LU_neutral} for various values of the kinetic mixing $\epsilon$ (and therefore of DM dilution).
We find also an excluded region for $\mV < 2.5$~GeV, but we do not show it as it is an artifact of the way we modeled dark photon decays in that mass region, see~\cite{Baldes:2017gzu} for more details. 
Our limits can be considered conservative in the sense that one should have summed over the neutrino fluxes from all final states (this would be possible if {\sc Antares} data were public), and aggressive in the sense that {\sc Antares} assumes a DM density profile which is rather peaked towards the Galactic Center.
We do not show the analogous {\sc Icecube} limits~\cite{Aartsen:2017ulx} because they are not provided for $\MDM > 1$~TeV, and are weaker than those of {\sc Antares}.
The latter aspect could be caused by the fact that {\sc Icecube} does not perform as good as {\sc Antares} for neutrinos coming from the Galactic Center, which drive the exclusions (see also~\cite{ElAisati:2017ppn} for a derivation of DM limits from IceCube data).

\subsubsection{Gamma rays}
\label{sec:ID_local_gamma}

We consider first constraints from the {\sc Fermi} satellite observations of several dwarf spheroidal galaxies (dSphs)~\cite{Ackermann:2015zua}, which are the strongest ones on our model among those derived from {\sc Fermi} data (see~\cite{Baldes:2017gzu}).
We use the results of~\cite{Profumo:2017obk}, which derived limits on secluded DM models starting from public {\sc Fermi} data from dSphs observations, and presented them for different assumptions on the decay products of the mediator and of its mass. In particular, we exclude a point in our parameter space if it does not satisfy
\beq
\langle \sigma_{\rm tot} \vrel \rangle {\rm BR}(V \to \bar{f} f) < 2\,\langle \sigma_{{\rm DM\,DM} \to f \bar{f}}\, \vrel \rangle_{\rm limit}^\gamma
\label{eq:ID_Gammas}
\eeq
for at least one of the final states $\bar{f} f = \bar{b} b, \bar{q}q, \tau^+\tau^-,\dots$, and where again the factor of 2 accounts for the fact that, in our model, DM is not self-conjugate.
The resulting exclusions are shown in figure~\ref{fig:LU_neutral}, for various values of $\epsilon$. We find that these exclusions do not significantly depend on the assumption on $\mV$, among the two presented in~\cite{Profumo:2017obk}.
As a cross-check we note that the excluded regions are in very good agreement with those derived in~\cite{Cirelli:2016rnw}, for the same model in the case of no dilution, and which made direct use of {\sc Fermi} data.

Moving to higher energy gamma rays, we now consider observations of gamma rays from the GC with the {\sc Hess} telescope~\cite{Abdallah:2016ygi}.
We use the results in~\cite{Profumo:2017obk} that, analogously to what done for {\sc Fermi} dSphs, derived {\sc Hess} limits on secluded DM models.
An important caveat is that they are derived from the public data used in the {\sc Hess} analysis~\cite{Abramowski:2011hc} (corresponding to 112 hours of observation time), but presented assuming a rescaling of the 112h limits with the observation time of 254h on which~\cite{Abdallah:2016ygi} is based (because these extra data are not public).
Therefore we only treat these limit as an indication of the sensitivity, to this model, that could be achieved by the {\sc Hess} telescope. In this spirit, we extend them up to $\MDM = 70$~TeV, because this is the maximal DM mass reached in the {\sc Hess} analysis~\cite{Abdallah:2016ygi}.
We derive the sensitivities as in eq.~(\ref{eq:ID_Gammas}), where now of course $\langle \sigma_{{\rm DM\,DM} \to f \bar{f}}\, \vrel \rangle_{\rm limit}^\gamma$ are the exclusions given in~\cite{Profumo:2017obk},  and show them in figure~\ref{fig:LU_neutral}.
We stress that the {\sc Hess} sensitivity, from observations of the GC, would completely evaporate if the DM density profile has a core of size larger than~500~pc, and that it would be substantially weakened for cores down to a few tens of pc~\cite{HESS:2015cda} (both possibilities are currently allowed both by simulations~\cite{DiCintio:2013qxa,Marinacci:2013mha,Tollet:2015gqa} and by data~\cite{Nesti:2013uwa,Cole:2016gzv,Wegg:2016jxe}).

Other ground-based telescopes have observed high energy gamma rays and cast bounds on annihilations of heavy DM, e.g. {\sc Veritas}~\cite{Archambault:2017wyh}, {\sc Magic}~\cite{Ahnen:2017pqx} and {\sc Hawc}~\cite{Albert:2017vtb,Abeysekara:2017jxs}.
We do not show the resulting limits here because they are derived assuming no steps in the DM annihilations, and because they are anyway weaker than the {\sc Hess} sensitivity that we show.

\subsubsection{Antiprotons}
\label{sec:ID_local_pbar}

We now turn to the constraints imposed by galactic antiproton observations. The method follows closely the previous analyses in~\cite{Boudaud:2014qra,Giesen:2015ufa,Cirelli:2016rnw,Baldes:2017gzu}. For completeness, however, we briefly review it here.
 
Consistently with the previous studies, we use the {\sc Ams} measurement~\cite{Aguilar:2016kjl} of the antiproton to proton ratio over the range from 1 to 450 GeV. The data are rather well explained by astrophysical antiproton sources, at least within the uncertainties related to $\bar p$ production in the interstellar medium, to their propagation process in the galactic environment and to the modulation of their flux due to solar activity~\footnote{Possible claimed excesses at low energies~\cite{Hooper:2014ysa, Cuoco:2016eej, Cui:2016ppb, Huang:2016tfo, Feng:2017tnz, Arcadi:2017vis} are now shown to have a limited statistical significance~\cite{Reinert:2017aga}.}. 
As for the DM contribution, we adopt an Einasto profile for the distribution in the MW galactic halo and we set the local dark matter density to $\rho_{sun}=0.42$~GeV/cm$^3$~\cite{Pato:2015dua}. We choose a `MAX' propagation scheme, as it is the one favored by the fit of the astrophysical background.
Upon imposing that the DM contribution does not worsens the astrophysics-only fit by more than $\Delta \chi^2 =9$, we obtain conservative DM bounds.

The excluded regions are highlighted in magenta in fig.~\ref{fig:LU_charged}. They consist of an area below $M_{\rm DM} \lesssim 1$ TeV, where the DM $\bar p$ contribution falls into the relatively low energy and high precision portion of the {\sc Ams} data, and a series of funnels at $M_{\rm DM} \gtrsim 1$ TeV, corresponding to the resonances of the annihilation and BSF cross-sections. The excluded regions terminate for $m_{\rm V} \lesssim 2.5$ GeV, consistently with the fact that the BRs of a light $V$ into protons and antiprotons are inhibited (see the discussion in sec.~\ref{sec:darkphoton} and appendix~\ref{app:DP}).

\subsubsection{Electrons and positrons}
\label{sec:ID_local_epem}

\begin{figure}[!t]
\begin{center}
\includegraphics[width=0.44\textwidth]{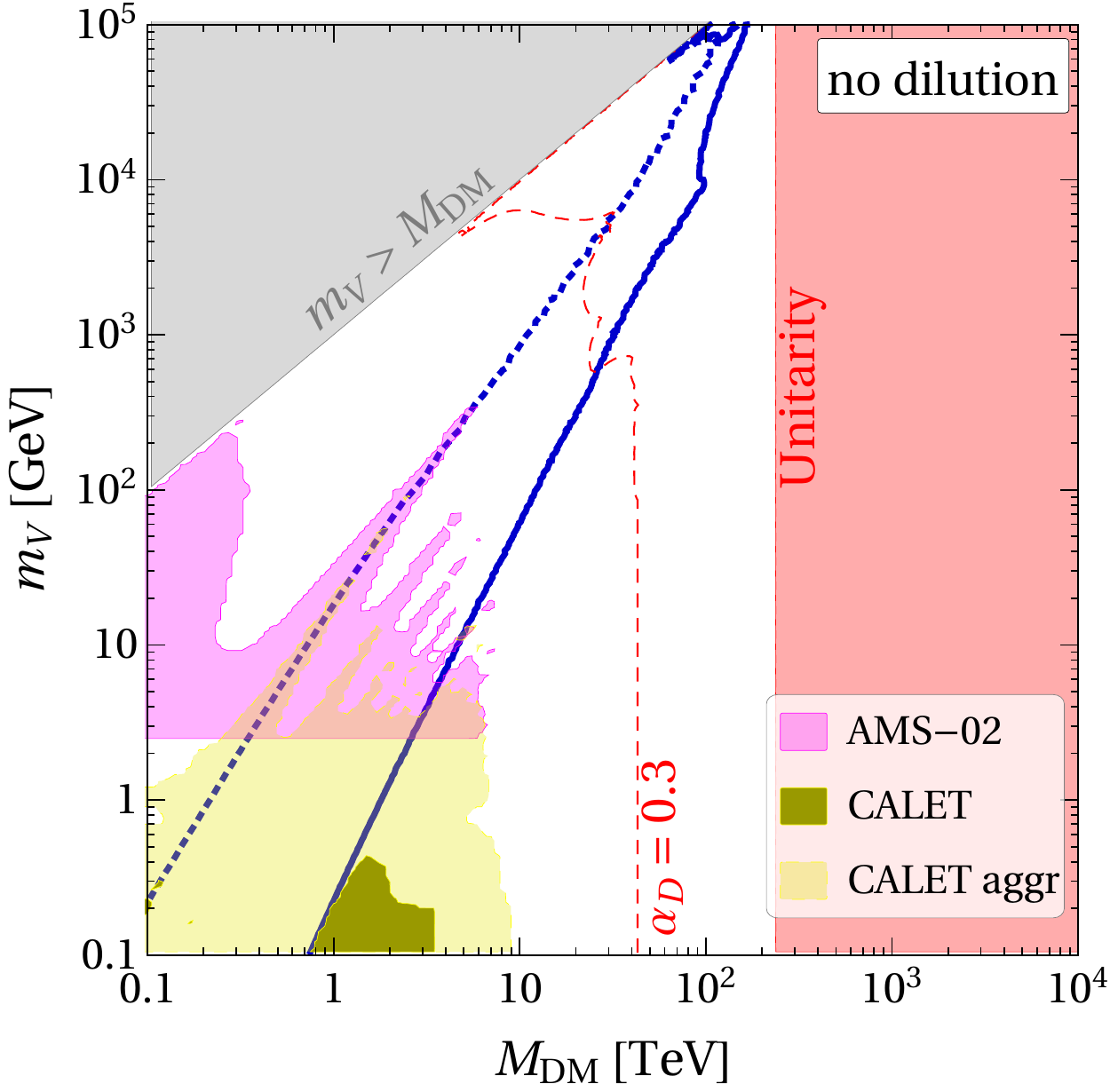}\quad
\includegraphics[width=0.44\textwidth]{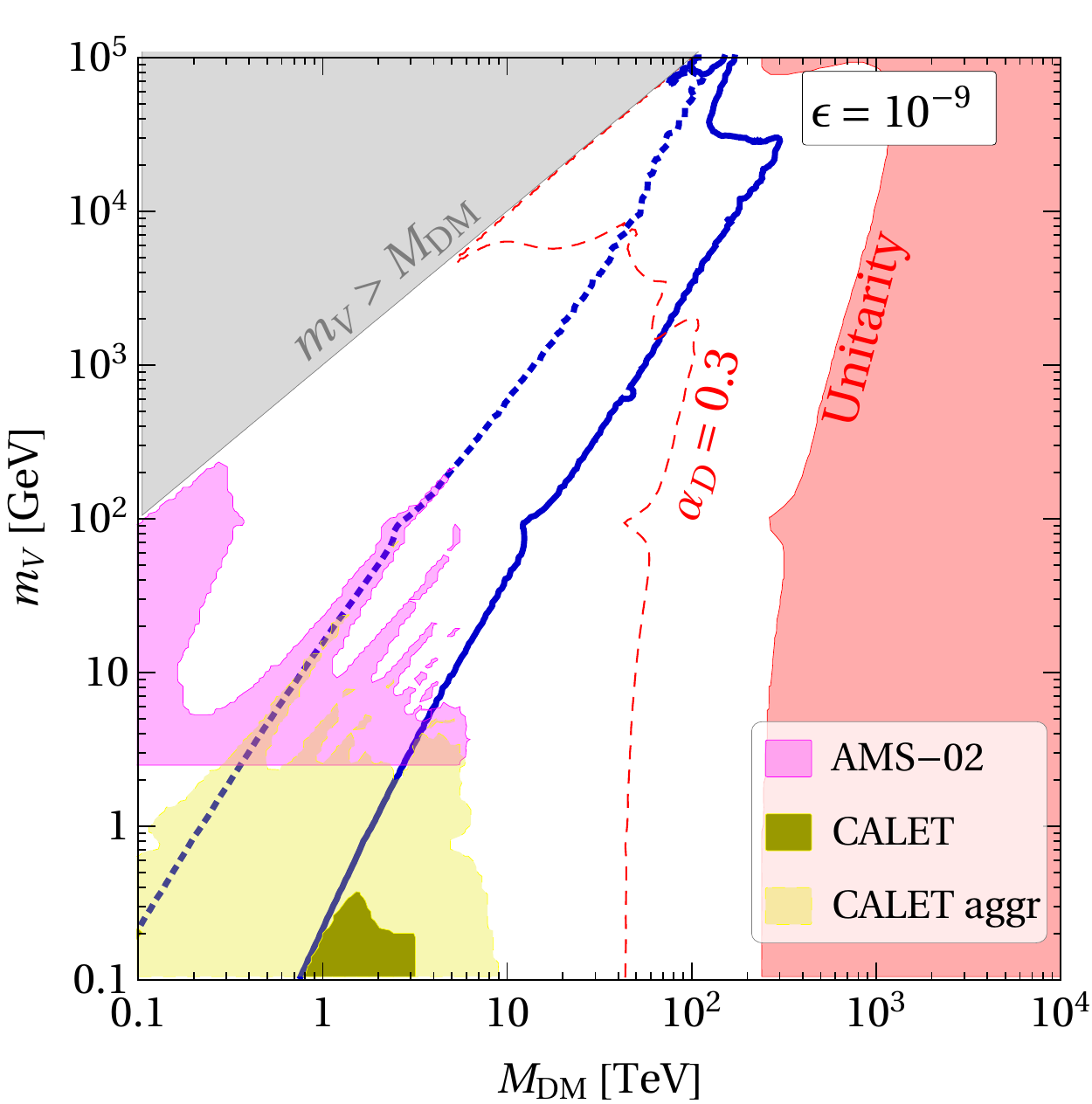}\\
\vspace{.3 cm}
\includegraphics[width=0.44\textwidth]{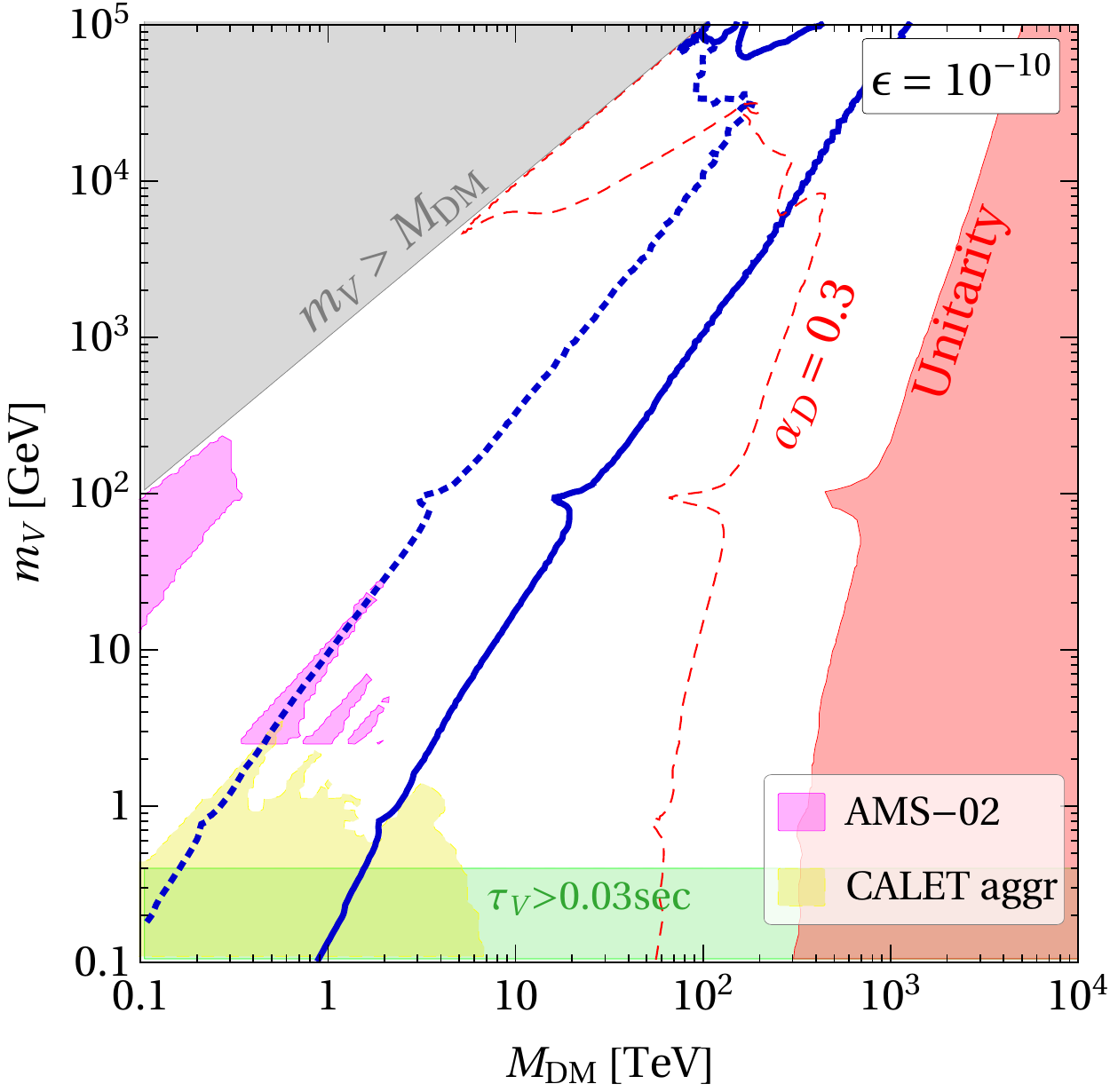}\quad
\includegraphics[width=0.44\textwidth]{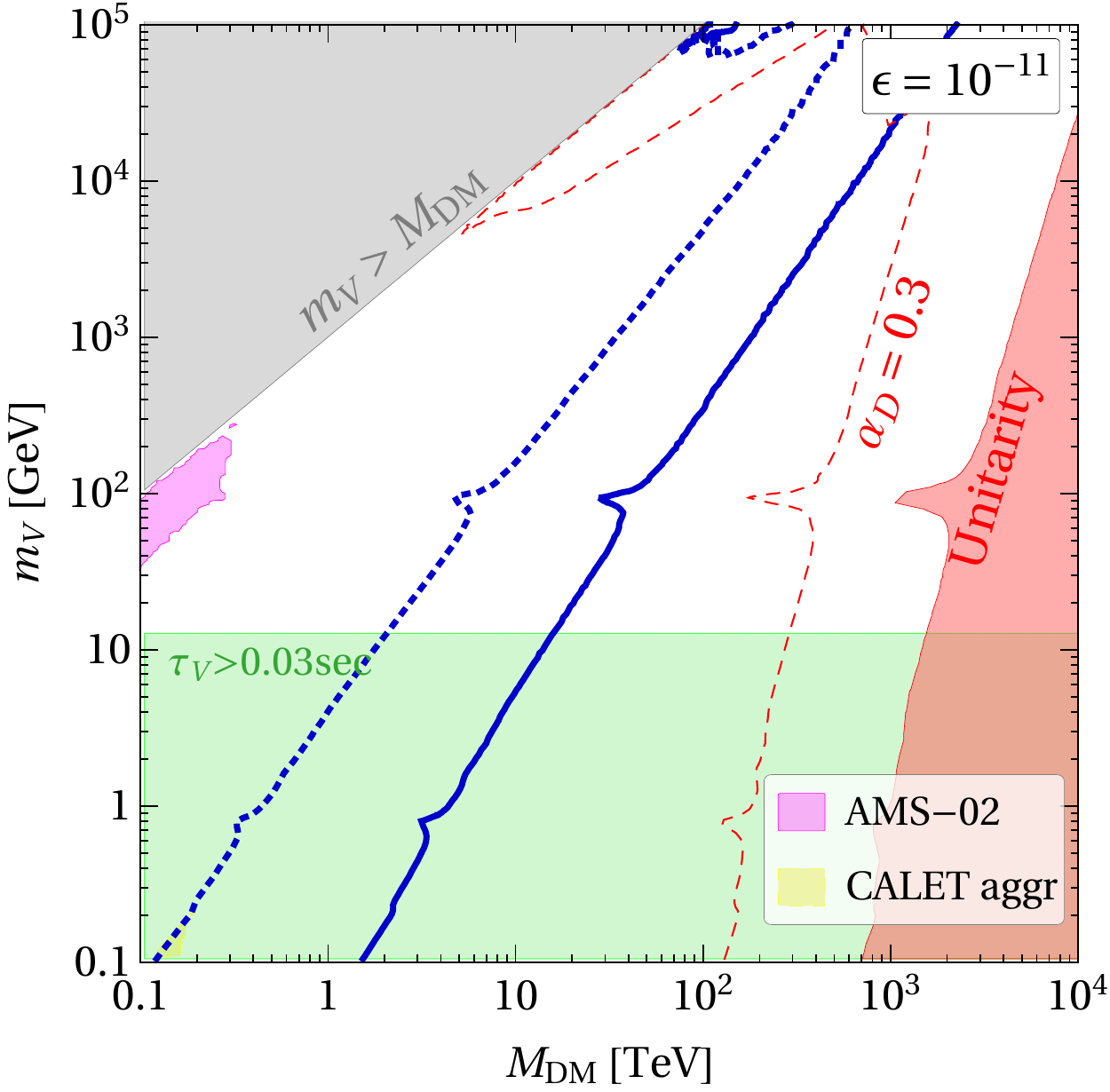}
\end{center}
\caption{\it \small  \label{fig:LU_charged} Local Universe limits from charged messangers (antiprotons and electrons plus positrons).
Shaded: excluded by {\sc Ams} (magenta), by {\sc Calet} assuming no background knowledge (ocra) and by {\sc Calet} assuming background knowledge (light yellow) 
The other shaded areas and lines are as in fig.~\ref{fig:ID_summary}.
}
\end{figure}

We derive constraints also from the measurements of the `all-electron flux' (i.e.: $e^++e^-$) by the {\sc Calet} experiment~\cite{Adriani:2017efm} onboard the International Space Station, which cover the range 10 GeV $-$ 3 TeV. Measurements of the same observable have also been produced by the {\sc Dampe} satellite~\cite{Ambrosi:2017wek}, by {\sc Ams}~\cite{Aguilar:2014fea} and by {\sc Hess}~\cite{HESSICRC2017}. The {\sc Calet} data agree well with {\sc Ams} but the former extend to higher energies: since our focus is on very heavy DM, we use {\sc Calet}. The data by {\sc Hess}, which are an update of former measurements~\cite{Aharonian:2008aa,Aharonian:2009ah} with extended energy range, have so far been presented only at conferences. They display large systematic uncertainties and we checked that they are currently non competitive (in terms of constraining power), so that we do not use them here.
Finally, the data by {\sc Dampe} are somewhat higher in normalization then those by {\sc Calet} and {\sc Ams}, in the range 50$-$1000 GeV, and have a harder spectrum. As this might be due to energy miscalibration in {\sc Dampe}, we use {\sc Calet} only.

The astrophysical fluxes of $e^++e^-$ are poorly modeled at the high energies under consideration. We therefore decide to proceed, as customary in these case, with two methods: a conservative one and a more aggressive one. In the conservative approach, we assume no astrophysical background, we compute the DM $e^++e^-$ and we impose that it does not exceed any {\sc Calet} data point by more than 2$\sigma$. This yields the excluded regions in ocra in fig.~\ref{fig:LU_charged}, which sit in the ballpark of $M_{\rm DM} \sim$ few TeV and $m_{\rm V} \lesssim$ 500 MeV. They shrink and disappear as soon as the dilution parameter $\epsilon \lesssim 10^{-10}$. 
In the aggressive approach, we include an astrophysical $e^++e^-$ background flux modeled with a power-law fitted on the {\sc Calet} data over the whole energy range. We find $E^3 \, \Phi^{e^++e^-}_{\rm astro} = 3.4 \times 10^{-2} E^{-0.18} \, {\rm GeV}^2/({\rm cm}^{2} \, {\rm s}\, {\rm sr})$, with $\chi^2 = 16.4$ for 38 d.o.f., in reasonably good agreement with the analogous power law fit quoted in \cite{Adriani:2017efm}.
Then, in analogy with the procedure followed for antiprotons, we impose that the DM contribution does not worsens the modeled-astrophysics-only fit by more than $\Delta \chi^2 =9$ and we obtain the regions drawn in yellow in fig.~\ref{fig:LU_charged}. Not surprisingly, these regions are much more extended than those derived in the conservative approach. We stress, however, that these regions should be interpreted more as a {\it reach} of the $e^++e^-$ measurements {\it under the assumption} that the astrophysical flux is well modeled by the {\it empirical} fit described above. In order to be conservative, we do not show them in the summary of Figure~\ref{fig:ID_summary}.

\subsubsection{On galactic subhalos and on DM limits from clusters of galaxies.}

In all the limits presented in this section~\ref{sec:ID_local}, we have not included a possible `boost' of the DM signals from the existence of DM clumps. We summarize their possible impact here.

In principle, the presence of DM subhalos is indeed expected to enhance the annihilation flux~\cite{Silk:1992bh,Bergstrom:1998jj,Ullio:2002pj,Berezinsky:2003vn}.
However, in the central regions of the Milky Way subhalos are typically tidally disrupted by both the GC and the Galactic disk, so that such a boost turns out to be negligible for signals, e.g.~in gamma rays, originating from the inner kiloparsecs (see~\cite{Stref:2016uzb} for a recent assessment).
Boost factors up to a few tens of percent are possible for signals coming from dwarf spheroidal galaxies, even though, for conservative assumptions on the mass index, they become again negligible~\cite{Moline:2016pbm}.
Boosts of signals coming from our neighborhood in the MW, relevant for DM searches in charged cosmic rays, are found not to exceed a few tens of percent for conservative assumptions~\cite{Lavalle:2006vb,Lavalle:1900wn,Pieri:2009je,Stref:2016uzb}.

On the basis of the above, we find it justified to neglect the impact of DM subhalos as far as limits are concerned. On the other hand, we underline that the interplay of a possible DM signal in charged cosmic rays with gamma observations could provide precious indications on DM subhalo existence and properties.

DM subhalos might have a more dramatic impact on DM signals from galaxy clusters~\cite{Moline:2016pbm,Stref:2016uzb}, leading to an interesting interplay with other signals from heavy annihilating DM~\cite{Murase:2012xs}.
However, DM limits from galaxy clusters are either negligible with respect to all those considered here~\cite{Abramowski:2012au}, or at most comparable for aggressive assumptions on the DM subhalo properties~\cite{Ackermann:2015fdi}. Therefore, while a possible ID signal from a cluster could in principle be ascribed to annihilating DM, we do not discuss these limits further here.

We finally comment that additional DM subhalos could arise in cosmological histories with a matter-dominated phase after DM (chemical and thermal) decoupling, see~\cite{Gelmini:2008sh,Erickcek:2011us,Erickcek:2015jza} for quantitative studies of this possibility. The investigation of the formation and survival of these subhalos in our model, while interesting, goes beyond the purpose of this paper. In the spirit of being conservative on this matter we neglect it in our ID constraints.

\subsection{On the dark photon emitted during the formation of bound states}
\label{sec:ID_DPemittedBS}

So far we have not discussed the signals from the low-energy dark photon emitted during BSF, which could potentially strengthen the indirect detection constraints. This $V$ carries away the binding energy $E_B = \aD^2 \MDM/4$, where we neglect the kinetic energy $\vrel^2 \MDM/4$ because $\aD \gtrsim \vrel$ when BSF is important.
It thus produces a spectrum of SM final states analogous to the one produced by a DM of mass $E_B/2$ annihilating directly into SM pairs (the same SM pairs that $V$ decays into). However, DM with such low mass would have number density larger by $\MDM/(E_B/2) = 8/\aD^2$. 

We may then recast the existing upper bounds, $\overline{\sigma v}_\mathsmaller{\text{rel}}$, on (non-self-conjugate) DM annihilating into SM pairs as follows
\beq
\langle \sigma_{\BSF} \vrel \rangle (\MDM, \mV) < 
\Big(\frac{8}{\aD^2}\Big)^2   \times 
\overline{\sigma v}_\mathsmaller{\text{rel}} (M = \aD^2 \MDM /8)\,.
\label{eq:BSFphoton_RecastID}
\eeq
Even for large values of the dark fine structure constant, the prefactor above is rather large ($64/\aD^4 \gtrsim 10^3$ for $\aD \lesssim 0.5$), and the existing indirect detection bounds (see e.g.~\cite{Ade:2015xua,Ackermann:2015zua} for some strongest ones) do not yield any further constraints on our parameter space -- except perhaps on top of the resonances. We shall therefore not consider this low-energy dark photon further.

Note that the CMB and 21~cm constraints scale proportionally to the DM mass [cf.~eqs.~\eqref{eq:CMB} and \eqref{eq:21cm}], which partly mitigates the loosening of the constraints seen in \eqref{eq:BSFphoton_RecastID}. However, as noted earlier, BSF via vector emission is a $p$-wave process, and becomes entirely negligible for the very low $\vrel$ during CMB.

\section{Summary and Outlook}
\label{sec:outlook} 

High energy telescopes constitute our unique direct access to energies much above a TeV, therefore they are in a privileged position to test BSM physics at those scales. Annihilating heavy Dark Matter is a particularly motivated example of such BSM physics.
The existing and upcoming telescope data at unprecedented energies, together with the growing theoretical interest in DM with masses beyond 10-100 TeV, make it particularly timely to explore the associated phenomenology.
In this paper we performed a step in this direction.

We pointed out that models where DM pairs annihilate mostly in `dark' mediators, themselves decaying into SM particles, allow to naturally evade the unitarity limit on the DM mass and, at the same time, to reliably compute cosmic ray signals. They therefore overcome the two main obstacles to the use of data of high energy telescopes to test annihilating heavy Dark Matter.
Encouraged by this observation, we then studied in detail several aspects of the cosmological history and of the indirect detection signals of these models, which we summarise as follows:
\begin{enumerate}
\item We computed the dilution of relics induced by i) the entropy injection from decays of the mediators after DM freeze-out, and ii) the possibility that the SM and dark sector were not in thermal equilibrium at early times.
We then determined how i) and ii) affect a) the thermal freeze-out cross section and, in turn, the maximal DM mass compatible with unitarity, see Figures~\ref{fig:sigmaFOandMuni} and \ref{fig:UnitaryLimit}; b) the maximal dilution compatible with BBN constraints, see Section~\ref{sec:max_homeopathy}. 
Our study is the first, to our knowledge, to systematically explore the combination of these effects, therefore we reported ready-to-use analytical results (on dilution, freeze-out cross section, BBN limits, maximal DM mass allowed by unitarity) throughout Section~\ref{sec:dilution}.

\item As a case study we then considered a model of fermion DM charged under a spontaneously broken dark $U(1)$, where the role of the mediator is played by the associated massive vector. 
We computed the values of the dark gauge coupling that results in the correct DM abundance, over a wide mass range and for different values of the kinetic mixing and therefore of dilution (see Figure~\ref{fig:alphaD} right). Our freeze-out computation  improves over previous ones in that it includes long-range interactions beyond the Coulomb approximation, which is particularly relevant for heavy DM and mediator masses (see Figure~\ref{fig:alphaD} left).
The resulting cross sections relevant for DM indirect detection are shown in Figure~\ref{fig:sigmav_MW_unitarity_mVD_10GeV}.

\item We finally studied the indirect detection phenomenology of the Dark $U(1)$ model, both with and without dilution, see Figure~\ref{fig:ID_summary} for a summary.
We found the encouraging result that multimessenger searches for heavy DM are possible, offering different lines of attack to test these models.
Our findings also give reasons to think that some telescopes (especially neutrino and gamma ray) have an unexplored potential to test annihilating DM with a mass of  $O(100)$~TeV. This motivates an experimental effort in this direction and, importantly, the grant of public access to experimental data in some form.
The latter would also overcome the problem, in interpreting experimental results, that some DM models (e.g. secluded ones) result in SM spectra different from the benchmarks used by the collaborations.

\end{enumerate}

Our results open the doors to many possible future investigation.
Concerning high energy cosmic rays, it would be interesting to determine the sensitivity of future experiments like {\sc Cta, Lhaaso, Taiga, Km3net, Herd, Iss-Cream} on secluded DM models, and to explore whether annihilations of heavy secluded DM could explain the highest energy neutrinos observed by {\sc Icecube}.
It would also be intriguing to study the interplay of indirect detection signals with other effects of entropy injections, e.g. on the baryon asymmetry.
On a more theory side, two exciting directions of exploration regard the connection, of the small portal of these models with the SM and/or of the heavy DM mass, with the solution of other problems of the SM.
We plan to return to some of these aspects in future work.


\medskip

\section*{Acknowledgements}

We thank Iason Baldes, Netra Gourlay, Paolo Panci, G\'eraldine Servant, Marco Taoso and Sebastian Wild for useful discussions. We are grateful to Andrea Mitridate and Alessandro Podo for discussions that led us to identify a mistake in our BBN treatment in the first version of this work.
Y.G. and F.S. thank the LPTHE Paris and M.C. thanks the IAP Paris for hospitality.
\paragraph*{Funding information}
This work has been done in part within the French Labex ILP (reference ANR-10-LABX-63) part of the Idex SUPER, and received financial state aid managed by the Agence Nationale de la Recherche, as part of the programme {\it Investissements d'avenir} under the reference ANR-11- IDEX-0004-02. 
The work of M.C.~is supported by the European Research Council ({\sc Erc}) under the EU Seventh Framework Programme (FP7/2007-2013) / {\sc Erc} Starting Grant (agreement n. 278234 - {\sc `NewDark'} project). 
The work of Y.G. and F.S. is partly supported by a PIER Seed Project funding (Project ID PIF-2017-72).
K.P.~was supported by the ANR ACHN 2015 grant (``TheIntricateDark" project), and by the NWO Vidi grant ``Self-interacting asymmetric dark matter".

\medskip

\appendix

\section{More on the computation of the dilution factor}
\label{app:dilution}

In this appendix we present the computation of the dilution factor in more detail, and study the validity of different approximations. 
We do not discuss the validity of the instantaneous decay approximation since this has already been done in \cite{Scherrer:1984fd}.

\subsection*{Computation of $\boldsymbol{\DSM}$ and validity of related approximations}

\begin{figure}[t]
\begin{center}
\includegraphics[width=0.48 \textwidth]{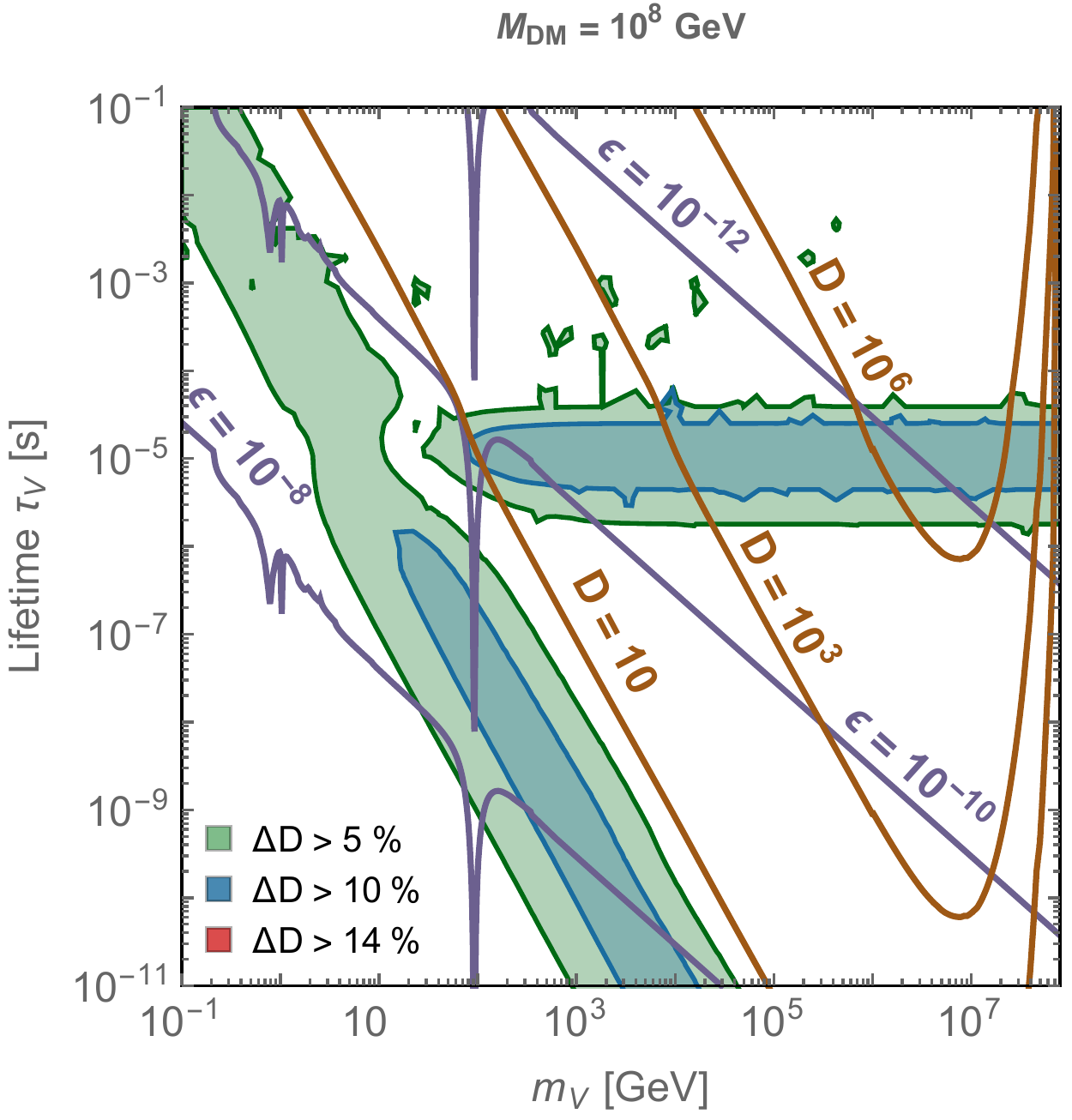}
\includegraphics[width=0.48\textwidth]{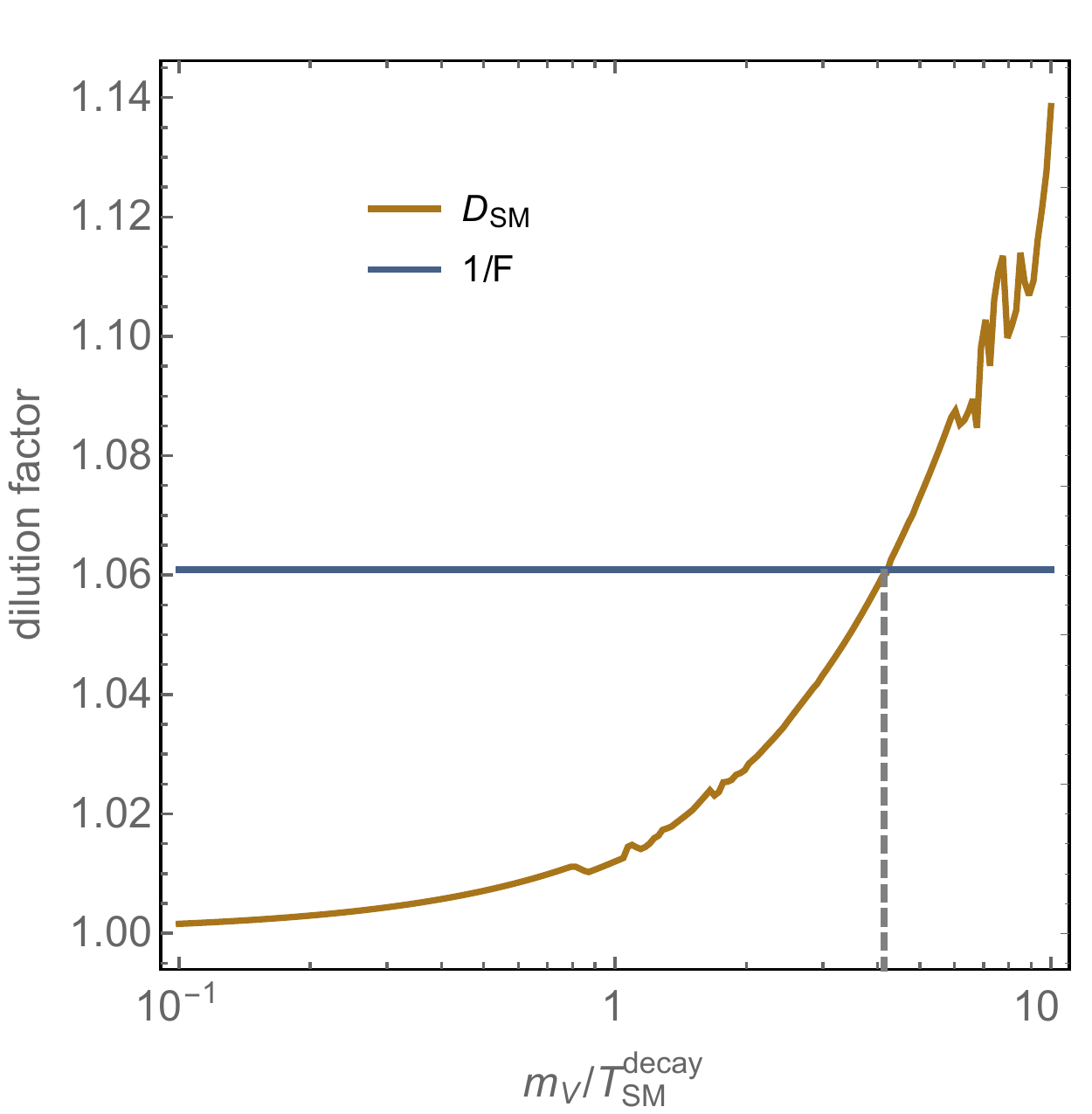}
\end{center}
\caption{\label{fig:dilution_factor_vs_mVDOverTdec} \it \small Left: $\Delta D = 200 |\frac{D_{\rm full} - D_{\rm approx}}{D_{\rm full} + D_{\rm approx}}|$ where $D_{\rm full}$ is defined in eq.~\eqref{eq:dilution_factor_expression} and where $D_{\rm approx}$ is defined in eq.~\eqref{eq:dilution_factor_expression_VD_dom_approx} together with eq.~\eqref{eq:gSM_approx1}. As expected, the errors arise in the limit when the mediator decays before dominating the energy density of the univers, corresponding to a dilution factor of order a few, and when the mediator decays around the QCD scale when the number of relativistic degrees of freedom is very sensitive to the temperature, corresponding to lifetimes around $10^{-5}$ s.
Right: SM dilution factor $\DSM$ under the assumption that the relic is non-relativistic at the time of its decay as specified in eq.~\eqref{eq:dilution_factor_expression} versus $\mV/\TSM^{\rm decay} $.
We can see that $\DSM$ is smaller than $1/F$ and then, according to the discussion in appendix~\ref{app:dilution_DP_relativistic}, that the 2nd law of thermodynamics is violated when $\mV/\TSM^{\rm decay} \le 4$. This corresponds to the limit where the relic is relativistic.
}
\end{figure}

In order to calculate the dilution factor \eqref{eq:dilution_factor_expression}, we must first compute the normalised scale factor  $z(x) \equiv a(x)/a^{\rm before}$, where $x \equiv \GammaV t$ is a time parameter. Following \cite{Scherrer:1984fd}, the Friedmann equation for $z(x)$ in presence both of the SM radiation and of the mediator decaying into SM radiation, reads
\begin{equation}
\frac{z'(x)}{z(x)} = 
\frac{1}{x_\text{before}} 
\left[
\frac{e^{-x}}{z(x)^3} 
+ \frac{1}{\,\gSM(x)^{1/3} \, z(x)^4} \, \int_{x_{\rm before}}^{x}\, \gSM^{1/3}(x')\,z(x')\,e^{-x'}dx'
+ \frac{1}{z(x)^4}\left(\frac{\rho_r^{\rm before}}{\rho_V^{\rm before}} \right)  
\right]^{1/2},
\label{eq:Friedmann_eq}
\end{equation}
where the energy densities of the SM bath and the dark mediator before the mediator starts to dominate the energy density of the universe are 
\begin{align}
\rho_{r}^{\rm before} &= (\pi^{2} / 30) \, \gSM^{\rm before} \, T_{\rm SM, before}^{4} \,,
\\
\rho_V^{\rm before} &= \mV \fV (2\pi^{2}/45) \gSM^{\rm before}T_{\rm SM, before}^{3} \,,
\end{align}
and where
\begin{equation}
\label{eq:xH}
x_\text{before} \equiv \GammaV \sqrt{3\MPl^{2}/\rho_V^{\rm before}}.
\end{equation}
In the parameter range of interest of the dark $U(1)$ model, we can choose $\TSM^{\rm before}=100$ PeV and we set the initial condition
 \begin{equation}
z(x_{\rm before})=1.
\label{eq:z_boundary}
\end{equation}
A quantity necessary to compute the dilution is the average of $\gSM$ over the decay period 
\begin{equation}
\gSMdec = \left[ \frac{\int_{0}^{x} \gSM^{1/3}(z) z(x') e^{-x'} d x'}{\int_{0}^{x} z(x') e^{-x'} d x'} \right]^{3}
\label{eq:averaged_g_SM}
\end{equation}
Since the function $z(x) \, e^{-x}$ in eq.~\eqref{eq:averaged_g_SM} is maximal for $x \approx 1$, as discussed in \cite{Scherrer:1984fd}, we can replace the number of degrees of freedom averaged over the decay period $\gSMdec$ in eq.~\eqref{eq:averaged_g_SM} by the number of relativistic degrees of freedom in the SM at the time of decay.  Moreover, as discussed in sec.~\ref{sec:max_homeopathy} the temperature of the Universe at the time of decay is identical to the temperature of a standard radiation-dominated Universe whose age is equal to the lifetime of the dark photon. So we can use the very good approximation
\begin{equation}
\label{eq:gSM_approx1}
\gSMdec \approx \gSM \left( T_{\rm rad}(\tau_V) \right) 
\end{equation}
where $T_{\rm rad}(t)=\frac{1\;{\rm MeV}}{\gSM\left(1\;{\rm MeV}\sqrt{\frac{2.42\;s}{t}}\right)^{1/4}}\sqrt{\frac{2.42\;s}{t}}$ is obtained from the expression of the Hubble factor for a radiation-dominated Universe $H=1/2t$. It is shown in fig.~\ref{fig:dilution_factor_vs_mVDOverTdec} that the approximation in eq.~\eqref{eq:gSM_approx1} as well as the assumption in eq.~\eqref{eq:dilution_factor_expression_VD_dom_approx} that the mediator dominates the universe before decaying gives less than $14\%$ error.

\subsection*{The case where the mediator is non-relativistic at chemical decoupling}
\label{app:dilution_DP_non_relativistic_FO}

The mediator relic density is fixed during the dark matter freeze-out. However when $m_V$ is close to $\MDM$, the mediator is non-relativistic at that time and eq.~\eqref{eq:fV} has to be replaced by
\begin{equation}
\fV = \fV^{\rm rel}  \frac{1}{2\zeta(3)} I(\mV / T_{D})
\end{equation}
with $I(x) = \int_{0}^{\infty} d\xi \; \frac{\xi^2}{\exp{\sqrt{\xi^2 + x^2}-1}}$ and where $\fV^{\rm rel}$ is defined by eq.~\eqref{eq:fV}. In the non-relativistic limit, $I(x) =\sqrt{\frac{\pi}{2}} x^{3/2} e^{-x}$ and the dilution factor is exponentially suppressed.

\subsection*{The case where the mediator is relativistic when it decays}
\label{app:dilution_DP_relativistic}

As stated in eq.~\eqref{eq:D_and_DSM}, the total dilution factor $D$ and the SM dilution factor $\DSM$ defined in eq.~\eqref{eq:dilution_factor_definition} are related through $ D = F\; \DSM$, 
with $F = 1/(1 + \frac{\tilde{g}_{D}}{\tilde{g}_{SM}} \rtilde^3)$. In the limit where the mediator is relativistic, the total entropy of the Universe, $S=s \, a^{3}$, must be conserved and the total dilution factor $D$ must be equal to 1.
Therefore, in the massless limit the SM dilution factor must be equal to 
\begin{equation}
\DSM= 1/F
\end{equation}
and for finite mass, $\DSM$ must be always larger than $1/F$
\begin{equation}
\label{eq:lower_limit_D_SM}
\DSM \ge 1/F.
\end{equation}
Having $\DSM$ smaller than $1/F$ corresponds to $D$ smaller than $1$ which is forbidden by the 2nd law of thermodynamics. 
However, taking the relativistic limit of $\DSM$, $m_{V}\to 0$ in eq.~\eqref{eq:dilution_factor_expression} leads to $\DSM \to 1$, which clearly violates eq.~\eqref{eq:lower_limit_D_SM}. But we already knew that the massless limit was unphysical since the non-relativistic expression for the energy density of the dark photon has been assumed for computing $\DSM $ in eq.~\eqref{eq:dilution_factor_expression}. For consistency, we checked that the violation of the 2nd law of thermodynamics due to the non-relativistic approximation in eq.~\eqref{eq:dilution_factor_expression} only occurs in the relativistic limit, usually defined as $m/T \lesssim 3$, e.g. \cite{Kolb:1990vq}. Indeed, in fig.~\ref{fig:dilution_factor_vs_mVDOverTdec} (right) 
we can see that $\DSM$ becomes lower than $1/F$ for $m/T< 4$. Then we can replace $\DSM$ by $\DSM^{\rm corr.}$
\begin{equation}
\DSM^{\rm corr.} = \text{Max}[\DSM, 1/F]
\end{equation}
where $\DSM^{\rm corr.}$ respects the 2nd law of thermodynamics\footnote{This prescription can be considered as ineffective when considering a dark sector with a lower number of relativistic degrees of freedom than the SM since then, in the relativistic limit the SM dilution factor is almost $1$. For instance, in the case of the $U(1)_{D}$ model discussed in section~\ref{sec:U1model}, $ \tilde{g}_{D} = 6.5$ and $\tilde{g}_{SM} = 106.75$ lead to $\DSM^{\rm corr.}= \rm Max[\DSM, 1.06]$} as stated in eq.~\eqref{eq:lower_limit_D_SM}. This is the procedure we have employed in the main text.

\section{Dark Photon Decay Widths}
\label{app:DP}
We have computed the decay width  of the dark photon $V$ into pairs of SM fermions, at tree-level, in the small $\epsilon$ limit. The final expression reads
\begin{equation}
\Gamma(V \rightarrow  \bar{f}f) = \frac{N_{f}}{24\pi} \mV \sqrt{1-4\delta_{f}^{-2}} \left[ g_{f_{L}}^{2}+g_{f_{R}}^{2} - \delta_{f}^{-2} (g_{f_{L}}^{2} + g_{f_{R}}^{2} - 6g_{f_{L}}g_{f_{R}}) \right]
\label{eq:VD_decay_rate_ff}
\end{equation}
with 
\begin{align}
&g_{f} = \epsilon e \left[ Q_{f} \frac{1}{1-\delta_{Z}^{2}}  - \frac{Y_{f}}{c_{w}^{2}} \frac{\delta_{Z}^{2}}{1-\delta_{Z}^{2}}  \right ] + O(\epsilon^{2}), \\
&\delta_{f} = \mV/m_{f},  \quad \delta_{Z}=\mV/m_{Z}, \\
&e=\frac{g_{2}g_y}{\sqrt{g_{2}^{2}+g_y^{2}}}, \quad c_{w} = \frac{g_{2}}{\sqrt{g_{2}^{2}+g_y^{2}}}.
\end{align}
$N_{f}$ stands for the color number and is equal to $3$ for  quarks and $1$ for leptons.
$Q_{f}$, $Y_{f}$, $g_{2}$ and $g_y$ are respectively the electric charge of the fermion $f$ in unit of $e$, the hypercharge of $f$, the gauge coupling constant of $SU(2)_{L}$ and that of $U(1)_{Y}$.
We have also computed the decay widths of the dark photon, at tree-level, into $Zh$
\begin{equation}
\Gamma_{V\rightarrow Zh} = \epsilon^{2} \left(\frac{g_y}{g_{2}} \right)^{2} \frac{g_y^{2}+g_{2}^{2}}{192\pi} \mV \left( \frac{\delta_{Z}^{2}}{1-\delta_{Z}^{2}} \right)^{2} \lambda_{Zh,V}^{3/2} \left( 1 + 12 \lambda_{Zh,V}^{-1} \delta_{Z}^{-2} \right)
\label{eq:VD_decay_rate_Zh}
\end{equation}
where the 2-body phase space function $\lambda_{Zh,V}$ is defined as
\begin{equation}
\lambda_{Zh,V} = \left(1 - \delta_{Z}^{-2} - \delta_{h}^{-2} \right)^{2} - 4 \left(\delta_{Z}\delta_{h}\right)^{-2}
\end{equation}
with  $\delta_{h}=\mV/m_{h}$, and into $WW$
\begin{equation}
\Gamma_{V\rightarrow WW} = \epsilon^{2} \left(\frac{g_y}{g_{2}} \right)^{2} \frac{g_y^{2}+g_{2}^{2}}{192\pi} \mV \left( \frac{\delta_{Z}^{2}}{1-\delta_{Z}^{2}} \right)^{2}  \left( 1 - 4\delta_{W}^{-2} \right)^{3/2} \left( 1 + 20\delta_{W}^{-2} + 12 \delta_{W}^{-4} \right).
\label{eq:VD_decay_rate_WW}
\end{equation}
with $\delta_{W} = \mV/m_{W}$.
We have verified that these expressions agree with the existing literature, e.g. \cite{Bai:2016vca} and \cite{Ekstedt:2016wyi}.

In our calculations, we use the expressions above in the region $\mV < 350$~MeV and $\mV > 2.5$~GeV. For  350~MeV$ < \mV < 2.5$~GeV we use the perturbative tree-level expression in eq.~(\ref{eq:VD_decay_rate_ff}) for the width into leptons, and for the width into hadrons we instead use measurements of $R(s) = \sigma(e^+e^- \to {\rm hadrons})/\sigma(e^+e^- \to \mu^+\mu^-)$, that we extract from~\cite{Curtin:2014cca}, and where $\sqrt{s}$ is the energy of the collision in the center of mass frame.
These scattering processes are dominated by a virtual photon exchange in $s$-channel, therefore the quantum numbers of the final state coincide with those of the dark photon, making their use justified for our purpose. Therefore we write
\beq
\Gamma(V \to {\rm hadrons}) = R(s = \mV^2)\times \Gamma(V \to \mu^+\mu^-),
\label{eq:Width_hadrons}
\eeq
where we take $ \Gamma(V \to \mu^+\mu^-)$ from eq.~(\ref{eq:VD_decay_rate_ff}).
The hadronic decay chains in $e^+ e^- \to$~hadrons end dominantly in charged pions~\cite{Whalley:2003qr,Buschmann:2015awa}, that have BR$(\pi^\pm \to \mu\nu_\mu)>99.9\%$. Therefore, following~\cite{Cirelli:2016rnw}, we assume for simplicity that the $V$ final states, for 350 GeV~$< \mV < 2.5$~GeV, consist $50~\%$ of $\mu\bar{\mu}$ and $50~\%$ of $\nu\bar{\nu}$.\\

\medskip
\small

\bibliographystyle{JHEP}
\bibliography{HeavyDarkU1}
\end{document}